\documentclass[pdflatex,sn-basic]{sn-jnl}

\usepackage{graphicx}%
\usepackage{multirow}%
\usepackage{amsmath,amssymb,amsfonts}%
\usepackage{amsthm}%
\usepackage{mathrsfs}%
\usepackage[title]{appendix}%
\usepackage{xcolor}%
\usepackage{textcomp}%
\usepackage{manyfoot}%
\usepackage{algorithm}%
\usepackage{algorithmicx}%
\usepackage{algpseudocode}%
\usepackage{listings}%
\usepackage{natbib}%
\usepackage{anyfontsize} 
\usepackage{dirtytalk}
\usepackage{subcaption}
\usepackage{lscape}
\usepackage{enumitem}
\usepackage{booktabs}
\usepackage{gensymb}
\usepackage{longtable}

\newcommand{\ariel}{\textsl{Ariel}}
\newcommand{\batman}{\texttt{batman}}

\newcommand{\corot}{\textsl{CoRoT}}
\newcommand{\earthtwo}{\textsl{Earth 2.0}}

\newcommand{\kepler}{\textsl{Kepler}}

\newcommand{\pine}{PINE}

\newcommand{\plato}{PLATO}
\newcommand{\platonium}{\texttt{Platonium}}
\newcommand{\platosim}{\texttt{PlatoSim}}

\newcommand{\psls}{PSLS}

\newcommand{\tess}{\textsl{TESS}}
\newcommand{\tlcm}{\textsc{TLCM}}
\newcommand{\deltanu}{\delta\nu_{\ell = 1 \mathrm{, max}}}

\newcommand*\aap{A\&A}

\newcommand*\ao{Appl.~Opt.}

\newcommand*\apjl{ApJ}

\newcommand*\apjs{ApJS}

\newcommand{\rearth}{R$_\mathrm{Earth}$}

\newcommand{\affWRno}{1}
\newcommand{\affWR}{\orgdiv{Institut f\"ur Weltraumforschung}, \orgname{Deutsches Zentrum f\"ur Luft- und Raumfahrt}, \orgaddress{\street{Rutherfordstr. 2}, \city{Berlin}, \postcode{12489}, \state{Berlin}, \country{Germany}}}

\newcommand{\affDLRno}{2}
\newcommand{\affDLR}{\orgname{Deutsches Zentrum f\"ur Luft- und Raumfahrt}, \orgaddress{\street{Markgrafenstra{\ss}e 37}, \city{Berlin}, \postcode{10117}, \state{Berlin}, \country{Germany}}}

\newcommand{\affFUBno}{3}
\newcommand{\affFUB}{\orgdiv{Institut f\"ur Geologische Wissenschaften}, \orgname{Freie Universit\"at Berlin}, \orgaddress{\street{Malteserstra{\ss}e 74-100}, \city{Berlin}, \postcode{12249}, \state{Berlin}, \country{Germany}}}

\newcommand{\affLIRAno}{4}
\newcommand{\affLIRA}{\orgdiv{LIRA}, \orgname{Observatoire de Paris, Universit\'e PSL, CNRS, Sorbonne Universit\'e, Universit\'e Paris Diderot, Sorbonne Paris Cit\'e}, \orgaddress{5 place Jules Janssen}, \postcode{92195}, \state{Meudon}, \country{France}}

\newcommand{\affPadovaOsservatoriono}{5}
\newcommand{\affPadovaOsservatorio}{\orgdiv{INAF}, \orgname{Osservatorio Astronomico di Padova}, \orgaddress{vicolo dell'Osservatorio 5}, \postcode{35122}, \state{Padova}, \country{Italy}}

\newcommand{\affPadovaUniversitano}{6}
\newcommand{\affPadovaUniversita}{\orgdiv{Centro di Ateneo di Studi e Attivit\`a Spaziali “Giuseppe Colombo” (CISAS)}, \orgname{Universit\`a degli Studi di Padova}, \orgaddress{via Venezia 1}, \postcode{35131}, \state{Padova}, \country{Italy}}

\newcommand{\affESTECno}{7}
\newcommand{\affESTEC}{\orgdiv{ESTEC}, \orgname{European Space Agency}, \orgaddress{Keplerlaan 1}, \postcode{2201, AZ}, \state{Noordwijk}, \country{The Netherlands}}

\newcommand{\affESACno}{8}
\newcommand{\affESAC}{\orgdiv{ATG Science \& Engineering}, \orgname{European Space Agency (ESA), ESAC}, \orgaddress{Camino bajo del Castillo, s/n} \postcode{28692}, \state{Urbanizaci{\'o}n Villafranca del Castillo}, \country{Spain}}

\newcommand{\affKULno}{9}
\newcommand{\affKUL}{\orgdiv{Institute of Astronomy}, \orgname{KU Leuven}, \orgaddress{Celestijnenlaan 200D}, \postcode{3001}, \state{Leuven}, \country{Belgium}}

\newcommand{\affOxfordno}{10}
\newcommand{\affOxford}{\orgdiv{Department of Physics}, \orgname{University of Oxford}, \state{Oxford}, \country{United Kingdom}}

\newcommand{\affMPSSRno}{11}
\newcommand{\affMPSSR}{\orgname{Max-Planck-Institut f\"ur Sonnensystemforschung}, \orgaddress{Justus-von-Liebig-Weg 3}, \postcode{37077}, \state{G\"ottingen}, \country{Germany}}

\newcommand{\affIAAno}{12}
\newcommand{\affIAA}{\orgname{Instituto de Astrof{\'i}sica de Andaluc{\'i}a (IAA-CSIC)}, \orgaddress{Glorieta de la Astronom\'ia s/n}, \postcode{18008}, \state{Granada}, \country{Spain}}

\newcommand{\affIASno}{13}
\newcommand{\affIAS}{\orgdiv{Institut d'Astrophysique Spatiale}, \orgname{UMR8617, Universit\'e Paris-Saclay}, \postcode{91405}, \state{Orsay Cedex}, \country{France}}

\newcommand{\affUWano}{14}
\newcommand{\affUWa}{\orgdiv{Department of Physics}, \orgname{University of Warwick}, \orgaddress{Gibbet Hill Road}, \state{Coventry} \postcode{CV4 7AL}, \country{UK}}

\newcommand{\affLiegeno}{15}
\newcommand{\affLiege}{\orgdiv{Space sciences, Technologies and Astrophysics Research (STAR)}, \orgname{CSL (Centre Spatial de Li\`ege)}, \orgaddress{avenue Pr\'e-Aily 19}, \postcode{4031}, \country{Belgium}}

\newcommand{\affKonkolyno}{16}
\newcommand{\affKonkoly}{\orgdiv{Konkoly Observatory}, 
\orgname{HUN-REN Research Centre for Astronomy and Earth Sciences, Konkoly Observatory, MTA Centre of Excellence}, 
\orgaddress{Konkoly Thege Mikl\'os \'ut 15-17.}, \postcode{H-1121}, \state{Budapest}, \country{Hungary}}

\newcommand{\affELTEno}{17}
\newcommand{\affELTE}{\orgdiv{Institute of Physics and Astronomy}, 
\orgname{E\"otv\"os L\'or\'and University}, 
\orgaddress{P\'zm\'any P\'eter s\'et\'any 1/A}, \postcode{H-1117}, \state{Budapest}, \country{Hungary}}

\newcommand{\affBhamno}{18}
\newcommand{\affBham}{\orgdiv{School of Physics and Astronomy}, \orgname{University of Birmingham}, \state{Birmingham}, \postcode{B15 2TT}, \country{United Kingdom}}

\newcommand{\affCAUPno}{19}
\newcommand{\affCAUP}{\orgdiv{Instituto de Astrof\'isica e Ci\^encias do Espa\c{c}o}, \orgname{Universidade do Porto, CAUP}, \orgaddress{Rua das Estrelas}, \postcode{PT4150-762}, \state{Porto}, \country{Portugal}}

\newcommand{\affIoAno}{20}
\newcommand{\affIoA}{\orgdiv{Institute of Astronomy}, \orgname{University of Cambridge}, \orgaddress{Madingley Rd}, \state{Cambridge} \postcode{CB3 0HA}, \country{UK}}

\newcommand{\affLAMno}{21}
\newcommand{\affLAM}{\orgdiv{Aix Marseille Univ, CNRS, CNES, LAM} \orgname{Aix Marseille University}, \orgaddress{38 rue Fr\'ed\'eric Joliot-Curie} \postcode{13388} \state{Marseille}, \country{France}}

\newcommand{\affIAPno}{22}
\newcommand{\affIAP}{\orgname{Leibniz-Institut f\"ur Astrophysik Potsdam (AIP)}, \orgaddress{An der Sternwarte 16}, \postcode{14482} \state{Potsdam}, \country{Germany}}

\newcommand{\affIACno}{23}
\newcommand{\affIAC}{\orgname{Instituto de Astrof{\'i}sica de Canarias (IAC)}, \orgaddress{c/V{\'i}a L{\'a}ctea s/n}, \postcode{38205}, \state{La Laguna (Tenerife)}, \country{Spain}}

\newcommand{\affINAFPlanetologiano}{24}
\newcommand{\affINAFPlanetologia}{\orgdiv{INAF} \orgname{Istituto di Astrofisica e Planetologia Spaziali}, \orgaddress{Via del Fosso del Cavaliere, 100} \postcode{00100} \state{Roma}, \country{Italy}}

\newcommand{\affSRONno}{25}
\newcommand{\affSRON}{\orgdiv{SRON}, \orgname{Netherlands Institute for Space Research}, \orgaddress{Niels Bohrweg 4}, \postcode{2333 CA}, \state{Leiden}, \country{The Netherlands}}

\newcommand{\affINGno}{26}
\newcommand{\affING}{\orgname{Isaac Newton Group of Telescopes}, \orgaddress{Apartado de correos 321}, \postcode{38700}, \state{Santa Cruz de La Palma}, \country{Spain}}

\newcommand{\affCataniano}{27}
\newcommand{\affCatania}{\orgdiv{INAF}, \orgname{Osservatorio Astrofisico di Catania}, \orgaddress{Via S. Sofia 78}, \postcode{95123}, \state{Catania}, \country{Italy}}

\newcommand{\affRIUno}{28}
\newcommand{\affRIU}{\orgdiv{Rheinisches Institut f\"ur Umweltforschung}, \orgname{Planetenforschung}, \state{Cologne}, \country{Germany}}

\newcommand{\affNiceno}{29}
\newcommand{\affNice}{\orgdiv{Observatoire de la C{\^o}te d'Azur}, \orgname{Universit{\'e} C{\^o}te d'Azur, CNRS, Laboratoire Lagrange}, \orgaddress{Bd de l'Observatoire, CS 34229}, \postcode{06304} \state{Nice cedex 4}, \country{France}}

\newcommand{\affTelespaziono}{30}
\newcommand{\affTelespazio}{\orgname{Telespazio UK for} \orgname{European Space Agency, ESAC}, \orgaddress{Camino Bajo del Castillo, s/n}, \state{Villanueva de la Ca\~nada}, \postcode{28692}, \country{Spain}}

\newcommand{\affGenevaObsno}{31}
\newcommand{\affGenevaObs}{\orgname{Observatoire Astronomique de l'Universit\'e de Gen\`eve}, \orgaddress{Chemin Pegasi 51}, \state{Versoix}, \country{Switzerland}}

\newcommand{\affPadovaFisicano}{32}
\newcommand{\affPAdovaFisica}{\orgdiv{Dipartimento di Fisica e Astronomia “Galileo Galilei”}, \orgname{Universit\`a degli Studi di Padova}, \orgaddress{Vicolo dell’Osservatorio 3}, \postcode{35122}, \state{Padova}, \country{Italy}}

\newcommand{\affINAFRomano}{33}
\newcommand{\affINAFRoma}{\orgdiv{INAF}, \orgname{Osservatorio Astronomico di Roma}, \orgaddress{Via Frascati, 33}, \postcode{00078}, \state{Monte Porzio Catone (RM)}, \country{Italy}}

\newcommand{\affASIno}{34}
\newcommand{\affASI}{\orgname{SSDC-ASI}, \orgaddress{Via del Politecnico, snc}, \postcode{00133}, \state{Roma}, \country{Italy}}

\newcommand{\affCABno}{35}
\newcommand{\affCAB}{\orgdiv{Centro de Astrobiolog\'ia (CAB)}, \orgname{CSIC-INTA, ESAC Campus}, \orgaddress{Camino bajo del Castillo s/n}, \postcode{28692}, \state{Villanueva de la Ca\~nada, Madrid}, \country{Spain}}

\newcommand{\affKeeleno}{36}
\newcommand{\affKeele}{\orgdiv{Astrophysics Group}, \orgname{Keele University}, \orgaddress{Staffordshire} \postcode{ST5 5BG}, \country{UK}}

\newcommand{\affBolognano}{37}
\newcommand{\affBologna}{\orgdiv{Dipartimento di Fisica e Astronomia}, \orgname{Universit{\`a} degli Studi di Bologna, INAF - Osservatorio di Astrofisica e Scienza dello Spazio di Bologna}, \orgaddress{Via Gobetti 93/2}, \postcode{40129}, \state{Bologna}, \country{Italy}}

\newcommand{\affSTARno}{38}
\newcommand{\affSTAR}{\orgdiv{Space Sciences, Technologies, and Astrophysics Research (STAR) Institute}, \orgname{Universit{\'e} de Li{\`e}ge, Quartier Agora, B{\^a}t B5c}, \orgaddress{All{\'e}e du 6 ao{\^u}t, 19c}, \postcode{4000}, \state{Li{\`e}ge}, \country{Belgium}}

\newcommand{\affSERCOno}{39}
\newcommand{\affSERCO}{\orgname{SERCO for} \orgname{European Space Agency, ESAC}, \orgaddress{Camino Bajo del Castillo, s/n}, \state{Villanueva de la Ca\~nada}, \postcode{28692}, \country{Spain}}

\newcommand{\affUVano}{40}
\newcommand{\affUVa}{\orgdiv{Departament d’Astronomia i Astrof{\'i}sica}, \orgname{Universitat de Val{\`e}ncia}, \orgaddress{Av. Vicent Andr{\'e}s Estell{\'e}s, 19}, \postcode{46120}, \state{Burjassot}, \country{Spain}}

\newcommand{\affOVano}{41}
\newcommand{\affOVa}{\orgdiv{Observatori Astron{\`o}mic}, \orgname{Universitat de Val{\`e}ncia}, \orgaddress{c/ Cat. Jos{\'e} Beltr{\'a}n 2}, \postcode{46980}, \state{Paterna}, \country{Spain}}

\newcommand{\affINTAno}{42}
\newcommand{\affINTA}{\orgdiv{Departamento de {\'O}ptica Espacial - Subdirecci{\'o}n General de Sistemas Espaciales}, \orgname{Instituto Nacional de T{\'e}cnica Aeroespacial (INTA)}, \orgaddress{Ctra. Ajalvir Km. 4}, \postcode{28850}, \state{Torrej{\'o}n de Ardoz, Madrid}, \country{Spain}}

\newcommand{\affUCLno}{43}
\newcommand{\affUCL}{\orgdiv{University College London}, \orgname{Mullard Space Science Laboratory}, \orgaddress{Holmbury St. Mary}, \state{Dorking}, \postcode{RH5 6NT}, \country{UK}}

\newcommand{\affInnsbruckno}{44}
\newcommand{\affInnsbruck}{\orgdiv{Universit{\"a}t Innsbruck}, \orgname{Institut f\"ur Astro- und Teilchenphysik}, \orgaddress{Technikerstra{\ss}e 25}, \postcode{6020}, \state{Innsbruck}, \country{Austria}}

\begin{document}

\title[PLATO Performance]{Assessment of PLATO Science Performance}



\author*[\affWRno]{\fnm{Juan} \sur{Cabrera}}\email{juan.cabrera@dlr.de}

\author[\affDLRno,\affFUBno]{\fnm{Heike} \sur{Rauer}}

\author[\affLIRAno]{\fnm{R\'eza} \sur{Samadi}}

\author[\affPadovaOsservatoriono,\affPadovaUniversitano]{\fnm{Valerio} \sur{Nascimbeni}}

\author[\affWRno]{\fnm{Anko} \sur{B\"orner}}

\author[\affWRno]{\fnm{Denis} \sur{Grie{\ss}bach}}

\author[\affWRno]{\fnm{Carsten} \sur{Paproth}}

\author[\affWRno]{\fnm{Martin} \sur{Pertena\"is}}

\author[\affESTECno]{\fnm{Sami-M.} \sur{Niemi}}

\author[\affWRno]{\fnm{Szil{\'a}rd} \sur{Csizmadia}}

% ...


\abstract{
The \plato~mission is scheduled for launch early 2027. 
In this paper we present an overview of the performance drivers 
for the mission at the time where all flight models of the cameras 
have been tested and integrated on the optical bench.
The \plato~consortium needs an estimate of the planet detection yield 
to dimension the ground-based radial velocity follow-up resources. 
We provide updated estimates on the yield of planet detections that 
can be expected from the mission under certain assumptions.
As of today, large uncertainties remain on the planet occurrence rates, 
especially for small planets in long-period orbits, and on our ability to 
detect these planets in the presence of stellar variability and instrumental 
noise. 
To partially overcome these limitations, we compare results using 
different planet occurrence rates, detectability rates, and we include
an estimate on the expected contribution of stellar variability 
to the noise budget.
The final detection yield of \plato~will provide constraints to planet 
occurrence rates which in turn will help constraining planet formation models.}




\maketitle

\section{Introduction}
\label{section:introduction}

\plato~\citep[PLAnetary Transits and Oscillations of stars,][]{catala2006} is 
the third mission of the medium class in the Cosmic Vision 2015-2025 program 
of ESA scheduled for launch early 2027~\citep{rauer2025}. 
The overall science goals of the \plato~mission are to answer the
following questions~\citep{rauer2014}:
\begin{itemize}
    \item How do planets and planetary systems form and evolve?
    \item Is our Solar System special or are there other systems like ours?
    \item Are there potentially habitable planets? 
\end{itemize}
The strategy chosen to answer these questions is the analysis of several tens of 
thousands of stellar light curves to characterise the stellar oscillation frequencies 
in such a way that precise and accurate stellar parameters can be derived, and 
to search for and precisely characterise transiting extrasolar 
planets, down to the size of the Earth, with orbital periods up to around 1 year.

The mission concept, the flight segment including payload and spacecraft, and 
the ground segment have been described in detail in~\citet{rauer2025}.
The mentioned paper provides an updated report of the science goals together with
the mission and instrument concepts at the time of the Critical Milestone Review 
of the project.

In this paper we review the main parameters driving the performance of the 
\plato~Mission and estimate the expected planet yield and the expected
accuracy and precision reached in planetary characterisation. 
We use information about the mission at the time where all flight models 
of the cameras have been tested and integrated on the optical bench.
There remain tests at spacecraft level whose results will be known in 
before the end of 2026.

\section{Drivers for Performance}
\label{section:drivers_for_performance}

The primary scientific drivers for the design of \plato~are i) the ability 
to detect terrestrial exoplanets, down to the size of the Earth, at orbits 
up to the habitable zone of solar-type stars and to characterise their bulk 
properties: radius from space-borne photometry and mass from additional 
ground-based spectroscopic observations, and ii) the ability to characterise 
solar-like stars with asteroseismology, determining accurate values for their 
masses, radii, and ages.

More concretely, the most demanding scientific drivers for mission performance 
are:
\begin{enumerate}
    \item The ability to measure a sample of more than 100 (goal 400) 
    exoplanets characterised for their orbits, radii (accuracy better 
    than 3\% for planets orbiting stars brighter than magnitude 10 and
    5\% for planets orbiting stars brighter than magnitude 11), and
    masses (accuracy better than 10\% for planets orbiting stars brighter than
    magnitude 11) over a wide range of physical masses and mean
    densities, including more than 5 (goal 30) (super-)Earths in the
    habitable zone of solar-like stars.
    \item The ability to derive accurate ages (10\% accuracy) for
    bright planet-hosting stars from asteroseismology~\citep{aerts2021}.
\end{enumerate}

\subsection{Photometric precision}
\label{subsection:photometric_precision}

The transit of the Earth around the Sun produces a drop in flux of about 80 ppm in 
the photometry which lasts around 13h.
For simplicity, we are considering that the transit depth is just the square of 
the radius ratio between the Earth and the Sun, a central transit, and ignoring 
the contribution of limb darkening. 
For reference, a central transit of the Earth around the Sun including limb 
darkening produces a signal of about 
100 ppm~\citep[e.g.][]{csizmadia2013a,heller2019}.
The number of planets found by \plato~will depend on the number of stars 
observed with such a signal-to-noise ratio that allows the detection of planets
of given characteristics.
Signal-to-noise ratios (S/N) larger than 7.1 are empirically required to have a 
reasonable chance (e.g. a detection rate of 50\%) of detecting the transit of a planet with 
unknown properties \citep{jenkins1996,fressin2013}. 
This same S/N threshold has been shown to limit the false positive rate to 
1/100,000 \citep{jenkins2002b} in white-noise dominated light curves.

In planet detection, the signal is the depth of the transit, which has to be 
compared with the noise in the light curve at the time-scale of the transit 
duration.
If the noise properties allow for it, the S/N increases with the square root of the 
number of transits 
observed~\citep[but see the impact of correlated noise, e.g. in][]{pont2006}.
However, the signal can decrease by, e.g., dilution from background sources 
or limited duty cycle.
Furthermore, stellar variability~\citep[see e.g.][]{jenkins2002a} and instrumental 
noise sources impact both detectability and our ability to validate genuine 
planetary signals~\citep{bryson2021}.
In the literature, metrics like the expected detection 
statistic~\citep[MES, see][]{christiansen2016} and the signal detection 
efficiency \citep[SDE, see][]{kovacs2002,2019A&A...623A..39H} are defined to quantify the 
detectability of a given signal.

The value of 80 ppm in 1h\footnote{Rigorously speaking, 80 ppm$\cdot$h$^{1/2}$ in the Fourier
  domain, as derived later.} is mentioned because it is representative of the photometric quality 
required to detect planets like the Earth in a 1-year orbit. 
But for the~\plato~design we have more demanding noise requirements derived from 
the need to precisely characterise planets and stars.
The benchmark chosen is to achieve a noise-to-signal ratio (NSR) of 50 ppm in 1h
for a G0V star of magnitude 11 (see Section~\ref{section:noise_budget}) in order 
to reach the planet characterisation requirements~\citep[see e.g.][]{csizmadia2023}.

\subsection{Field of view}
\label{subsection:field_of_view}

\begin{figure}
  \centering
  \includegraphics[%
    width=0.9\linewidth,%
    height=0.5\textheight,%
    keepaspectratio]{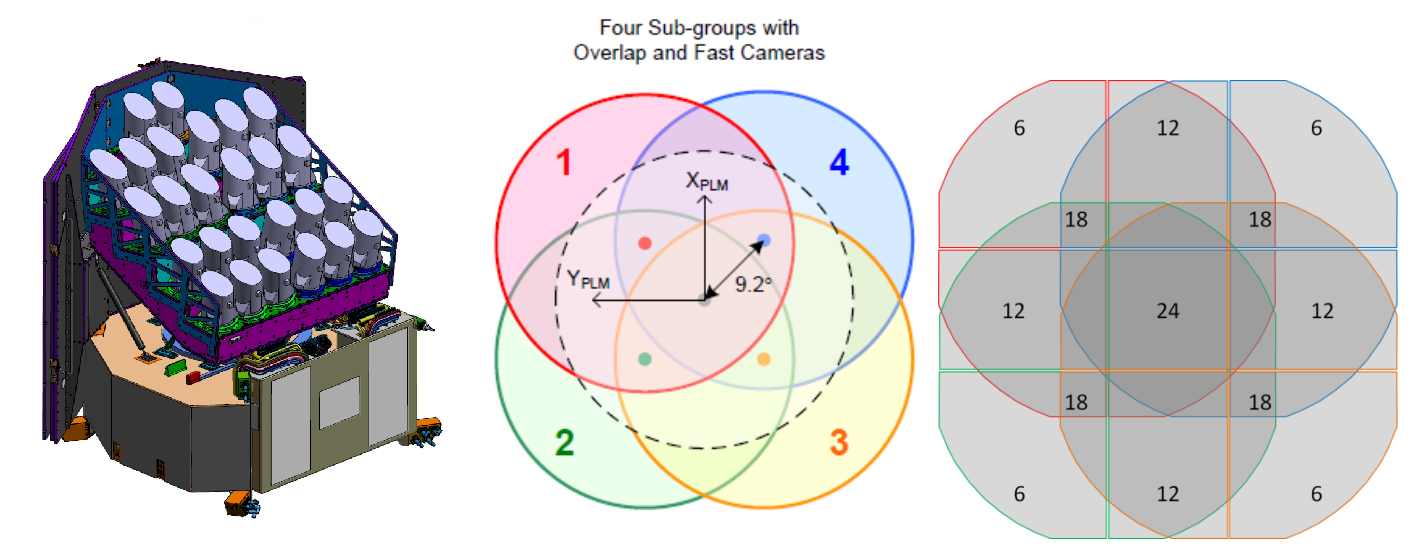}
  \caption{Sketch of the camera arrangement on the optical
    bench. The groups of N-CAMs are 9.2\degree\,offset from 
    the Z axis of the payload, which is co-aligned with the F-CAMs. Credit: ESA.}
  \label{figure:plato_fov}
\end{figure}

Reflecting telescopes are mostly limited to fields of view of maximum 
10$^{\circ}$ diameter. 
Therefore, a very large number of them would be needed to reach the 
required field of view for~\plato. 
Using refractive lenses instead offers the possibility for a much larger 
field of view per camera ($>35^{\circ}$ diameter, \citealt{ragazzoni2016,magrin2018}). 
However, using a single big refractive camera was also not a realistic 
option as it would lead to both a very large detector and extremely heavy 
lenses, neither of which are compatible with the constraints of a space 
mission. 
The \tess~mission~\citep{ricker2015} and the~\earthtwo~mission~\citep{ge2022} 
have also chosen a concept of cameras, in contrast to \corot~or \kepler, 
for the same reasons.
However, the novelty of the \plato~concept is that the photometric
accuracy is enhanced by observing stars simultaneously with several
cameras (between 6 and 26, depending on the position of the star in
the instrument field of view).
The 26 cameras of the payload are divided into two fast-cameras (F-CAMs),
responsible for the fine pointing (see Section~\ref{subsection:sampling_cadence}), 
and 24 normal-cameras (N-CAMs) which in turn are divided into four groups 
of 6 cameras each.
The F-CAMs point along the Z axis of the payload and control the
instantaneous pointing direction of the instrument.
Each group of six N-CAMs points to a slightly different direction with
respect to the instrument, with an offset of 9.2 degrees (see
Fig.~\ref{figure:plato_fov}). 
The offset is a compromise between the instrument field of view and
the size of the region where the 24 N-CAMs (and the two F-CAMs) overlap,
acquiring the maximum photometric precision.

The field of view per camera is a compromise between the scientific needs 
and the feasibility constraints, including focal length and detector size, 
but also manufacturing of lenses, thermal properties, power budget, mass budget, 
and 
schedule~\citep[e.g.][]{ragazzoni2016,magrin2010,magrin2016b,magrin2016a,magrin2020}.
Each PLATO camera features four CCDs with 4510x4510 pixels of 18 micron size each.
The CCD270, especially designed and manufactured by Teledyne e2V for the mission, has 
a large format (8 cm x 8 cm), is back-illuminated, and operates at 3 MHz 
pixel rate~\citep{verhoeve2016}.
The CCDs of the N-CAMs will be read in full-frame mode while the CCDs of the 
F-CAMs will be read in frame-transfer mode, complying to the latency requirements
of the fine guidance system (FGS) and attitude and orbit control system (AOCS) 
performances.
The pixel scale is 15.0 arcsec/pixel on-axis, constraining the field of view 
size for a given focal length.
The focal length of the \plato~cameras is 247.5 mm and the F-number is f/2, 
mostly driven by feasibility constraints, and the entrance pupil has a size 
of 12 cm.
Additional performance drivers for the optics are the point spread function
size~\citep{borsa2022} and the alignment of the lenses~\citep{novi2022}.
The results of the vacuum test campaign have been reported in~\cite{greggio2024}
and the focus performance can be followed in~\citet{pertenais2025}.
The achieved total field of view is 2132\,deg$^2$, resulting from the 
combination of the four N-CAM groups, each camera providing a field of view 
of 1037\,deg$^2$. 
The F-CAM field of view is effectively half of this value, as their CCDs are 
operated in frame-transfer mode~\citep{pertenais2021}. 
Table~\ref{table:parameters}~includes a useful list of parameters 
characterising the payload~\citep[see also][]{rauer2025}.

The stars in the field of view have been classified into 4 different samples
(P1, P2, P4, P5) depending on their nature and the performance achieved by
the payload, which depends on the magnitude and the position on the field
of view. 
More information on the stellar samples can be found in the \plato~Input 
Catalogue~\citep[PIC,][and Montalto et al. in prep.]{montalto2021,prisinzano2026}.

\subsection{Absolute pointing error}
\label{subsection:ape}

\plato\,has a multi-telescope approach; therefore the final number of stars 
observed brighter than magnitude V 11 (R-SCI-090) observed with a given noise 
limit is a compromise between the total size of the instrument FOV and the 
degree of overlap between camera FOVs. 

There is an additional hard constraint related to the spacecraft power budget 
and Sun exclusion angle requirements. 
In order to keep the solar panels aligned towards the Sun and prevent 
solar-illumination of the payload, the satellite is rotated around the mean 
line of sight by 90 degrees every approximately 3 months (quarterly roll). 
The science case of \plato\,requires observing the same stars for long periods 
of time with minimum interruptions (see below). 
In order to avoid stellar losses every quarterly roll, the payload must keep 
90 degree symmetry. 
The final size of the instrument FOV is given i) by shape chosen to fulfil 
the the 90 degree requirement for the symmetry of the payload around the mean 
line of sight; and ii) by the amount of overlap between the camera FOVs, 
which is measured as the aperture angle of the groups of N-CAMs with respect 
to the mean line of sight (see Fig.~\ref{figure:plato_fov}.
The positioning of the cameras on the optical bench and the alignment of 
the camera interface with the optical bench is responsibility of the 
industrial prime contractor. 
The tolerances of the alignment of the cameras on the optical bench are 
described by the absolute pointing error requirements, which are of 
the order of 4.5 arcmin half-cone angle at 99.7\% confidence level in
the transverse direction and 9 arcmin half-cone angle at 99.7\% confidence
level around the line of sight. 
The tolerances in the repointing performance after the quarterly rolls 
are given by the relative pointing error, which shall not exceed 3 arcsec
half-cone angle in the transversal direction and 6 arcsec around the 
line of sight.

The NSR computation for the PIC assumes that stars fall on silicon for 
perfect alignment (absolute pointing error set to zero). 
We account for uncertainties on the size of the field of view and on the 
pointing performance by providing NSR values for targets that do not fall 
on silicon with nominal pointing (zero absolute pointing error), but which 
could fall on silicon considering possible in-flight camera alignment 
(accounting for some margins). 
To indicate that these stars might not fall on silicon, we indicate 
that they are observed with zero cameras (the parameter 'EOLnCameraObsNCAM\_R' 
is set to zero).

\begin{figure}
  \centering
  \includegraphics[%
    width=0.9\linewidth,%
    height=0.5\textheight,%
    keepaspectratio]{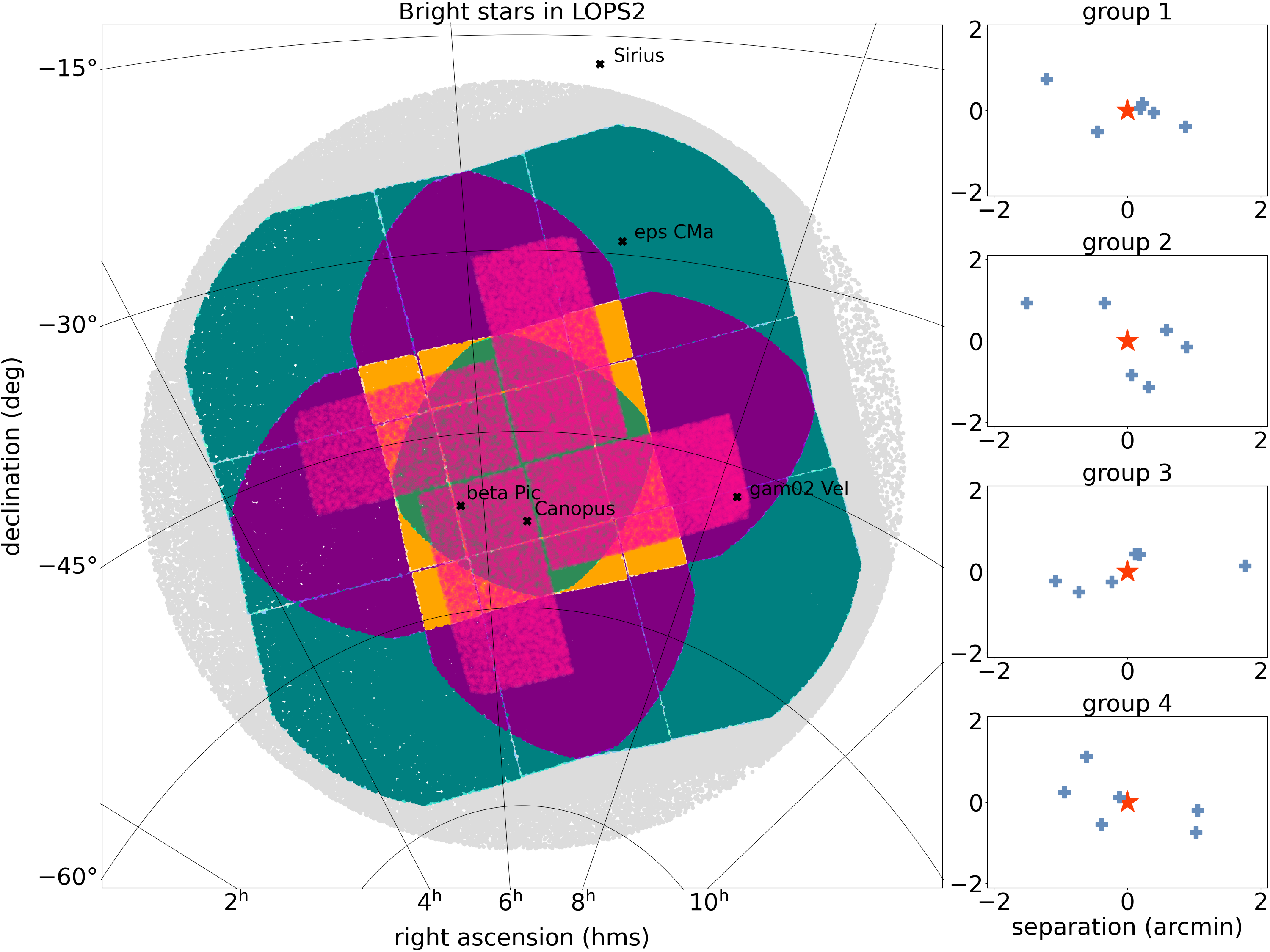}
  \caption{Camera footprint on the sky, including the expected in-flight absolute pointing error.
  Different colours represent different number of cameras observing the same region of the sky.
  Grey stars belong to the PIC, but they will not be observed in LOPS2 with the nominal 
  pointing error budget.
  On the right panel there is a detail of the misalignment of the N-CAMs with respect to the 
  commanded attitude.}
  \label{figure:footprint}
\end{figure}

The absolute pointing error has been measured on the optical bench and
the expected in-flight values are reported, as quaternions in Table~\ref{table:ape}, 
where the quaternions follow the convention in~\citet{jannsen2024}.
The mean boresight of all N-CAMs coincides with LOPS2 as defined in~\citep{nascimbeni2025},
see Fig.~\ref{figure:footprint}.

\begin{landscape}
\begin{longtable}[c]{cccc}
\caption{Expected pointing direction of the camera boresight in-flight.\label{table:ape}}\\
\hline
camera & right ascension & declination & quaternion \\ 
\hline\hline
\multicolumn{4}{l}{Group 1 of N-CAMs}                  \\
FM5    & 108.164043      &  -51.943770 & [0.13271421,-0.63482432,0.70050065,0.29779835]  \\
FM1    & 108.166884      &  -51.947503 & [0.13246165,-0.63430144,0.70098847,0.29787707]  \\ 
FM2    & 108.174795      &  -51.953277 & [0.13247320,-0.63448197,0.70084723,0.29781980]  \\ 
FM3    & 108.163490      &  -51.945812 & [0.13263489,-0.63465098,0.70066554,0.29781524]  \\
FM4    & 108.140078      &  -51.933928 & [0.13272242,-0.63445404,0.70079826,0.29788358]  \\
PFM    & 108.152804      &  -51.955366 & [0.13240118,-0.63406386,0.70123354,0.29783296]  \\ \hline
\multicolumn{4}{l}{Group 2 of N-CAMs}                  \\
FM25   & 101.439703      &  -39.806375 & [0.21306021,-0.60117055,0.67726771,0.36675299]  \\
FM12   & 101.479714      &  -39.824287 & [0.21273255,-0.60121368,0.67731806,0.36677951]  \\
FM13   & 101.470184      &  -39.840820 & [0.21270110,-0.60120463,0.67740788,0.36664669]  \\
FM14   & 101.459204      &  -39.806401 & [0.21313044,-0.60153092,0.67694779,0.36671195]  \\
FM11   & 101.465939      &  -39.835751 & [0.21283605,-0.60135417,0.67725006,0.36661469]  \\
FM10   & 101.474586      &  -39.817449 & [0.21291878,-0.60143006,0.67709210,0.36673392]  \\ \hline
\multicolumn{4}{l}{Group 3 of N-CAMs}                  \\
FM17   &  84.570880      &  -42.621982 & [0.28204342,-0.57825947,0.71006700,0.28613344]  \\ 
FM18   &  84.576715      &  -42.610576 & [0.28183041,-0.57761941,0.71053623,0.28647112]  \\
FM20   &  84.562629      &  -42.626034 & [0.28134965,-0.57650089,0.71151381,0.28677028]  \\
FM15   &  84.604311      &  -42.615310 & [0.28172392,-0.57777379,0.71043210,0.28652280]  \\
FM16   &  84.556807      &  -42.621647 & [0.28182621,-0.57754053,0.71065037,0.28635114]  \\
FM21   &  84.577804      &  -42.610717 & [0.28190383,-0.57781730,0.71037595,0.28639729]  \\ \hline
\multicolumn{4}{l}{Group 4 of N-CAMs}                  \\
FM6    &  86.942900      &  -55.515981 & [0.20141800,-0.61069991,0.73429401,0.21745969]  \\ 
FM7    &  86.947362      &  -55.504925 & [0.20148568,-0.61075588,0.73421025,0.21752261]  \\
FM24   &  86.939089      &  -55.488411 & [0.20170730,-0.61105474,0.73390593,0.21750486]  \\ 
FM9    &  86.966999      &  -55.510306 & [0.20144016,-0.61096562,0.73405384,0.21750364]  \\
FM8    &  86.933579      &  -55.502773 & [0.20166715,-0.61114718,0.73387732,0.21737884]  \\ 
FM22   &  86.966520      &  -55.519247 & [0.20140457,-0.61102507,0.73403445,0.21743503]  \\ \hline
\multicolumn{4}{l}{Group F-CAMs}                  \\
FM19 (blue) & 95.292701 &   -47.900669 & [0.20773163,-0.60768874,0.70831224,0.29301149]  \\
FM23 (red)  & 95.330310 &   -47.893590 & [0.20770997,-0.60800622,0.70801049,0.29309751]  \\ \hline
\end{longtable}
\end{landscape}

\subsection{Duty cycle}
\label{subsection:duty_cycle}

Finally, the detection of planets in a 1-year orbit requires long observation 
baselines and high duty cycles. 
Naively, the probability to detect $N$ planetary transits can be calculated by 
$p_N = (d_f)^N$ with $d_f$ being the duty cycle.
The threshold of 80\% probability of detection is achieved with 93\% duty cycle
(for N=3) or 7\% gap allocation, which is the baseline for~\plato.
For comparison, the global duty cycle of~\corot~was 90\%~\citep{baglin2006}, 
and a similar value was reached by~\kepler~\citep{garcia2014b,lissauer2024}.
The impact of gaps on science depends on the duration of the 
gaps, which can range from few minutes interruptions caused by e.g. rotation
of solar panels to a few days caused by software failures and spacecraft safe 
modes~\citep[see also the discussion in][]{ballot2011}.

Uninterrupted, long observation baselines with excellent photometric quality 
require observing from space.
The need to observe a large number of bright stars requires a large field of 
view, which prevents observing from a low-Earth orbit (because of straylight 
and orbital constraints).
\plato~has chosen to operate in a halo orbit around the L2 point (the second 
Lagrange point of the Earth-Sun system, about 0.01 au away from our planet).
To avoid Sun-illumination of the cameras while keeping the solar panels 
aligned to the Sun, the spacecraft rotates 90 degrees every quarter of a year
(the actual length might range between 84 and 97 days). 
This solution requires a payload design with 90 degree rotation symmetry, 
which is the same concept under which~\kepler~operated.

\subsection{Saturation}
\label{subsection:saturation}

The need to obtain precise photometry of bright stars (V$<$11), 
mitigating the impact of CCD saturation, constrains the size of the entrance 
pupil and the camera cadence.
The brightest non-saturated (or moderately saturated) targets observed by the 
N-CAMs have magnitude 8 and magnitude 4 for the F-CAMs. 
However, there is a requirement that photometric extraction of the light curves 
(on-ground) must be possible regardless of the saturation of the target.
This will be achieved using enlarged apertures for stars between magnitudes
4 and 8 and by alternative methods for the very brightest stars~\citep[e.g.][]{white2017}.

\subsection{Sampling cadence}
\label{subsection:sampling_cadence}

Solar-like oscillations have characteristic frequencies corresponding to
time scales of a few minutes~\citep{garcia2019}. 
Planetary transit durations last for several hours (depending on the
orbital period and the impact parameter), but the ingress and egress
phases, which are critical for planet characterisation, also last only
for a few minutes.

Therefore, a reasonable choice for the sampling rate for PLATO is
25\,s for the N-CAMs (normal cadence, NC).
This provides a Nyquist frequency of 1/50\,s$^{-1}$, enough to constrain solar-like 
oscillations of a few minutes, and to sample the ingress
and egress phases of transiting planets, in order to constrain 
precisely and accurately their radius \citep[see, e.g.][]{csizmadia2013a}.

Planet detection can be achieved with sampling rates of few minutes. 
The faint star channel of~\corot~worked at 512\,s whereas~\kepler~long cadence 
was approximately 30 min.
In~\plato~the cadence of 600\,s (long cadence, LC) is used for most of the
stars in the statistical sample (P5, FGK stars brighter than magnitude 13), 
though it is possible to observe a significant fraction 
of them at a cadence of 50\,s (short cadence, SC, for more than 10\% of the sample).

The main performance driver for the fast cameras is the fine
guidance system performance~\citep[FGS, see][]{griessbach2021}, information 
provided by the payload to the attitude and orbit control system (AOCS) in 
the spacecraft, maintaining a stable pointing in flight and mitigating the 
impact of jitter motion on the photometric 
performance~\citep{boerner2024,bowling2026}.
In order to achieve the required performance, the fast cameras operate at a
cadence of 2.5\,s (high cadence, HC), a sub-multiple of the cadence of the N-CAMs, 
which simplifies the synchronization concept of the payload.
The high cadence of the F-CAMs, together with the readout time and FGS 
on-board processing time, results in a short enough latency required by 
the AOCS performance.
The FGS relies on a pre-selected sample of bright stars in the field of 
view of the F-CAMs. This sample is called the fine guidance \plato\,Input Catalog, 
or fgPIC for short~\citep{cPICfgPIC2026}.

\subsection{Instrument Response}
\label{subsection:instrument_response}

The instrument response is defined as the fraction of photons (input signal) 
that are converted to digital units (output signal) by a given instrument. 
For optical instruments like \plato, it is typically the product of the optics 
transmission (including contributions from particulate and molecular 
contamination, but also coatings and filter responses) and the CCD quantum 
efficiency. 

Instrument responses can be built using different sets of system options, which 
refer to \say{as required}, \say{as designed}, \say{as built}, and 
\say{as simulated} parameters. 
These are collectively called Mission Realizations (MR). The parameters comprising the 
\say{as built} system are derived from spacecraft and payload test campaigns and 
characterisation data. Multiple \say{as built} MRs can be 
included based on different models, e.g. \say{as built EM} (engineering model), 
\say{as built QM} (qualification model), and \say{as built FM} (flight model).
The definition of the \plato~magnitude is done with the \say{as simulated} 
realization while the NSR values in the PIC are computed \say{as required}.
The reason behind these choices is that users should be aware of 
\plato~magnitudes in a realistic realization, expected beginning of life,
while the NSR values are used to verify requirements in realistic worst-case
scenarios, following the minimum requirements.
Finally, \say{as built} is based on test results which become available only 
at a later stage of the mission, while derivation of requirements and
assessment of their feasibility have to be studied in earlier phases of the 
project.

\begin{figure*}
  \centering
  \includegraphics[%
    width=0.9\linewidth,%
    height=0.5\textheight,%
    keepaspectratio]{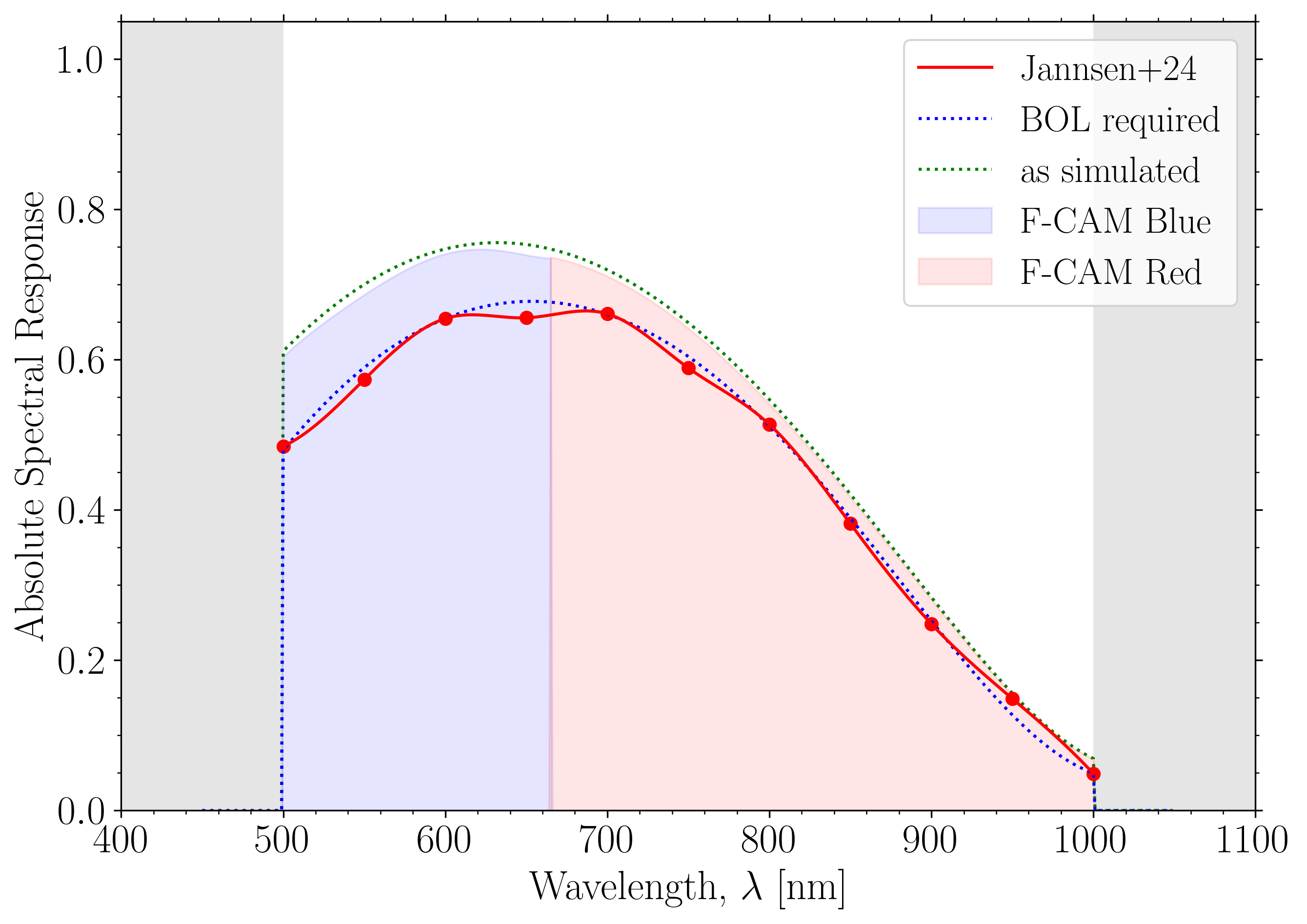}
  \caption{Instrument spectral response for a PLATO camera for the ``as required BOL'' 
  and ``as simulated'' realizations and comparison with the results published 
  in~\citet{jannsen2024}. The in-flight performance of the instrument is expected
  to be better than the requirements. The bandpass of a N-CAM goes from 500 nm to about 1000 nm
  while the F-CAMs have the same optical train, but include filters with a cut-off 
  around 700 nm. Adapted from~\citet{jannsen2024}.}
  \label{figure:instrument_response}
\end{figure*}

The response function of \plato~was presented for the first time 
in~\citet{marchiori2019} with an instrument design that was representative of 
Phase B of the project. 
The study used performance parameters based on end-of-life requirements 
for the normal cameras. 
At the time writing there are measurements of flight 
hardware and we need to review the assumptions used in Phase B. 
A new definition of the PLATO magnitude has been proposed in order to fulfil 
the requirements for the generation of the PIC (Montalto et al. in prep.).
The instrument response provided here is the one that has been used 
for the computation of the \plato~magnitude delivered with the PIC 
(Montalto et al., in prep.).
Figure~\ref{figure:instrument_response} shows the comparison with the
values provided by the camera end-to-end simulator 
\platosim~\citep[][see also next section]{jannsen2024}.
The only difference is on the overlap at 700 nm 
between the blue and red filters.

The requirement for the \plato~filters is to have the ability to observe in 
two spectral bands, red and blue, with a total spectral band overlap of less 
than 30\% and a combined throughput greater than 85\%.
This was translated into requirements such as the effective spectral range 
of the blue F-CAM shall start at $505\pm10$\,nm and shall end at $700\pm10$\,nm.
The effective spectral range of the red F-CAM shall start at $665\pm10$nm and 
shall end with the detector response (around 1000 nm).
The measured throughput of the filters are actually approximately 98.7\% 
in the blue and 99.0\% in the red.

In the \say{as designed} scenario there is no overlap between both
while in the \say{as built} we expect a certain overlap. 
The difference in flux is minimal (few percent in flux, comparable 
with the uncertainties in other parameters, like the CCD quantum
efficiency). 
The in-flight determination of the photometric throughput of the
\plato\,cameras will be done using reference stars distributed across 
the entire field of view. These stars are defined in the calibration 
\plato\,Input Catalog, or cPIC for short~\citep{cPICfgPIC2026}.

In the appendix we provide tabulated instrument response functions for 
the N-CAMs, F-CAM blue and red, in the realizations \say{as required}
for beginning of life (BOL) and end of life (EOL), but also 
\say{as simulated} BOL (the one used to generate the PIC).

\section{Noise budget}
\label{section:noise_budget}

The \plato~Mission Consortium (PMC) needs to demonstrate and verify the 
overall performance of the \plato~project during the different mission phases 
and during instrument development and calibration. 
It must be shown that the science requirements of the mission can be 
met, including instrument verification and mission profile (e.g. 
observing sequences, calibrations).
The PLATO Performance Team (PPT) was formed to coordinate the 
performance study activities across the PMC. 
The PPT maintains and develops simulators to verify the instrument 
performance requirements and to estimate the noise budget in the 
\plato~light curves:
\begin{itemize}
\item The \plato~Simulator\footnote{https://ivs-kuleuven.github.io/PlatoSim3/} (\platosim) 
is a camera end-to-end software tool designed to perform realistic 
simulations of the expected observations starting at pixel 
level~\citep{jannsen2024}. 
Built upon \platosim, the \platonium~toolkit provides a 
mission-level simulator combining the individual camera simulations 
consistently with the mission design and performance~\citep{jannsen2025}.
\item The PLATO Solar Light-Curve Simulator\footnote{https://sites.lesia.obspm.fr/psls/} (\psls) 
generates light curves representative of typical \plato~targets. 
It includes the capability to model the instrumental response: systematic 
errors and random noises representative for the PLATO instrument. 
It allows to simulate solar-like oscillations, stellar granulation, 
magnetic activity, and planetary transits~\citep{samadi2019}.
\item The software \pine~(\plato~Instrument Noise Estimator) produces 
noise-to-signal ratio (NSR) calculations used for signal and noise 
relevant investigations~\citep{boerner2024}. This software models the signal 
flow from a target star to a digital output considering the main optical, 
mechanical, thermal and electrical effects and considers all known noise 
sources. The needed parameters are copied or derived from requirements.
The noise budget values provided to the community with the PIC were 
computed with \pine.
\end{itemize}

\subsection{General Considerations}
\label{subsection:noise_general_considerations}

\begin{table*}
\caption{Equivalent noise values.}             
\label{table:noise_equivalent_values}
\centering
\renewcommand{\arraystretch}{1.2}
\begin{tabular}{p{1.4cm}ccccccc}
\hline
metric     & NSR   & NSR    & NSR     & NSR     & ASD                             & PSD                               & ASD                 \\
           & {\footnotesize ppm 1h} & {\footnotesize ppm 25\,s} & {\footnotesize ppm 2.5\,s} & {\footnotesize ppm 1s} & {\footnotesize ppm$\cdot\mu\mathrm{Hz}^{-1/2}$} & {\footnotesize ppm$^2\cdot\mu\mathrm{Hz}^{-1}$} & {\footnotesize ppm$\cdot$h$^{1/2}$} \\
\hline\hline
{\footnotesize 24 cameras} &  50 &    600 &  1\,900 &  3\,000 &  3.0 &   9.0 &  50 \\
{\footnotesize 1 camera}   & 245 & 2\,900 &  9\,300 & 15\,000 & 14.7 & 216.0 & 245 \\
\hline
\end{tabular}
\end{table*}

At system level, the instrument performance is driven by the
capability of the payload to obtain light curves of a G0V star of
V magnitude 11 with an NSR of 50 ppm in 1h. To be more precise, the \plato~noise 
budget is driven by two requirements, a requirement in the time domain 
defined as an NSR value in a given timescale for a star of a given brightness (50 ppm 
integrating measurements during 1h for a star of V magnitude 11), 
and a requirement in the Fourier domain specified as the maximum 
value for the residual error in the amplitude spectral density (ASD) 
of the light curve (0.68 ppm$\cdot\mu\mathrm{Hz}^{-1/2}$) within a given 
interval of frequencies.

The requirement in the time domain has to be understood as follows. 
We assume that the flux values obtained from a stellar light curve follow a 
discrete random variable of a Gaussian distribution with mean
$<z>$ and standard deviation $\sigma_z$.
Let's suppose that we have obtained $N$ measurements sampled at a
constant cadence $\Delta t$ so the total baseline for the measurement
is $T=N\Delta t$. 
The NSR value is defined as:
\begin{equation}
    \mathrm{NSR}_z = \frac{\sigma_z}{<z>}.
\end{equation} 
We can build new time series from the original by averaging the
measurements over different time scales. 
If the NSR$_z$ value of the distribution is $3\,000$ ppm when
$\Delta t=1$\,s, then the NSR value would be $1\,897$ ppm when averaging
the signal over 2.5\,s, $600$ ppm when averaging over 25\,s, and $50$ ppm
when averaging over 1h. 

The fast Fourier transform of the time series $z$ is a random complex 
variable $Z$ whose real and imaginary parts have standard deviation 
$\sqrt{\frac{\sigma_z^2}{2N}}$, where one has to pay attention to the
exact definition of the fast Fourier transform, as it is not unique.
The power spectral density (PSD) of the time series $z$ is defined as  
$\mathrm{PSD}(z) = |Z|^2\,\Delta t$.
The expected value $\mathrm{E}$ and the square root of the 
variance $\mathrm{VAR}$ of $\mathrm{PSD}(z)$ have the same value, namely 
\begin{equation}
  \mathrm{E}[\mathrm{PSD}(z)] = \sqrt{\mathrm{VAR}[\mathrm{PSD}(z)]}=\Delta t\,\sigma_z^2.
\end{equation}
The amplitude spectral density (ASD) of the time series $z$ is 
defined as the square root of the PSD and follows a Rayleigh
distribution characterised by an expected value and variance
(respectively):
\begin{equation}
  \mathrm{E}[\mathrm{ASD}(z)] = \sqrt{\frac{\pi}{2}\,\Delta t\,\frac{\sigma_z^2}{2}}; \;
  \mathrm{VAR}[\mathrm{ASD}(z)] = \frac{4-\pi}{2} \Delta t \frac{\sigma_z^2}{2}.
\end{equation}
These definitions implicitly assume that we are using the double-sided
definitions for PSD and ASD.
If $z$ follows the random variable described above with $50$ ppm in 1h, 
then the expected value of the PSD is 
$9$~ppm$^2\cdot\mu\mathrm{Hz}^{-1}$ and
the expected value of the ASD is $2.7$~ppm$\cdot\mu\mathrm{Hz}^{-1/2}$
or, as expected, 50~ppm$\cdot\mathrm{h}^{1/2}$.
Table~\ref{table:noise_equivalent_values} provides equivalent values
for the different metrics averaging over 1 or 24 identical cameras 
(under the assumption above of a discrete random variable of a Gaussian distribution).

We have designed \plato~such that the noise budget is dominated by
photon noise from the star for the P1 sample.
Therefore, the requirement is that the total residual error, after all
corrections have been applied (gain correction, temperature drift 
correction, jitter correction, etc.), is less than one third of the
random noise associated with a star of V magnitude 11. 
In the frequency domain we express this requirement as follows:
\begin{itemize}
\item In the range between $40$ mHz (25~s) and $20 \mu$Hz (50 ks or approx. 13.9 hours) the ASD of the
  residual error shall remain below $0.68$~ppm$\cdot\mu\mathrm{Hz}^{-1/2}$.
  The frequency range goes from the sampling rate of the N-CAMs to the
  typical duration of the transit of the Earth around the Sun in
  the habitable zone (around 13h).
\item In the range between $20 \mu$Hz and $3 \mu$Hz (333 ks or approx. 3.9 d) the ASD of the
  residual error shall increase monotonically up to a maximum of
  $50$~ppm$\cdot\mu\mathrm{Hz}^{-1/2}$. 
  This relaxation is needed for the technical feasibility of the
  mission and is justified by the ability of data correction tools to
  correct long-term trends in the data.
\end{itemize}

\subsection{Quick noise model}
\label{subsection:quick_noise_model}

As described above, the noise budget values provided with the 
PIC are computed with \pine~\citep{boerner2024}.
This software models the signal flow from a target star to a digital 
output considering the main optical, mechanical, thermal and electrical 
effects and considers all known noise sources. 
However, in some circumstances we might want to make use of a simple 
approach to have a quick estimate of the NSR of a given target. 
For example, for targets which are not in the PIC (e.g. OBA stars)
or to estimate \plato~performance beyond the nominal pointing 
direction~\citep[long pointing observation in the south direction, LOPS2,][]{nascimbeni2025}.

The analysis of the \pine\,results shows that the expected NSR value of
a \plato~target can be approximated with reasonable accuracy with a model 
that includes jitter, dominating the noise budget in the bright end, 
background and readout noise, dominating the budget in the faint end, and 
photon noise elsewhere. 
This approximation has already been used in the 
literature~\citep{matuszewski2023,eschen2024,rauer2025} for the PIC 1.0,
an older version of the input catalogue. 
Here we provide tabulated values for the model parameters computed with 
PIC 2.2 for reference scenarios BOL and EOL, when the performance of the 
instrument is degraded because of ageing and radiation impact.

Equation~\ref{eq:nsr} shows the model, including the three uncorrelated
noise components for jitter, photon, and background (and readout) noise, 
and the flux at system level $f_s$; the total is expressed in parts per 
million (ppm):
\begin{equation}
\label{eq:nsr}
NSR = \frac{\sqrt{ \sigma_\mathrm{jitter}^2 + \sigma_\mathrm{photon}^2 + \sigma_\mathrm{background}^2}}{f_s} \cdot 10^6.
\end{equation}

The jitter contribution is set to a constant value of $k_j$ in 1 hour
of integration time ($t$, expressed in seconds), but independent of the 
brightness of the star and the number of cameras ($n$) observing the same 
star at system level~\citep[see the assumptions in][]{boerner2024}:
\begin{equation}
\label{eq:sigma_jitter}
\sigma_\mathrm{jitter} = k_j  f_s \sqrt{ 3600/t}.
\end{equation}

The photon noise is, per definition, the square root of the flux
measured at system level:
\begin{equation}
\label{eq:sigma_photon}
\sigma_\mathrm{photon} = \sqrt{f_s}.
\end{equation}

Finally, the background and readout noise is a constant $k_r$ 
that is proportional to the size of the point spread function
(background level) and to the square root of the number of 
exposures integrated (assuming a cadence time of 2.5 s for the
F-CAMs and 25\,s for the N-CAMs):
\begin{equation}
\label{eq:sigma_readout}
\sigma_\mathrm{background} = k_r \sqrt{ t/\mathrm{cadence}} \sqrt{ n}.
\end{equation}

The flux at system level is (the flux collected with $n$ cameras in 
an integration time $t$):
\begin{equation}
\label{eq:flux}
f_s = n \, \frac{t}{\mathrm{cadence}} \, f_\mathrm{ref} \, 10^{ -0.4 ( m - m_\mathrm{ref})},
\end{equation}
with the value of the reference flux $f_\mathrm{ref}$ fixed
at the value computed by \citet{boerner2024}.
The fitting plots are displayed in Fig.~\ref{figure:quick_noise_model_bol} 
for BOL.
In the appendix~\ref{appendix:response_functions} we include the scenarios for 
EOL (see Fig.~\ref{figure:quick_noise_model_eol}) and for the F-CAMs.
The main difference between the BOL and EOL scenarios is caused by 
the increase of the background and readout noise component due to 
ageing of the electronics.
The parameters of the fit are tabulated in Table~\ref{table:quick_noise_model},
where we also indicate the magnitude range where the fit is valid.
For very faint targets, the model with three components is not accurate
anymore because of the impact of charge transfer inefficiency and 
ultimately digitalisation noise beyond \plato~magnitude~17-17.5.

\begin{table*}
\caption{Fitted parameters for the quick noise model. Uncertainties are 68\% confidence levels.}             
\label{table:quick_noise_model}
\centering
\renewcommand{\arraystretch}{1.2}
\begin{tabular}{r*{2}{r@{$\;\pm\;$}l}}
\hline
 & \multicolumn{4}{c}{N-CAM}   \\ 
 & \multicolumn{2}{c}{BOL} & \multicolumn{2}{c}{EOL} \\
\hline\hline
$f_\mathrm{ref}$ & \multicolumn{4}{c}{177\,000 (fixed)} \\
$m_\mathrm{ref}$ & 10.80 & 0.01 & 10.69 & 0.03          \\
$k_j$ (ppm)      &   8.9 & 0.1  & 9.4   & 0.5           \\
$k_r$ (ppm)      &   161 & 3    & 290   & 9             \\   
magnitude range  &  \multicolumn{4}{c}{4 - 15}          \\
\hline
 & \multicolumn{4}{c}{F-CAM blue}   \\ 
 & \multicolumn{2}{c}{BOL} & \multicolumn{2}{c}{EOL} \\
\hline\hline
$f_\mathrm{ref}$ & \multicolumn{4}{c}{750\,000 (fixed)} \\
$m_\mathrm{ref}$ & 6.41 & 0.02 & 6.64 & 0.12            \\
$k_j$ (ppm)      &  8.4 & 0.6  & 13.4 & 1.7             \\
$k_r$ (ppm)      &  460 & 10   &  890 & 110              \\   
magnitude range  & \multicolumn{2}{c}{3 - 10} & \multicolumn{2}{c}{3 - 9} \\
\hline
 & \multicolumn{4}{c}{F-CAM red}   \\ 
 & \multicolumn{2}{c}{BOL} & \multicolumn{2}{c}{EOL} \\
\hline\hline
$f_\mathrm{ref}$ & \multicolumn{4}{c}{420\,000 (fixed)} \\
$m_\mathrm{ref}$ & 7.04 & 0.02 & 7.27 & 0.13            \\
$k_j$ (ppm)      &  8.4 & 0.6  & 13.5 & 1.7             \\
$k_r$ (ppm)      &  460 & 10    &  890 & 110              \\   
magnitude range  & \multicolumn{2}{c}{3 - 10} & \multicolumn{2}{c}{3 - 8.5} \\
\hline
\end{tabular}
\end{table*}

\begin{figure*}
  \centering
  \includegraphics[%
    width=0.9\linewidth,%
    height=0.5\textheight,%
    keepaspectratio]{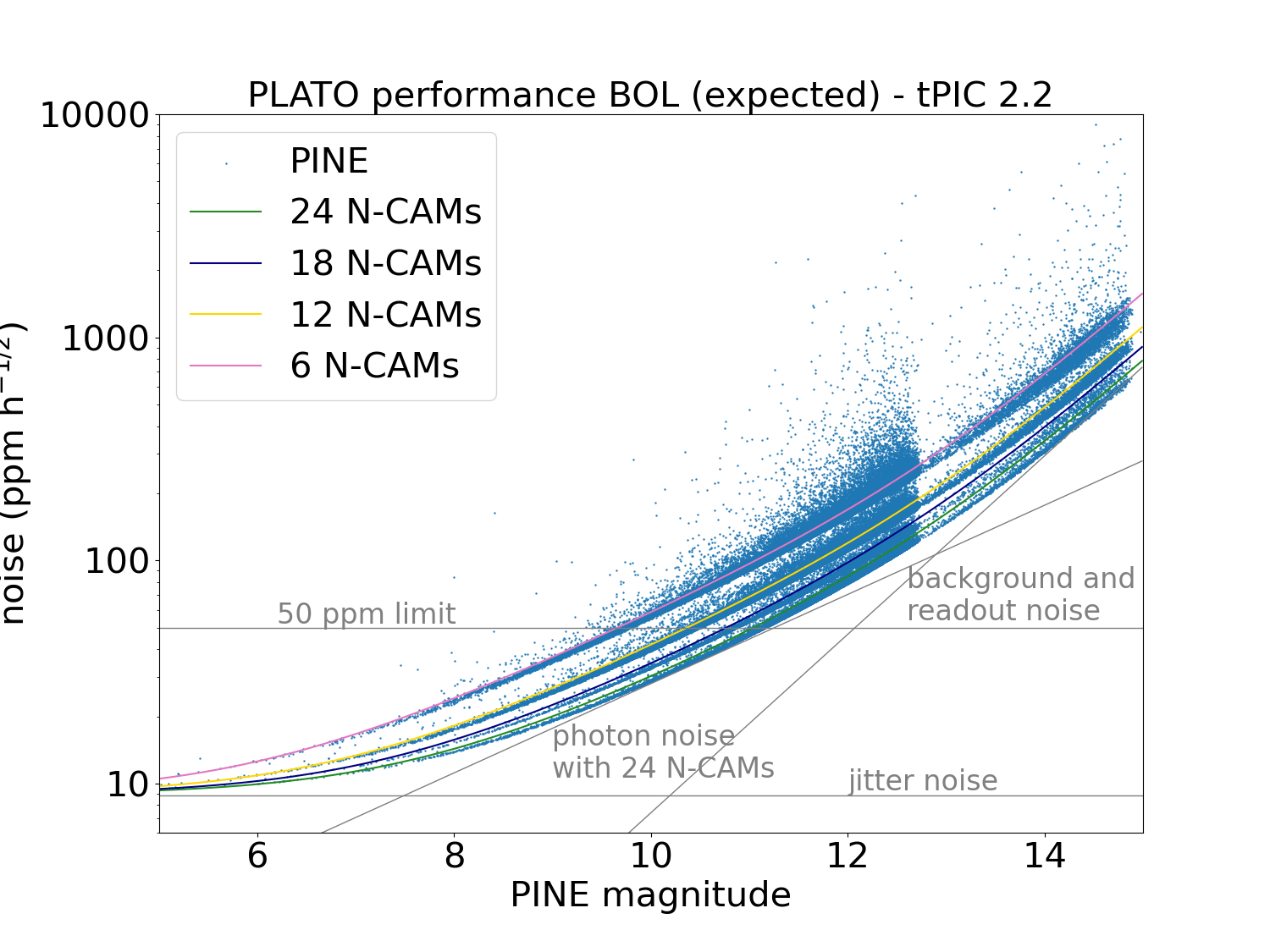}
  \caption{Noise to signal ratio (NSR) for the PIC 2.2 computed with \pine~for N-CAMs BOL.
  Overplotted lines are the quick noise model values for the beginning of life scenario (see Table~\ref{table:quick_noise_model}).}
  \label{figure:quick_noise_model_bol}
\end{figure*}

\section{The Expected Planet Yield}
\label{section:planet_yield}

The planet yield of a photometric transit survey like \plato\,is the result of: 
\begin{enumerate}
    \item The stellar population observed, which in this study we fix to be
    the stars in the PIC 2.2.0.1, including targets from the tPIC samples P1, P2, P4, and P5.
    The properties of the stellar samples are described in Montalto et al., in prep. 
    and in~\citet{prisinzano2026} for the M dwarfs.
    \item The planet occurrence rate, which is highly uncertain as of today. 
    Actually, most of the uncertainty of the results in this study comes
    from the occurrence rates and the impact of stellar variability. 
    In this study we will consider values from the literature. 
    For FGK stars we use~\citet{fressin2013,hsu2019,kunimoto2020} 
    while for M dwarfs we follow~\citet{dressing2013}. 
    \item The transit probability, which is determined by the inclination of the orbit.
    \item The detection efficiency, where we will use the studies from
    ~\citet{fressin2013,hsu2019,kunimoto2020,christiansen2020}. The assumptions
    on the impact of stellar variability are detailed below.
    \item The noise performance, which we take from 
    the \pine~estimates in the PIC.
    \item The observing strategy, where we will compare the nominal
    mission operations~\citep[see][]{rauer2025} with some other
    scenarios.
\end{enumerate}

This approach is appropriate for a first order assessment of the detection capability
of a photometric transit survey before launch. 
After launch, when light curves including stellar variability and instrumental
systematics are available, it is more appropriate to use signal injection 
studies~\citep[e.g.][]{christiansen2013,christiansen2016,christiansen2020}.

Residuals from instrumental systematics compromise the assessment of the
completeness of a transit survey~\citep[see e.g.][]{bryson2021}.
We ignore the impact of these effects on the yield estimates below because 
we do not know which kind of residuals will have the largest impact on \plato.
However, because of the multi-camera approach of \plato, we expect that 
instrumental systematics will be easier to mitigate, as it is unlikely 
that two independent cameras will experience the same systematic effect
simultaneously.

Finally, we also ignore the impact of astrophysical false positives, as 
we are not measuring here the completeness of the survey, but only the
planet yield.
The analysis of the completeness of the \plato~survey will be the subject of 
future studies by the PMC.
We only note here that, as a consequence of its design, we expect \plato~to be 
less subjected to astrophysical false positives than missions like \corot~or 
\kepler~\citep{santerne2013,bray2023,bray2025}.
Additionally, the \plato~observing strategy is designed to be robust
against the presence of these false positive scenarios~\citep{gutierrezcanales2025}.

For the definition of the habitable zone we are using the estimates in \citep{kopparapu2014}, 
but see also e.g. \citet{kasting1993,vonparis2013,leconte2013,kopparapu2013a,godolt2016}.

\subsection{Detection efficiency}
\label{subsection:detection_efficiency}

The detection efficiency addresses which fraction of the (transiting) planets are 
detected by transit detection algorithms.
For reference, the \plato~pipeline will use CETRA~\citep{lsmith2025} as detection 
algorithm.
The detection efficiency
\begin{itemize}
\item increases with number of transits observed $N$ as $N^{0.5}$; or, in other 
words, proportionaly to the square root of the duration or baseline of the survey,
\item increases with radius ratio ($R_p/R_s$), or more appropriate, with the 
transit depth as $\delta = (R_p/R_s)^2$, where we ignore the impact of limb darkening,
\item decreases with noise $\sigma$ (random noise and correlated, e.g. instrumental).
\end{itemize}

\begin{equation}
\label{eq:mes}
S/N = \sqrt{N} \frac{\delta}{\sigma}.
\end{equation}

More information on the subtleties of the signal-to-noise ($S/N$) ratio and how it 
depends on the different factors can be found in studies done by the \kepler~team, 
see~\citet{jenkins2002a,christiansen2016,christiansen2020}, etc.

The detection efficiency is not a linear function of the $S/N$ of
the detection, but it has a more complex behaviour.
Previous studies~\citep[see][]{fressin2013,kunimoto2020,hsu2019} show that the 
detection efficiency can be expressed as a cumulative gamma distribution function:
\begin{equation}
\label{eq:detection_efficiency}
P_\mathrm{det} (S/N) = \frac{c}{b^a \left( a-1 \right)!} \int_{0}^{S/N} dx \; x^{a-1} \; e^{-x/b},
\end{equation}
where the parameters $a$, $b$, and $c$ can be empirically recovered and
might be different for short- and long-period planets~\citep[e.g.][]{christiansen2020}.

\begin{figure}[ht]
\begin{subfigure}{0.5\textwidth}
\includegraphics[width=0.9\linewidth, height=6cm]{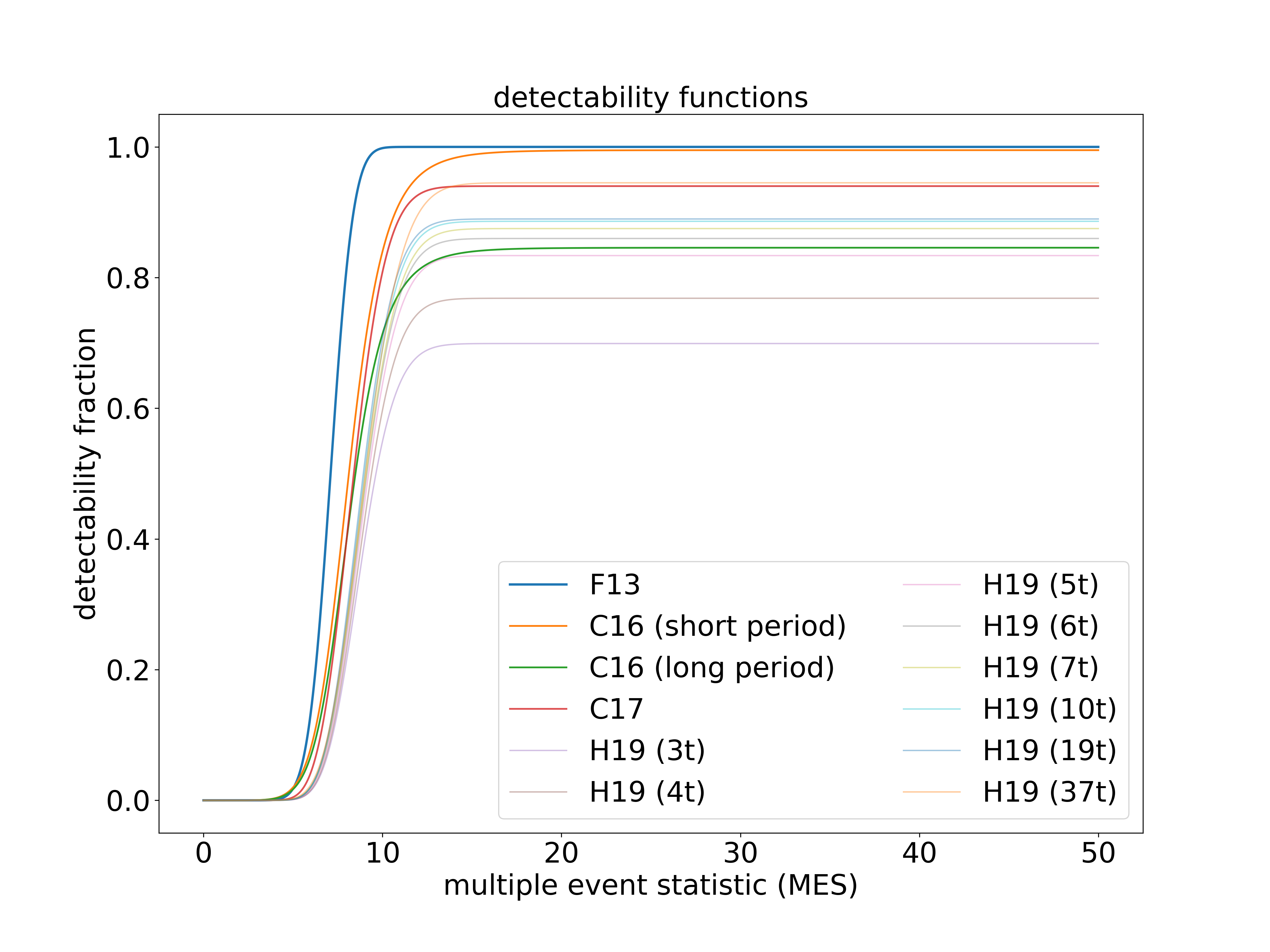} 
\end{subfigure}
\begin{subfigure}{0.5\textwidth}
\includegraphics[width=0.9\linewidth, height=6cm]{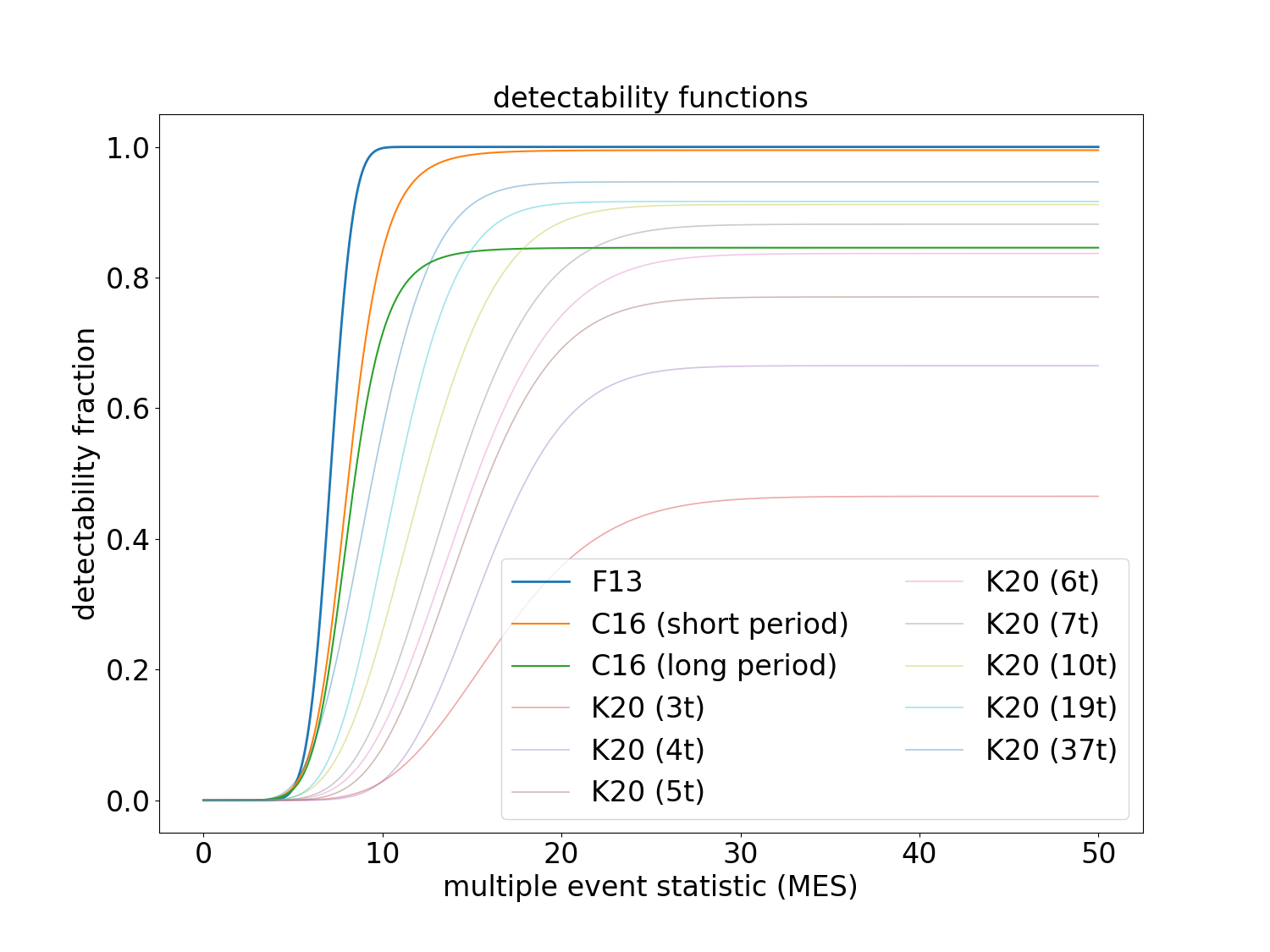}
\end{subfigure}
\caption{
Left: Detectability functions in~\citet{hsu2019} (labelled as H19 and $t$ indicating the number of
transits observed) and comparison with other studies, labelled F13 for~\citet{fressin2013}, C16 
for~\citet{christiansen2016}, and C17 for~\citet{christiansen2017}.
Right: Detectability functions in~\citet{kunimoto2020} (labelled as K20) and comparison with other studies.}
\label{figure:detectability_functions}
\end{figure}

\subsection{Stellar variability}
\label{subsection:stellar_variability}

We describe under stellar variability a series of phenomena apparent in the photometric 
light curves of stars which have different physical origins and different time scales.
Stellar variability impacts our ability to detect and precisely characterise 
planets~\citep[e.g.][]{barros2020}.
Magnetic activity and spot induced variability affect light curves at the time scales 
of stellar rotation. 
There are different strategies to mitigate their impacts~\citep[e.g.][]{cabrera2012}. 
We refer the reader to studies applied to the \plato~case~\citep[][Talens et al. submitted]{canocchi2023}.
Because we do not do signal injection, we will ignore the impact of magnetic activity and
assume that its impact is already included in the detection efficiencies, which do not always 
reach 100\% recovery even for large $S/N$ values (see Fig.~\ref{figure:detectability_functions}).

Granulation, however, will increase the noise budget estimates done with~\pine.
In order to include the impact of granulation, we take as first order approximation
the value of the flicker measured in 8h (F8), which correlates with the stellar surface 
gravity~\citep{bastien2013,bastien2016}.
We use equation 4 in~\citet{bastien2016} to increase the noise value provided by PINE~for
the time scale of the transits. 
Because the flicker F8 value is given in time scales of 8h, for transit durations $d$ shorter
than 8h we increase the noise by $\sqrt( 8/d)$ while we leave the full value of F8 for 
transit duration longer than 8h. This is a compromise, because it is unlikely
that F8 will scale as white noise for $d<8$h, but it also is the worst case for $d>8$h.
Ideally, one should again use signal injection and more representative estimators
of stellar granulation like, e.g. FliPer~\citep{bugnet2018}.

\subsection{Planet yield predictions}
\label{subsection:planet_yield_results}

For each star in the PIC 2.2 belonging to samples P1, P2, P4, and P5 we compute the 
likelihood that it hosts a planet of a given size at a given orbital period using the 
occurrence
rates in~\citet{fressin2013},~\citet{hsu2019}, and~\citet{kunimoto2020}
for FGK stars and~\citet{dressing2013} for M dwarfs.
We draw a number from a random uniform distribution between -1 and 1 for
the cosine of the inclination of the orbit and we only record planets
in transit (with impact parameter $b \leq 1$).
The transit epoch is computed using a uniform distribution in orbital phase.
We compute the $S/N$ of the detection using the simulated
planetary parameters (including the impact of the inclination and the
transit duration), the stellar parameters from the PIC, and the
noise in the light curve computed by PINE. 
We use the total noise EOL including instrumental systematics and
adding quadratically stellar variability as described in the section above.
Given the value of the $S/N$, we compute the probability that the planet
is detected in a given baseline of the observations (e.g. 2 years) following 
the appropriate detectability function (see equation~\ref{eq:detection_efficiency}).
We include a duty cycle of 93\% in the computation of the $S/N$, as per 
requirement.
We run 100 times the simulations for each case to have an idea of the
uncertainty of the numbers.

Since the uncertainty in the occurrence rate of planets in the habitable 
zone is high, we do a sensitivity study and generate simulations where no 
terrestrial planet in the habitable zone occurs (following strictly 
the reported occurrence rates) but we also generate simulations where
40\% of stars host planets in the habitable zone. 

The planet counts are presented in Table~\ref{table:planet-yield-fressin13hz040}
for occurrence rates and detectability criteria following~\citet{fressin2013} while
in the appendices~\ref{appendix:yield_estimates} we present the results 
for~\citet{hsu2019} and~\citet{kunimoto2020} in 
Tables~\ref{table:planet-yield-hsu19hz040} 
and~\ref{table:planet-yield-kunimoto20hz040} respectively.
All M dwarf numbers have been computed with~\citet{dressing2013}.

We use~\citet{fressin2013} as reference not because we think it is the most
recent or accurate estimate, but because it allows a homogeneous comparison
with previous values presented by the \plato~team.
For example, the methodology used here and in~\citet{rauer2025} is identical.
The main change for the numbers is the change from PIC 1.0 to PIC 2.2.
The tables in the appendix section show that, for the same methodology, there 
is a large uncertainty in the expected yield depending on the assumptions
taken on planet occurrence rates and detectability functions.
The values in~\citet{fressin2013} lie in between the results from
~\citet{hsu2019} and~\citet{kunimoto2020}.

\begin{table}[ht]
\tabcolsep=1pt
    \begin{tabular}{c|c|c|c|c|c|c} \hline
                               & known transiting & \multicolumn{5}{c}{2+2 scenario}   \\
 Samples                       & planets   & Red Book        & Heller  & Rauer         & This work     & Matuszewski    \\ \hline
all planets orbiting stars     &           &                 &         &               &               &                \\
$<$13 mag in P1+P5 samples     & 1\,550    & $\approx$4\,600 & n/a     & 6\,800-7\,100 & 6\,300-6\,600 & 4\,500-46\,000 \\ \hline
all planets orbiting stars     &           &                 &         &               &               &                \\
V$<$11 mag in P1+P5 samples    & 520       & $\approx$1\,200 & n/a     & 1\,200-1\,350 &       850-960 & 1\,700-11\,000 \\ \hline
planets $<$2 \rearth\, in HZ   &           &                 &         &               &               &                \\
orbiting P1+P5 stars $<$11 mag &      0    & 6 - 280         & 11 - 34 & 0 - 95        & 0 - 60        & $\approx$45    \\ \hline
         \hline
                               & known transiting & \multicolumn{5}{c}{3+1 scenario}      \\
 Samples                       & planets   & Red Book         & Heller  & Rauer           & This work     & Matuszewski     \\ \hline
all planets orbiting stars     &           &                  &         &                 &               &                 \\
$<$13 mag in P1+P5 samples     & 1\,550    & $\approx$11\,000 & n/a     & 10\,100-10\,700 & 8\,900-9\,500 & 12\,000-68\,000 \\ \hline
all planets orbiting stars     &           &                  &         &                 &               &                 \\ 
V$<$11 mag in P1+P5 samples    & 520       & $\approx$2\,700  & n/a     & 2\,200-2\,500   & 1\,400-1\,700 & 4\,000-42\,000  \\ \hline
planets $<$2 \rearth\, in HZ   &           &                  &         &                 &               &                 \\
orbiting P1+P5 stars $<$11 mag &        0  & 3-140            & 8-25    & 0 - 60          & 0 - 35        & $\approx$30     \\ \hline
    \end{tabular}
    \caption{Estimated PLATO planet yields. Red Book: ESA-SCI(2017)1; Rauer:~\citet{rauer2025}; Heller:~\citet{heller2022}; This work: using occurrence rates and detectability criterion as per \citet{fressin2013} on PIC 2.2; Matuszewski: \citet{matuszewski2023}. 2+2 means 2 long pointings of 2 years duration; 3+1 means one 3-year observation followed by one year with six target fields for 60 days each, as in the Red Book. Known (confirmed) transiting planets are taken from the NASA exoplanet archive in Feb.\,2026 (https://exoplanetarchive.ipac.caltech.edu/) for all planet radii and orbits.}
    \label{table:planet-yield-fressin13hz040}
\end{table}


As discussed above, the uncertainty on the occurrence rate of planets
in the habitable zone is not properly quantified in the empirical 
occurrence rate tables used here. 
Therefore, we assume that 40\% of stars have planets smaller than 2 Earth 
radii (assumed to be rocky) in the habitable zone.
We take the definition for the habitable zone from~\citet{kopparapu2014} and 
simulate planets with log-uniform distribution in radius and period in that range.

\begin{figure*}
  \centering
  \includegraphics[%
    width=0.9\linewidth,%
    height=0.5\textheight,%
    keepaspectratio]{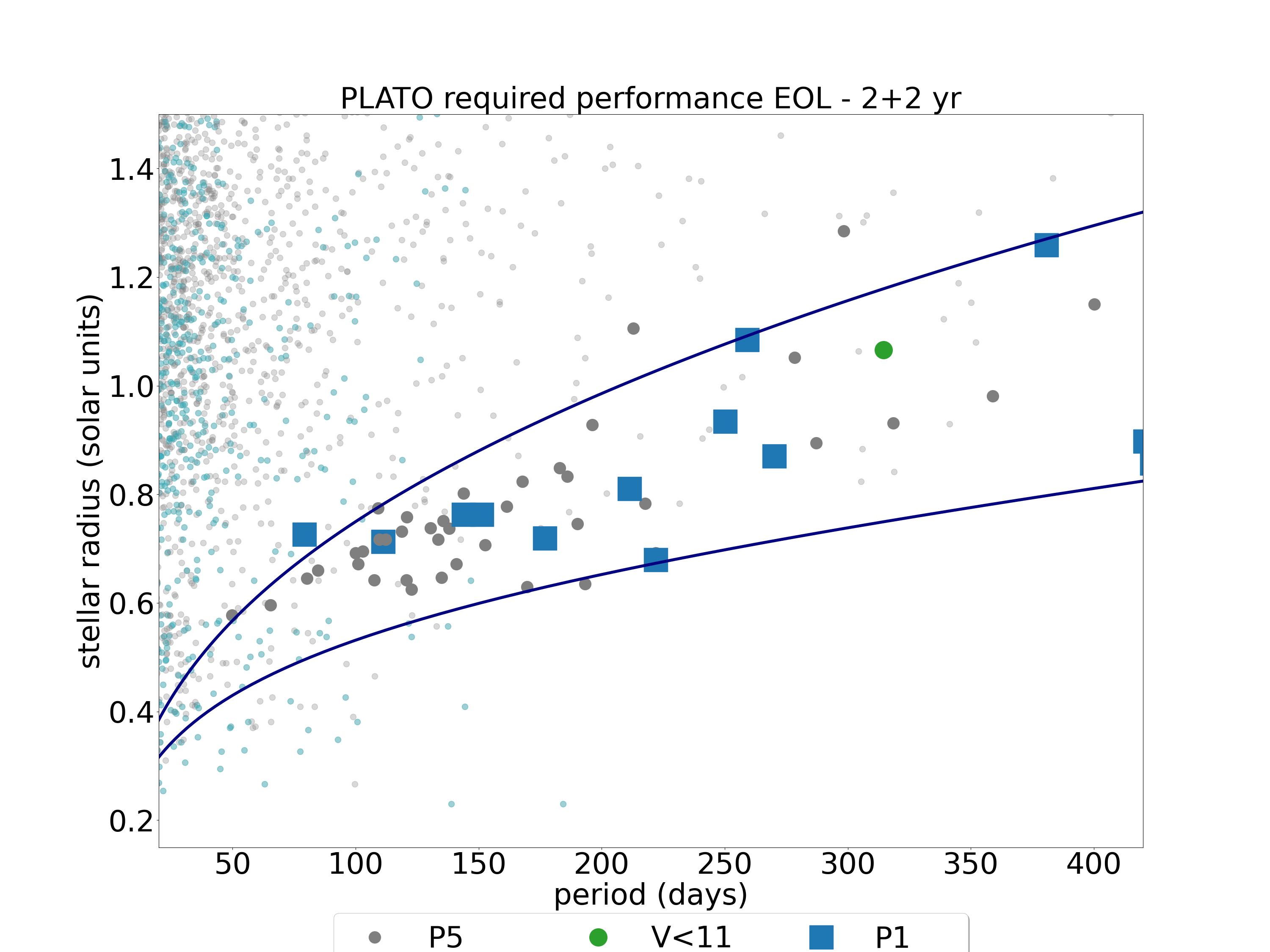}
  \caption{Distribution of detected planets in the habitable zone assuming 40\% occurrence rate 
  and detectability criteria as per~\citet{fressin2013}. The habitable zone is computed for each star
  according to its stellar properties in the PIC (mass, radius, effective temperature). In the 
  habitable zone, grey dots design are stars from the P5 sample fainter than magnitude 11, where
  follow-up efforts will be challenging. Green dots design stars from the P5 sample brighter than 
  magnitude 11, where follow-up efforts might be feasible. Blue squares design stars of the P1 
  sample, where full characterization shall be possible. Here we present two realizations of a
  2 year simulation (representative of a 2+2 scenario). To account in a more realistic way 
  for the dispersion of values in the number of planets expected in the habitable zone, refer to 
  Table~\ref{table:planet-yield-fressin13hz040_table4kunimoto2022}.}
  \label{figure:yield_hz_fressin13hz040}
\end{figure*}

When using this approach (same value for the occurrence rate in the habitable zone),
one can directly compare the impact on the counts of the detectability criteria, 
because it is the only parameter explaining the differences between the results 
shown in Fig.~\ref{figure:yield_hz_fressin13hz040} and 
Figs.~\ref{figure:yield_hz_hsu19hz040} and~\ref{figure:yield_hz_kunimoto20hz040}.


The science goals of \plato~and \tess~are different and cannot be compared just by looking at 
the number of planets.
What we want to highlight is the amount of planets that will require follow up efforts and 
to show that, in general, \plato~will be more sensitive to smaller planets in longer period 
orbits, as per design.
The PMC will coordinate the the Ground-based Observing Program (GOP) to perform the follow-up 
needed to confirm a fraction of the candidate planets photometrically detected by~\plato
and to measure their masses through spectroscopic radial velocity.
These planets will orbit stars from the Prime Sample (PS) which is identified as a subset
of the PIC~(Nascimbeni et al. submitted). 
We refer the reader to Nascimbeni's paper for the definition of the selection criteria for 
the PS.
Our estimates show that the radial velocity ground-based follow-up efforts for \plato\,will be 
comparable to those of \tess~(see Table~\ref{table:planet-yield-fressin13hz040_PS}).

\begin{table}[ht]
\renewcommand{\arraystretch}{1.1}
\begin{tabular}{r|*{2}{c|}} \hline
\multicolumn{3}{l}{2 years}\\\hline\hline
after 2 years in LOPS2       & Total       & smaller than 2\rearth \\\hline
Planets in the Prime Sample  & 395 - 449   & 255 - 297             \\
\end{tabular}
    \caption{Estimated planet yields on the Prime Sample using occurrence rates by~\citep{fressin2013}.}
    \label{table:planet-yield-fressin13hz040_PS}
\end{table}

We provide in table~\ref{table:planet-yield-fressin13hz040_table4kunimoto2022}
a comparison between \plato~and~\tess\,so the community understand the different strengths of both 
missions.
The reference to \tess~is to make the community aware of what to expect from \plato~in the first
years of operations. 
Note that the emphasis of \plato~is not as much on the absolute number of planets, but on the
relative abundance of smaller planets in longer orbital periods orbiting bright stars
among \plato~candidates compared to previous missions. Stars that are better characterized
thanks to asteroseismology and for which we can derive estimates on the ages, opening a 
new window on planetary studies.

The numbers presented in this work include the impact of stellar variability, as discussed. 
However, the assumptions made are quite simplistic and do not reflect the whole complexity of 
stellar activity patterns. 
We also anticipate instrumental noise sources that will only be evident in flight, as 
has happened with previous missions (e.g. straylight is a known problem as it is extremely 
difficult to model in an accurate way before flight).
Therefore, we will re-evaluate these figures after payload commissioning.

\begin{table}[ht]
\renewcommand{\arraystretch}{1.1}
\begin{tabular}{r|*{5}{c|}} \hline
\multicolumn{6}{l}{2 years}\\\hline\hline
by spectral type     & Total       & F                & G                    & K                     & M                \\ \hline
\tess~Prime Mission  & 4\,719      &           1\,209 &               2\,134 &                   859 &              261 \\
\plato               & 3\,698      &           1\,413 &               1\,479 &                   342 &              464 \\ \hline\hline
by size (values in \rearth) & Total       & $R_p < 2$ & $2 < R_p < 4$ & $ 4 < R_p < 8$ & $R_p > 8$ \\ \hline
\tess~Prime Mission  & 4\,719      &              152 &                  770 &                   673 &           3\,124 \\
\plato               & 3\,682      &           1\,593 &               1\,699 &                   214 &              176 \\ \hline\hline
by period            & $>$ 20 days &     $>$ 100 days & \multicolumn{3}{c}{} \\ \cmidrule{1-3}
\tess~Prime Mission  &         398 &               48 & \multicolumn{3}{c}{} \\
\plato               &      1\,151 &              170 & \multicolumn{3}{c}{} \\ \cmidrule{1-3}
\end{tabular}
    \caption{Estimated PLATO planet yields compared with the values in Table 4 of~\citet{kunimoto2022} using occurrence 
    rates by~\citep{fressin2013}.
    In the table we do not give explicit uncertainties in the values and refer to the text for details. 
    The \tess~total numbers include A stars, that we do not include for \plato~because they do not belong to the PIC.}
    \label{table:planet-yield-fressin13hz040_table4kunimoto2022}
\end{table}

\begin{figure*}[ht]
\includegraphics[%
    width=0.9\linewidth,%
    height=0.5\textheight,%
    keepaspectratio]{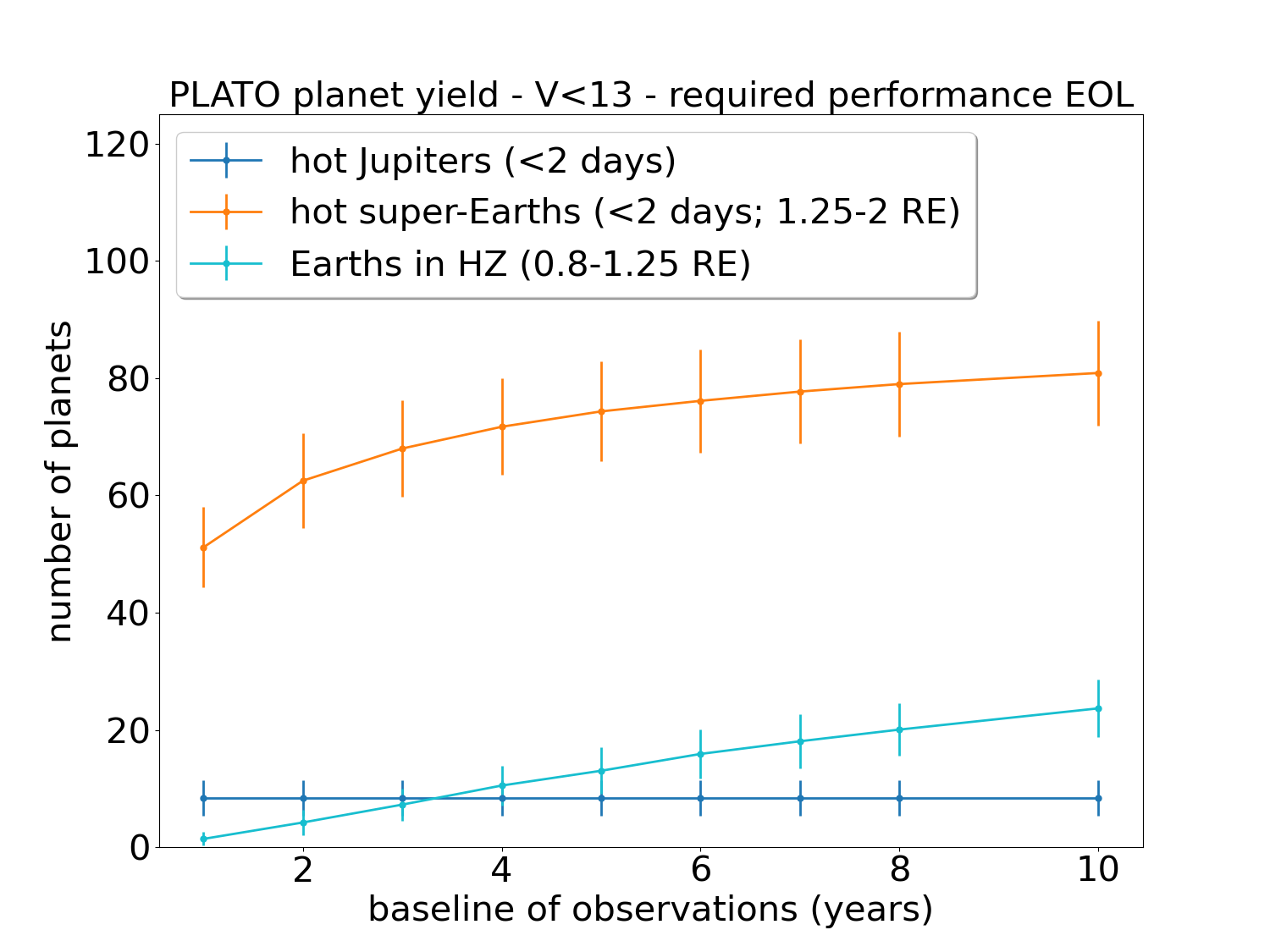}
    \caption{Number of planets anticipated to be found, for a single long pointing using LOPS2 PIC 2.2, as a function 
    of the observing baseline for hot-Jupiter planets 
    (defined as planets with 6 to 22 \rearth\, and orbital period $<2$ days), hot super-earths 
    (defined as planets with 1.25 to 2 \rearth\, and orbital period $<2$ days), and temperate Earths 
    (defined as planets with 0.8 to 1.25 \rearth\, and orbital period between 245 and 418 days). 
    The vertical lines represent the expected uncertainty in the number of planets. 
    We have considered the end-of-life (EOL) with PIC 2.2 and occurrence rates and detectability
    criteria as per~\citet{fressin2013}.}
\label{figure:yield_time_fressin13hz040}
\end{figure*}

The current nominal duration of the \plato\,mission is 4 years. 
The first pointing will be of at least two years duration towards LOPS2.
The in-flight performance will affect the final planet yield and might 
influence the selection of the observing strategy during the rest
of the nominal mission.
Possible operational scenarios include 4 years in LOPS2 and then 4 years extension 
in a different field; 3 years LOPS2 followed by 1 year step-and-stare phase~\citep[3+1, see][]{PLATORedBook2017};
or up to 8 years in the same field (LOPS2).
The planet yield might be one of the criteria used to take a decision.
With the approach presented here we can do sensitivity studies that predict
how many planets can be found as a function of the observing baseline
(see Fig.~\ref{figure:yield_time_fressin13hz040}).

We independently compute the yields of the \plato\ mission using the detection sensitivities of \plato\ presented in \citet{eschen2024}, 
who applied the Transit Investigation and Recoverability Application \citep[\texttt{TIaRA};][]{rodel2024} to determine how sensitive 
\plato\ is to detect a planet of given radius and period orbiting a given star. 
We compute these sensitivities for 2 years of \plato\ observations of the P1, P5 and Prime Sample (Nascimbeni et al. submitted) using 
the noise reported in PIC 2.2 for P1 and the Prime Sample and PIC 2.1 for P5. 
We bin the computed detection sensitivities into the radius and period bins presented in \citet{fressin2013}, \citet{hsu2019} and \citet{kunimoto2020}. 
Accounting for the transit probability we multiply each sensitivity bin with the respective occurrence rate for each star. 
As \citet{fressin2013} does not report occurrence rates for small planets of long orbital periods, these are not included in our yield estimates 
and hence the yields computed with the \citet{fressin2013} occurrence rates are lower limits. 
Since \citet{hsu2019} and \citet{kunimoto2020} present upper limits in some radius-period spaces, the yields obtained using these occurrence rates are 
upper limits. 
Finally, we sum up the computed yield of each bin of each star of the sample resulting in the total yield which we present in \autoref{tab:sensitivity_yields_total}.
Additionally, we compute the yields of planets with a radius below 2\,\rearth\ orbiting their star in its habitable zone. 
To do so, we compute the period boundaries of the habitable zone following \citet{kopparapu2014} 
for each star. We bin the detection sensitivity into one bin covering the radii from 0.5-2.0\,\rearth\ and the computed habitable zone period boundaries.
Since the occurrence rate for planets in this regime is not properly quantified as discussed above, we multiply this bin with the 40\% from above. 
Summing up this result for all stars of a given sample, we report the yield for habitable zone planets below 2\,\rearth in \autoref{tab:sensitivity_yields_hz}. 

\subsection{Synergies with Ariel}
\label{subsection:ariel}

\begin{figure*}
  \centering
  \includegraphics[%
    width=0.9\linewidth,%
    height=0.5\textheight,%
    keepaspectratio]{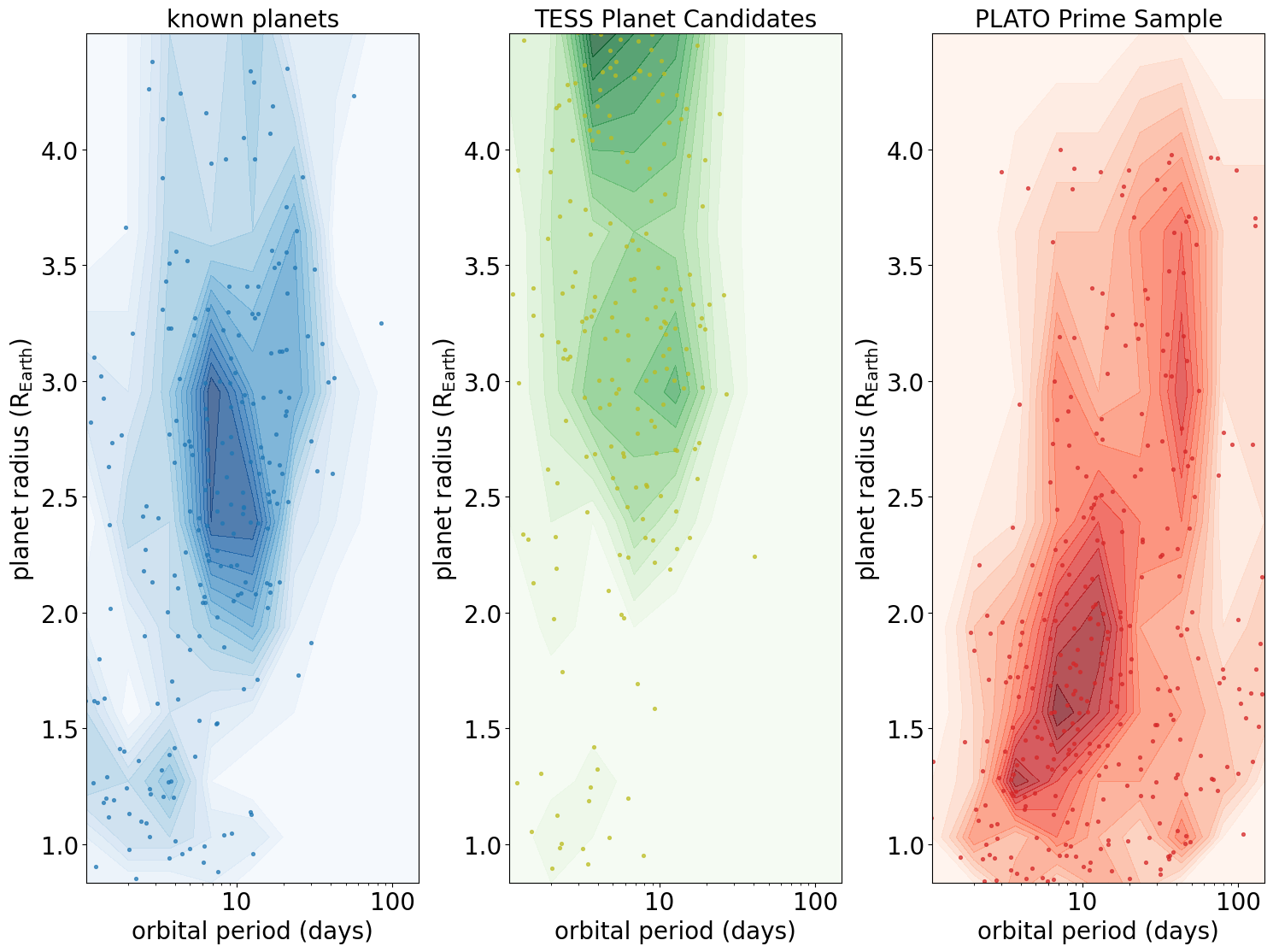}
  \caption{Density plots showing the distribution of known planets considered for follow-up with \ariel,
  \tess\,planet candidates considered for \ariel, and the distribution of Prime Sample targets expected
  to be detected with \plato~for occurrence rates by \citet{fressin2013}.}
  \label{figure:ariel_target_list}
\end{figure*}

In order to show the impact that \plato~will have on future missions, we 
have taken the current list of targets for the \ariel~mission~\citep{tinetti2022}.
We have used the \ariel~target list~\citep[see][]{edwards2019,mugnai2020,edwards2022},
which includes known planets and \tess~planetary candidates, and over-plotted the 
yield of small planets with \plato~(see Fig.~\ref{figure:ariel_target_list}). 
We highlight the planets in the Prime Sample (Nascimbeni et al. submitted).
They occupy a region of the parameter space of small planets with longer 
period orbits.

\section{Planet characterisation}
\label{subsection:planet_characterisation}

\begin{figure}
\centering
\includegraphics[width = \columnwidth]{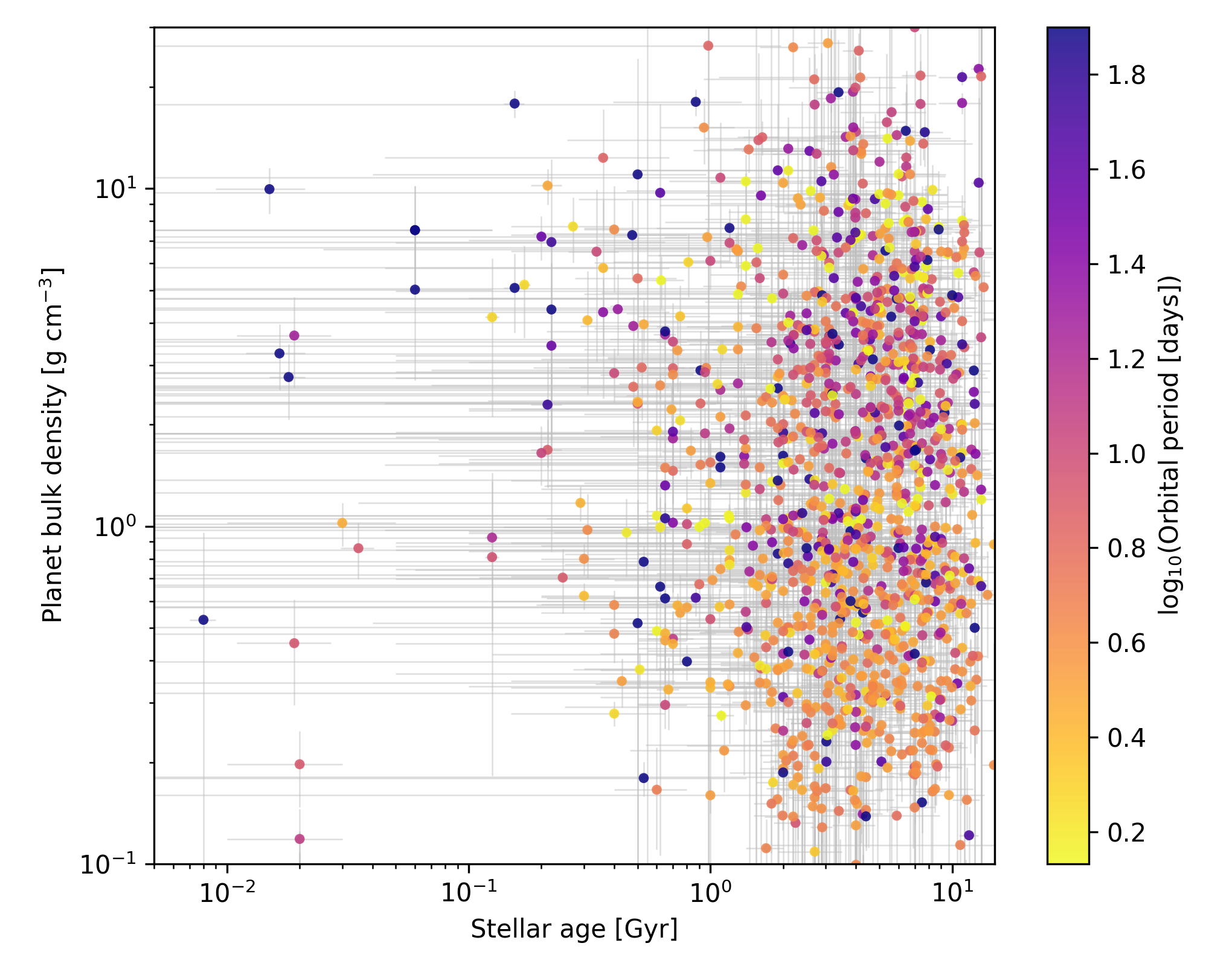}
\caption{Diagram showing planet density versus known age of the planetary system.
The uncertainties in the plot are large and it is not straightforward to see
correlations that could reveal the consequences of planetary evolution 
processes like, e.g., atmospheric erosion.}
\label{figure:ages}
\end{figure}

The design driver for the \plato\,payload is to have the ability to reach an 
uncertainty in the planetary radius better than 3\% for planets orbiting 
stars brighter than V magnitude 10 and 5\% for planets orbiting stars 
brighter than V magnitude 11. The uncertainty shall be precise, achieved 
by excellent photometry, and accurate, achieved by obtaining stellar 
parameters from asteroseismology. 

Beyond providing planetary masses and enabling the detection of non-transiting 
planets, the radial velocity technique allows the full set of orbital 
parameters to be determined, in particular the eccentricity. 
It also extends the range of accessible orbital separations, 
probing periods up to that of Jupiter, thanks to time series spanning 
nearly three decades~\citep[e.g.][]{bonomo2025}.

The choice of \plato\,to begin observations in the southern hemisphere, 
combined with its focus on bright stars, will further extend the accessible 
range of orbital periods. 
It will also benefit from targets already observed by high-contrast 
imaging and interferometric programs, thereby opening access to a broader 
diversity of systems, including younger objects.

In the intermediate orbital period regime, Gaia will provide new constraints 
and bridge the gap between the transit and radial velocity domain and that 
of directly imaged planets. 

A particularly novel aspect of these synergies concerns the determination 
of planetary ages. 
At present, stellar ages are derived using a variety of methods and, as 
shown for instance by~\citet{lebreton2014}, these methods often yield 
discrepant results with large uncertainties (see Fig.~\ref{figure:ages}). 
The observation of a large sample of stars, analysed in a homogeneous 
and comprehensive way through asteroseismology, will enable the 
determination of precise and consistent stellar ages.

Since \kepler\,observations, the population of small planets has 
revealed an unexpected diversity~\citep[e.g.][]{batalha2013}. 
Even in systems where multiple detection methods provide strong 
constraints on planetary properties, the lack of precise stellar 
ages prevents these systems from being placed along a well-defined 
evolutionary sequence. Obtaining precise and homogeneous ages for 
a significant fraction of small planets with well-constrained 
parameters should provide key insights into the interplay between 
formation and evolution processes.

Finally, it is important to emphasize that even for stars without 
well-characterized planetary systems, asteroseismic analysis of a 
large stellar sample will enable the construction of a new generation 
of stellar models. 
These models can then be used to determine the ages of planetary 
systems not directly observed by \plato. 
In this sense, \plato\,will provide a crucial theoretical framework 
for stellar physics and, more broadly, for many areas of astrophysics.

\subsection{Precise planet characterization}
\label{subsection:precise_planet_characterization}

The dependence of the precision on the radius ratio ($k$) from photometric 
transits in presence of stellar variability is discussed in several 
papers~\citep[e.g.][]{barros2014,barros2020,morris2020,sulis2020} but 
here we will follow the approach of the Transit and Light Curve Modeler 
(\tlcm) as described in \citet{csizmadia2020,csizmadia2023}. 
We refer to these papers for the description of the details, here we just 
summarize the main features of it. 

\tlcm~is able to fit the photometric light curve only, or to 
perform a joint fit of the radial velocity and light curve of a 
transiting exoplanet. 
The transit, occultation and phase curve (beaming, reflection, ellipsoidal 
effect are included) with or without a gravity darkened star can be modelled,
as well as the Rossiter-McLaughlin effect. 
Simultaneously to the light curve fit, a wavelet-based noise model 
can be fitted together with the transit-occultation-light curve model 
to remove the correlated noise from photometry. 
Circular and eccentric orbits are considered.

The wavelet-based model of \citet{carter2009} was extended by a penalty 
function to avoid overfitting of the photometric data~\citep{csizmadia2020}. 
This model has only two free parameters: the white-noise level $\sigma_w$ 
and a red-noise factor $\sigma_r$ while the power-spectrum of the noise 
($1/f^\gamma$, $f$ is the frequency) is fixed at $\gamma=1$~\citep{carter2009}. 
This approach was widely tested in \citet{csizmadia2023} and it was successfully 
applied e.g. in \citet{kalman2023,kalman2024,bernabo2025,asmith2025}.

As an example of the planet characterisation performance that can 
be achieved with \plato, we present in the section below a study of 
TOI-500b.

\subsection{TOI-500}
\label{subsection:toi-500}

TOI-500 is a planetary system consisting of 4 planets with orbital periods 
of 13 hours and 6.6, 26.2, and 61.3 days. 
The mass of the innermost planet is $1.42 \pm 0.18 M_\mathrm{Earth}$ and 
the minimum masses of the other three planets are $5.03 \pm 0.41$, 
$33.12 \pm 0.88$ and $5.0 5^{+1.12}_{-1.11} M_\mathrm{Earth}$, respectively. 
The outermost planets were detected in radial velocity variations of the 
star while the innermost planet - denoted by TOI-500b - is a transiting 
Ultra-Short period Planet (USP) with a radius of 
$1.166^{+0.061}_{‒0.058}~R_\mathrm{Earth}$~\citep{serrano2022}.
We selected this system because the star/planet radius ratio is typical for a 
future \plato~primary target system, the spectral type of the host star 
(K6V) which allows us to characterise its mass and radius via 
asteroseismology, its apparent brightness (V=10.5 magnitude) is 
the closest of host stars of the known exoplanets in the LOPS2 field to 
the magnitude where the \plato~requirements are defined 
($\mathrm{rms} = 27 \mathrm{ppm}/\sqrt{\mathrm{hr}} $ at V=10.2 magnitude), 
and it will be observed by 24 normal cameras and the 2 fast cameras in 
the LOPS2 field. 
Therefore, it serves as a good comparison, calibrator object and test 
object of the performance of \plato~relative to \tess.

The light curve of TOI-500 was simulated using the Plato Solar-like Light 
curve Simulator~\citep[\psls][]{samadi2019}, taking into account all known 
\plato~instrumental noise as well as a realistic description of stellar 
variability.
\psls\,calculates random instrumental noise determined using the expected 
NSR-magnitude relation for \plato, while systematic instrumental noise is 
based on simulated imagettes. 
In addition to instrumental noise, we simulate TOI-500's granulation 
spectrum using \psls's implementation of the scaling relationships 
of~\citet{kallinger2014}. 
To this, we added realistic spot activity, generated using~\citet{talens25}'s 
implementation of the analytic spot model of~\citet{kipping2012a}. 
We then injected transits of TOI-500b into this light curve using a quadratic 
limb darkening law with~\batman~\citep{kreidberg2015}. 
All transit and system parameters were taken from~\citet{serrano2022}, except 
the epoch which was arbitrarily set to $T_0 = 2.0$ days. 

The simulated light curve covers exactly 28 days of simulated 
observations with the same duty cycle as \tess\,and consists of 89,421 points 
at a cadence of 25 seconds. 
We modelled the data with \tlcm~as described above. 
The free parameters of the fit were:
\begin{itemize}[label={-},leftmargin=2em,topsep=0pt]
    \item Scaled semi-major axis ratio ($a/R_\mathrm{star}$).
    \item Planet-to-star radius ratio ($R_\mathrm{planet}/R_\mathrm{star}$).
    \item Impact parameter ($b$).
    \item Epoch ($T_0$).
    \item Period ($P$).
    \item Normalization constant ($h$).
    \item White noise level and red noise factor ($\sigma_w$, $\sigma_r$).
    \item Quadratic limb darkening coefficients ($A$ and $B$).
\end{itemize}

\noindent
Four modelling runs were carried out:
\begin{itemize}[leftmargin=4em,topsep=0pt]
\item[M1:] no prior applied.
\item[M2:] prior is applied on stellar radius: $N(0.678, 0.016)$ solar radii.
\item[M3:] prior is applied only on limb darkening coefficients: $A$: $U(1.27, 1.37)$, $B$: $U(1.69, 1.74)$.
\item[M4:] priors are applied on both the stellar radius and limb darkening.
\end{itemize}

\begin{figure}
\centering
\includegraphics[width = \columnwidth]{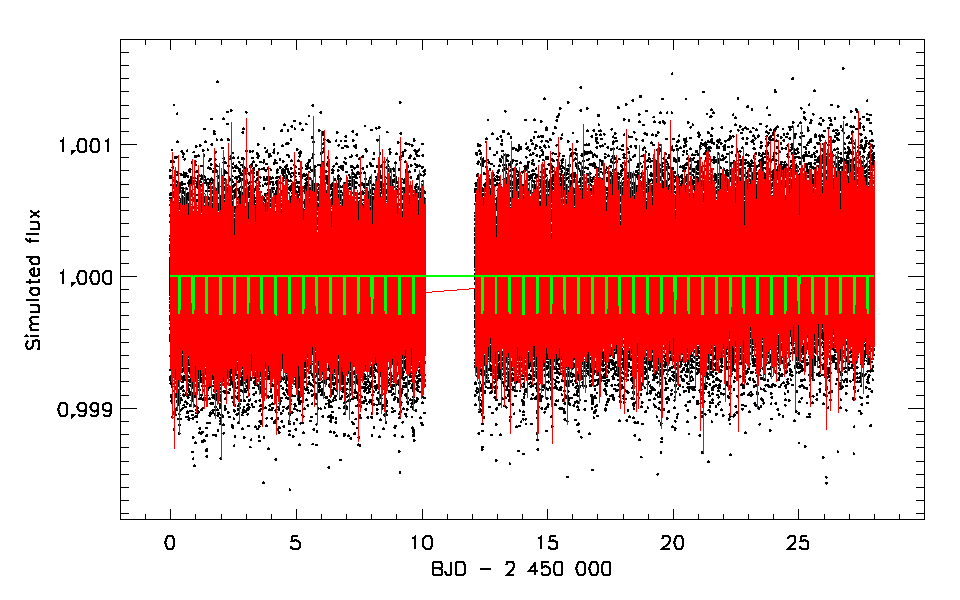}
\caption{28 days of simulated \plato~light curve of the TOI-500b system (black dots), 
the transit model (green curve) and the red-noise model (red curve). The same 
duty cycle as in \tess\,observations was used.}
\label{figure:toi500_simulation_model}
\end{figure}

\begin{figure}
\centering
\includegraphics[width = \columnwidth]{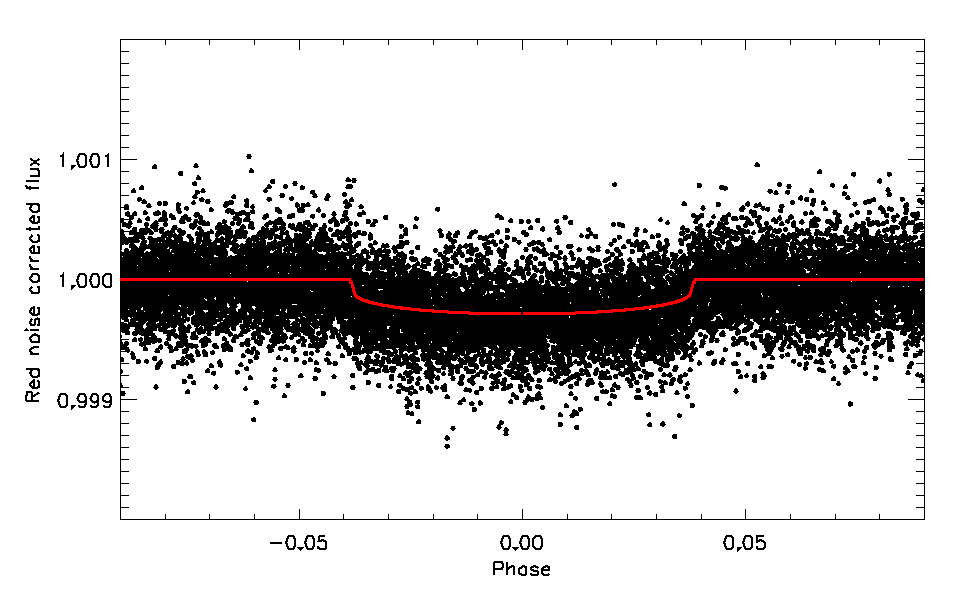}
\caption{Phase folded, red noise corrected fluxes (black dots) and the transit model 
(red curve) to TOI-500b simulated data (28 days observational segment).}
\label{figure:toi500_folded}
\end{figure}

The standard approach to model \plato~light curves is like M4, as 
the stellar radius will be obtained by combining Gaia parallax measurements with stellar 
asteroseismology and using priors for limb darkening  from theoretical 
calculations~\citep{rauer2025}. 
For the sake of simplicity, we fitted a circular orbit because~\citet{serrano2022} 
did not find any significant eccentricity for planet b. 
\tlcm~carried out a Genetic Algorithm optimization first with 100 individuals 
and 320 generations to get starting values. 
Then, a Differential-Evolution MCMC analysis was carried out with 20 chains 
and no thinning. 
The number of steps was at least 2000 in each chain; \tlcm~automatically 
extended the chains if the convergence criteria were not met 
(effective sample size $>200$ and Gelman-Rubin convergence parameter $R \approx\,1.1$). 
The results can be found in Table~\ref{table:toi_500_results} and they are illustrated
in Figs.~\ref{figure:toi500_simulation_model} and~\ref{figure:toi500_folded}.

As one can see from  Table~\ref{table:toi_500_results}, the M4 model is able to 
retrieve the input values and it verifies the \plato-approach of light curve 
modelling. 
One can see that \tess, which for the discovery paper observed the system for 
$\sim$three months in its Sectors 6, 7 and 8, resulted in a relative error in 
the planet-to-star radius of ca. 4\% while the \plato~simulated light curve yielded 
$\sim$ 1\% in the M4 solution, well within the prescribed 2\% requirement, on
just 1 month of data. 

\begin{landscape}
\begin{table*}[ht]
	\centering
	\caption{Results of the light curve solution of TOI-500b; simulated \plato~data. 
    See main text for explanation of the models and for the priors. 
    Note that epoch was set to $T_0 = 2.0$ days arbitrarily.
    The preferred solution is M4.}
	\label{table:toi_500_results}
	\begin{tabular}{lccccc}
		\hline
		Parameter & \citet{serrano2022} & M1  & M2 & M3 & M4 \\
        \hline
        $a/R_\mathrm{star}$                  & $3. 769\pm 0.090$      & $4.03\pm0.33$      & $4.02\pm0.02$      & $4.04\pm0.26$      & $3.71^{+0.07}_{-0.03}$ \\
        $R_\mathrm{planet}/R_\mathrm{star}$  & $0.01568\pm0.00068$    & $0.01539\pm0.0003$ & $0.01531\pm0.0002$ & $0.01526\pm0.0004$ & $0.01559\pm0.00018$    \\
        $b$                                  & $0.51^{+0.12}_{-0.17}$ & $0.299\pm0.23$     & $0.299\pm0.027$      & $0.285\pm0.210$    & $0.40^{+0.02}_{-0.03}$ \\
        Epoch ($T_0$)                        & -                      & $1.9997\pm0.0003$  & $1.9997\pm0.0003$  & $1.9997\pm0.0003$  & $1.9997\pm0.0002$      \\
        Period ($P$)                         & $0.548172(19)$         & $0.548178(8)$      & $0.548176(7)$      & $0.54818(7)$        & $0.54817(8)$ \\
        $-\log L$                              & - & -871921.09 & -871920.90 & -871920.55 & -871919.58\\
        \hline
	\end{tabular} \\
\end{table*}
\end{landscape}

\section{Stellar characterisation and seismic yields}
\label{section:stellar_characterisation}

One of the main scientific goals of the \plato\,mission is the precise 
and accurate characterisation of solar-like stars, in particular those 
hosting planets. 
For the brightest \plato\,targets, this objective will be achieved through 
asteroseismology, which involves the detection, precise measurement, and 
analysis of stellar oscillation modes. 
The analysis of \plato\,light curves is also expected to provide an 
in-depth characterisation of the rotation and activity of solar-like 
stars~\citep[e.g.][]{breton2024}. 
As such, \plato\,will deliver scientific results for all observed targets 
and explore a wide range of physical processes, thereby challenging nearly 
every domain of stellar physics. 
To achieve this, a dedicated pipeline is currently being developed to 
generate, in an automated way, scientific data products for each 
core-programme target observed by the spacecraft. This pipeline is briefly 
described in Sect.~\ref{stellar_pipeline}. 
Since the most precise determination of stellar global properties will be 
obtained through asteroseismic measurements, estimating the number of stars 
for which oscillations will be detectable -- and quantifying the associated 
precision -- is critical for assessing the overall performance of the 
\plato\,mission when it comes to stellar science. 
Using the latest version of the PIC, such estimates are provided in 
Sect.~\ref{stellar_yields}.

\subsection{The PLATO Stellar Analysis System}
\label{stellar_pipeline}

From the Level 1 light curves, the preparatory data, and stellar models, 
including oscillation frequencies, the stellar pipeline will generate 
Level 2 data products (DP3 to DP5). 
It will process the P1, P2, P4, and P5 samples, covering stars with 
spectral types ranging from F5 to K7, as well as M dwarfs, regardless 
of whether they exhibit planetary transits or oscillations. 
All stars will be fully characterised, including their global stellar 
properties (mass, radius, and age), as well as their rotation and 
activity properties. 
For stars exhibiting solar-like oscillations, the pipeline will provide 
significantly more precise fundamental properties, which in turn will 
enable a more precise characterisation of any potential candidate 
transiting object. 
By design, many stars in the P1 and P2 samples are excellent targets 
for asteroseismology based on solar-like oscillations. 
Additionally, a fraction of the P5 sample, observed at short cadence, 
will also be amenable to asteroseismic analysis.

The pipeline is structured in different modules, each of which is 
described in the rest of this Section. 
First, analysis-ready light curves and power spectra are computed 
from the Level 1 light curves. 
The pipeline will then perform the asteroseismic analysis (production 
of DP3) and the long-term variability analysis (production of DP4) 
in parallel. 
Classical constraints from atmospheric parameters are then added, 
and all these constraints are gathered to provide the best possible 
estimate of the mass, radius, and age of the star (production of DP5). 
The production of DP3 is expected for a large number of stars. 
If oscillations are not detected, the pipeline will still generate DP4 and DP5.

\subsubsection{Generation of analysis-ready light-curves}

The stellar pipeline starts by producing analysis-ready stellar time 
series and associated Power Spectral Density (PSD). 
It does so by processing Level 1 photometric light curves (stitched 
and detrended), together with auxiliary inputs, including transit 
models from the Exoplanet Analysis System (EAS) and internally 
generated event masks. 
This phase is carried out so as to generate homogeneous and 
well-documented data products that preserve the intrinsic stellar 
variability while removing signals that could bias subsequent stellar 
analysis.

The processing follows a sequential yet modular architecture. 
First, the light curve is regularised onto a uniform temporal grid, 
establishing a consistent basis for further analysis. 
Transient events such as stellar flares are then detected and 
flagged, so that they may be removed in the next step. 
Data points with significantly elevated flux values are flagged 
as potential flare events if a consensus is found amongst several 
independent flare detection algorithms. 
In addition, these candidate flares are validated by comparison 
to a standard flare template, and physical parameters of the 
confirmed flares are calculated (Binks et al., in prep.). 
The following processing stage subsequently combines several 
dedicated algorithms to remove known transits or transit-like 
signals -- using both external models and data-driven approaches 
such as those described in \citet{KADACS, KASOC, RER} -- and to 
eliminate flare signatures and reconstruct missing data through 
gap-filling techniques \citep{KADACS}. 
This stage produces multiple cleaned versions of the light curve 
tailored for different scientific uses.

Finally, PSDs are computed from the various processed light curves, 
including optional binning steps for the analysis of rotation and 
activity, or removal of long term variability for asteroseismic 
analysis \citep{KASOC}. Throughout the pipeline, masks and 
metadata are propagated and updated to ensure full traceability 
of all transformations. In this way, the stellar pipeline provides 
a robust interface between calibrated photometry and higher-level 
stellar characterisation, delivering consistent inputs optimised 
for asteroseismology and variability analysis.

\subsubsection{Measurement of rotation and activity}
\label{rotation_activity}

The long-term variability of stellar light curves carries the 
signature of both the rotation of the star and its magnetic activity. 
PLATO photometry can be exploited to measure stellar surface rotation 
periods from the light modulation produced by photospheric active 
regions as a star rotates \citep[see][for details and expected performances]{breton2024}. 
Knowing stellar rotation is fundamental to mitigate the impact of
stellar activity on radial velocity measurements acquired during 
the \plato\,ground-based follow up in order to improve the measurement 
of the masses of transiting planets and detect additional non-transiting 
planets in a system. 
Moreover, stellar rotation can be used to estimate the age of a star 
by means of gyrochronology \citep[e.g.,][]{Barnesetal16}. 
The precision is generally of the order of 20-25\%, that is, lower 
than in the case of asteroseismology, but the method can be applied 
to late-type stars lacking detection of p-mode oscillations that will 
be the majority of the targets in the PLATO P5 statistical sample. 
In addition to stellar rotation, \plato\,time series will be analyzed 
to extract information on the stellar activity level and its modulation 
by stellar activity cycles. 
Yearly long cycles require observations extending over at least 3-4
years \citep[see][]{breton2024}. 
Nevertheless, short-term activity cycles, so-called Rieger cycles 
can occur in the Sun and in solar-like stars with periods of the order 
of a few months \citep[e.g.,][]{Gurgenashvilietal26}. 
Such short-term cycles are relevant because they can produce modulations 
in the radial velocity time series that could be misinterpreted as due 
to additional non-transiting planets in a system. 
The analysis of the light fluctuations due to photospheric convection 
(granulation) will provide a measurement of the stellar surface gravity 
through the Fliper algorithm \citep{bugnet2018}. 
Such a measurement can be combined with an asteroseismic measurement of 
the star radius to get the star mass, thus providing another method for 
stellar mass determination that is the basic parameter upon which the 
determination of planetary masses relies.

\subsubsection{Determination of oscillation frequencies}
\label{seismic_properties}

The first objective of the asteroseismic analysis module is to identify 
light curves with detectable signatures of solar-like oscillations in cool 
main-sequence and sub-giant stars, as well as the low-luminosity red-giant 
stars included in the samples P1 to P5 
\citep[e.g., see][]{Chaplin2013b,garcia2019}. 
In the event of a positive detection, we will measure the global 
asteroseismic quantities $\nu_{\mathrm{max}}$ (i.e. the frequency of 
maximum oscillation power), and $\Delta\nu$ (i.e. the average large frequency 
separation); determine the radial orders and angular degrees of the detected 
modes; and finally measure the frequencies and additional parameters of these modes.

Detection of the oscillations is based on the method described by 
\citet{Nielsen2022} and it consists in essentially comparing the power 
density spectrum of the light curve with what we expect to see based on 
scaling relations calibrated on data from the \corot, \kepler\,and 
\tess\,missions. 
When a detection is made we follow the methodology described 
by~\citet{Nielsen2021}: an algorithm samples the posterior distribution 
consisting of the product of a likelihood function given the observed 
power density spectrum and a model that is largely based on the predictable 
pattern of mode frequencies, mode peak widths and heights that depends 
on the radial order and angular degree of the modes. 
The model parameters are drawn from a prior distribution informed by 
thousands of previous detections for other stars. 
As detections on more stars are made, and their parameters are incorporated, 
so the inference on the priors is improved.

The mode parameters are then used to construct a more detailed model of the 
oscillation spectrum. 
This allows for the highest precision estimates of the mode frequencies, 
which are then used to infer fundamental stellar properties. 
A model is constructed that is compared to the observed spectrum. 
However, the model parameters are all left as free variables, only 
subject to constraints from the prior distributions on each parameter. 
The priors on each parameter are set based on the mode identification 
previously performed. 
The models also include the effect on the modes due to rotation and 
the inclination of the stellar rotation axis relative to the line of 
sight to the observer. 
We also separate out high signal-to-noise (SNR) sub-giants and 
low-luminosity red giants, which are computationally more demanding 
for sampling-based Bayesian inference due to the presence of 
so-called `mixed' p and g modes, compared to low-SNR main-sequence 
stars which have detectable p modes only. 
We therefore process these targets using a maximum likelihood 
estimation-based method, which is computationally far more efficient. 

The exact performance characteristics are still to be determined. 
However, the algorithms for measuring the mode frequencies are based 
largely on literature methods like those used for the \kepler\,LEGACY 
sample \citep{Davies2016, Lund2017}. 
For similar stars which are targeted by the \plato\,mission \citep[see][]{goupil2024}, 
we can therefore expect a comparable precision on the mode 
frequencies on the order of $\sim 0.1-0.5~\mu$Hz (see also Sect.~\ref{stellar_yields}).

\subsubsection{Estimation of stellar parameters}

Inferring precise and accurate stellar parameters requires highly 
accurate light-curves, as expected to be observed by PLATO, but 
also accurate preparatory data. 
Spectroscopic observations are particularly important to infer 
the much needed effective temperature, surface gravity, and 
metallicity \citep{gent2022}. 
To ensure the availability of this spectroscopic data, by agreement, 
most of the core-programme FGK targets are expected to be observed 
by 4MOST \citep{deJong2019,walcher2019}. 
The processing of these preparatory spectra will build upon the 
work carried out in the context of the 4MIDABLE-HR 
survey \citep{Storm2025,Ksoll2026} and the 4MOST SPV phase. 
However, as is now common practice for seismic targets 
\citep[e.g.][]{Lund2024}, a constrained spectroscopic analysis 
will be performed to benefit from the fairly unbiased surface 
gravity derived from PLATO data. 
It could either be from the granulation properties 
(see Sect.~\ref{rotation_activity}), or from asteroseismology 
(see Sect.~\ref{seismic_properties}). 
In addition, other analysis techniques (i.e. IRFM, SBCRs, 
interferometry) are implemented within a Bayesian framework 
to constrain the classical parameters further, in particular 
the luminosity. 
The spectroscopic pipeline and its performance are 
discussed in \citet{gent2022}, but an update will be found in 
Lee et al. (in prep.). 
The M dwarfs pipeline shares the same features as much as 
possible, except that it has to accommodate the lack of seismic 
constraints (seismic oscillations cannot be detected in M dwarfs),
or the fact that the spectroscopic constraints are obtained from near-IR spectra. 
Preliminary performance tests are reported by \citet{olander2025}.

Ultimately, the seismic and non-seismic constraints extracted 
from \plato\,light curves and complementary data will be used 
to infer the stellar mass, radius, and age, enabling the precise 
characterization of \plato\,exoplanet properties. 
Because the quality of the seismic data will not be the same 
for all stars observed by \plato\,(cf. Sect.~\ref{stellar_yields}), 
the exploitation of these data is structured in different levels. 
These are designed to provide both a homogeneous set of stellar 
properties across all stars for which seismic detections are 
possible and, simultaneously, the best possible stellar properties 
for each individual star. 
Specifically, all stars with a detection of the global 
asteroseismic properties $\nu_{\rm max}$ and $\Delta\nu$ will 
have their mass, radius, and age inferred through a Bayesian 
grid-based modelling approach \citep{2022MNRAS.509.4344A} using 
these seismic constraints, along with constraints on effective 
temperature, metallicity, as well as luminosity when available. 
Stars for which individual mode frequencies are detected are 
then further analysed to extract more precise stellar properties.
Here, the pipeline branches out according to whether mixed modes 
are detected, as explained below.
          
Main sequence stars will have their properties inferred through 
two additional Bayesian grid-based modelling approaches: the 
first -- frequency fitting -- uses the individual mode frequencies, 
rather than the global seismic properties, as seismic constraints, 
and the second -- surface independent -- uses the mode frequencies 
to construct seismic constraints that are less sensitive to the 
near-surface layers and uses these combinations as seismic constraints. 
The motivation for the latter approach comes from the difficulty 
in modelling the near-surface layers of stars. 
Using a surface-independent approach enables the determination 
of stellar properties that are less affected by surface effects, 
but it is more data-demanding.
          
For subgiant stars, which exhibit mixed modes of oscillation, 
the surface-independent approach cannot be applied. 
This is because non-radial modes are affected by their g-mode 
character in the core (while radial modes are not) and the 
frequency combinations considered in the surface-independent 
approach no longer suppress the impact of the outer layers. 
Hence, for stars with mixed modes only the approaches based 
on the fitting of the global asteroseismic properties and 
the frequency fitting will be employed. 
          
Finally, for the very best cases where numerous individual 
mode frequencies are observed, stellar properties will, 
in addition, be determined following the frequency fitting 
approach, including additional constraints derived from 
inversions and/or stellar structural variation analyses. 
These sophisticated methods enable the inference of 
nearly model-independent constraints that can be 
incorporated in the Bayesian grid-based modelling 
approach to improve the precision on the stellar 
property determinations.

For targets without any seismic detection, it will
still be possible to infer the mass, radius, and age 
from non-seismic constraints. 
A minima, the fundamental properties of every star 
will be inferred from grid-based modelling based only 
on classical constraints -- effective temperature, 
metallicity, surface gravity, and luminosity when 
available. 
In addition, stellar ages will be provided from 
gyrochronology for targets whose rotation period 
could be determined \citep{GodoyRivera2021}, as 
well as from age-activity relations when activity 
levels could be inferred from the light curve \citep{Mathuretal2023}.
          
The different stellar parameter determinations produced 
by the pipeline will be delivered as intermediate data 
products. 
These will be important not only to ensure that ensemble 
studies using homogeneous stellar properties can later 
be performed by the community, but also to flag potential 
inconsistencies in the determinations themselves. 
Ultimately, the selection of the stellar mass, radius, 
and age for the pipeline data product (DP5) will follow a 
hierarchical ranking, starting on the most reliable 
inference procedure available for any given star. 
While the exact performance of the inference of stellar 
parameters remains to be determined by ongoing end-to-end 
tests of the stellar pipeline, hare-and-hounds exercises 
assuming 2 years of \plato\,observations and using an 
algorithm similar to the frequency fitting implemented 
in the stellar pipeline provide guidance on expected 
performance. 
Across stars of different masses and stellar physics, 
these exercises achieved relative accuracies (i.e., 
inferred values compared to the ground truth) better 
than 4\%, 1.5\%, and 10\% for mass, radius, and age, 
respectively \citep{Cunhaetal2021}.

\subsection{Expected seismic yield of PLATO}
\label{stellar_yields}

The ability to detect oscillation modes and the accuracy 
with which their frequency can be measured and analysed 
depend on several factors that are of astronomical, physical, 
instrumental, or operational origin: (i) stellar magnitudes, 
(ii) expected oscillation mode amplitudes and line-widths, 
(iii) noise, including photon noise, background noise, 
detector noise, jitter noise, etc..., (iv) total duration of 
the monitoring. 
\citet{goupil2024} have proposed a method to assess the 
probability of global detection of oscillation modes for the 
stars of the \plato\,Input Catalogue, as well as to estimate 
the achievable accuracy on the frequencies of individual modes. 
From these estimates, the expected precision on the mass, 
radius, and age of the stars can be derived, as shown in \citet{goupil2024}.

We have revisited Goupil et al's results using the new 
version of the PIC, in which we have considered all stars 
in samples P1, P2 and P5, while stars of sample P4 (M dwarfs) 
are too faint (or else have too low oscillation amplitude)
for their oscillations to be detectable with PLATO, and 
were therefore discarded from the study.

The detection probability $P_{\rm det}$ is defined as the 
probability to detect globally the oscillations in the power 
spectrum, not to be confused with the probability to detect 
and measure the properties of individual oscillation modes. 
The calculation is performed separately for samples P1, P2, P5, 
and for several values of the total duration $T_{\rm obs}$ of 
the monitoring: 30 days, 90 days, 180 days, 2 years, 4 years. 
It depends on the expected seismic  properties of each 
individual star, on the duration of the photometric monitoring 
and on the expected total noise level in the light curve. 
Seismic properties are computed from the physical characteristics 
of the stars (effective temperature, mass, radius) as available 
in the PIC, which also provides the expected total noise level. 
In all calculations, we have assumed a false alarm probability 
(probability that a peak appearing in the power spectrum is due 
to noise only) of 0.1\%. 
We have also calculated the expected uncertainty on the 
measurement of frequencies of individual modes with angular 
degree $\ell = 1$ in the region of maximum power of the 
spectrum, $\deltanu$. Each of these steps were performed as 
in \citet{goupil2024}.

For each target sample of \plato, we express the results 
in terms of (i) the number of stars for which the achieved 
probability of detection is $P_{\rm det} \geq 0.99$, and 
(ii) the subset of those for which an accuracy of individual 
mode frequencies $\deltanu \leq 0.2~\mu$Hz is also obtained. 
As discussed in \citet{goupil2024}, the latter criterion is 
empirically defined to yield a final precision better than 
10\% on the stellar ages, after modelling. 
All results are presented in Tables \ref{table1} and \ref{table2} 
for samples P1, P2 and P5. 
The gain of increasing the duration of monitoring is 
particularly obvious for the dwarfs in samples P1 and P5. 
Note in particular the important gain from 229 to 1253 
dwarfs with $P_{\rm det} \geq 0.99$ and $\deltanu \leq 0.2~\mu$Hz 
in sample P5, giving access to a precise characterisation of 
the stars and in particular to a good age estimate, 
when increasing the monitoring from 2 to 4 years.

The \plato\,Input Catalogue, in its version 2.2.0.1, includes 
stars which fall slightly outside the nominal \plato\,field for 
the planned first long pointing, as well as stars falling in 
the small gaps between CCDs. 
This is to allow for a potential slight misalignment of the 
line of sight with respect to the exact nominal field centre. 
These stars are easily identifiable in the catalogue because
they have the parameter 'EOLnCameraObsNCAM\_R' set to zero
(see Marrese et al. in prep.).
In the assessment presented here, we have considered these 
stars in the calculation of the oscillation detection probability 
and of the expected accuracy in the oscillation frequency 
measurements, but we have discarded them from the star counts 
presented below.

\begin{table}
          \centering
          \begin{tabular}{|c|c|c|c|c|c|c|c|}
               \hline\hline       
               \multirow{2}{*}{Sample} & \multirow{2}{*}{Total number in sample} & \multicolumn{6}{c|}{Star counts}               \\
                                       &                                         & Duration  & 30d  & 90d  & 180d & 2y    & 4y    \\
               \hline\hline
               \multirow{3}{*}{P1}     & \multirow{3}{*}{11018}                  & total     & 1977 & 3711 & 4962 & 7604  & 8704  \\
                                       &                                         & dwarfs    & 90   & 490  & 1043 & 3161  & 4246  \\
                                       &                                         & subgiants & 1887 & 3221 & 3919 & 4443  & 4458  \\
               \hline
               \multirow{3}{*}{P2}     & \multirow{3}{*}{712}                    & total     & 458  & 574  & 627  & 695   & 701   \\
                                       &                                         & dwarfs    & 172  & 279  & 332  & 399   & 404   \\
                                       &                                         & subgiants & 286  & 295  & 295  & 296   & 297   \\
               \hline 
               \multirow{3}{*}{P5}     & \multirow{3}{*}{157772}                 & total     & 1412 & 3908 & 6662 & 15660 & 22319 \\
                                       &                                         & dwarfs    & 0    & 0    & 0    & 314   & 1398  \\
                                       &                                         & subgiants & 1412 & 3908 & 6662 & 15346 & 20921 \\
               \hline\hline    
          \end{tabular}
          \caption{Number of stars with $P_{\rm det} \geq 0.99$}
          \label{table1}
\end{table}

\begin{table}
          \centering
          \begin{tabular}{|c|c|c|c|c|c|c|c|}
               \hline\hline       
               \multirow{2}{*}{Sample} & \multirow{2}{*}{Total number in sample} & \multicolumn{6}{c|}{Star counts}               \\
                                       &                                         & Duration  & 30d  & 90d  & 180d & 2y    & 4y    \\
               \hline\hline
               \multirow{3}{*}{P1}     & \multirow{3}{*}{11018}                  & total     & 1    & 1676 & 3745 & 7449  & 8691  \\
                                       &                                         & dwarfs    & 0    & 88   & 629  & 3009  & 4233  \\
                                       &                                         & subgiants & 1    & 1588 & 3116 & 4440  & 4458  \\
               \hline
               \multirow{3}{*}{P2}     & \multirow{3}{*}{712}                    & total     & 1    & 467  & 600  & 695   & 701   \\
                                       &                                         & dwarfs    & 0    & 182  & 305  & 399   & 404   \\
                                       &                                         & subgiants & 1    & 285  & 295  & 296   & 297   \\
               \hline 
               \multirow{3}{*}{P5}     & \multirow{3}{*}{157772}                 & total     & 0    & 723  & 4214 & 14215 & 21119 \\
                                       &                                         & dwarfs    & 0    & 0    & 0    & 229   & 1253  \\
                                       &                                         & subgiants & 0    & 723  & 4214 & 13986 & 19866 \\
               \hline\hline    
          \end{tabular}
          \caption{Number of stars with $P_{\rm det} \geq 0.99$  and $\deltanu \leq 0.2 ~ \mu$Hz}
          \label{table2}
\end{table}

We have also identified the stars in samples P1 and P2 for which 
we obtain an accuracy on individual mode frequencies 
$\deltanu \leq 0.1~\mu$Hz. 
These stars are expected to be best characterised and will 
certainly constitute the stellar seismic \plato\,legacy sample. 
Indeed their high frequency precision is expected to lead to 
strict constraints on the physics of the stellar models, 
improving the age characterisation of all stars of similar 
type. 
For both samples, a comparison of the numbers of stars for 
which $\deltanu \leq 0.2~\mu$Hz and $\leq 0.1~\mu$Hz is 
given in Table~\ref{table3}. 
For sample P2, we find that the best accuracy on individual 
mode frequencies is reached for almost all stars, while 
the number of such stars in sample P1 is roughly divided by 
two when going from $\deltanu \leq 0.2~\mu$Hz to $0.1~\mu$Hz.

\begin{table}
          \centering
          \begin{tabular}{|c|c|c|c|c|c|c|c|}
               \hline\hline       
               \multirow{3}{*}{Sample} & \multirow{3}{*}{Total number in sample} & \multicolumn{5}{c|}{Star counts} \\
                                   &                        & Duration                      & 2y          & 2y          & 4y          & 4y          \\
                                   &                        & $\deltanu \leq$               & $0.2~\mu$Hz & $0.1~\mu$Hz & $0.2~\mu$Hz & $0.1~\mu$Hz \\          
               \hline\hline
               \multirow{3}{*}{P1} & \multirow{3}{*}{11018} & total                         & 7449        & 3984        & 8691        & 6279        \\
                                   &                        & dwarfs                        & 3009        & 782         & 4233        & 2080        \\
                                   &                        & subgiants                     & 4440        & 3202        & 4458        & 4199        \\
               \hline
               \multirow{3}{*}{P2} & \multirow{3}{*}{712}   & total                         & 695         & 635         & 701         & 694         \\
                                   &                        & dwarfs                        & 399         & 340         & 404         & 398         \\
                                   &                        & subgiants                     & 296         & 295         & 297         & 296         \\
               \hline\hline    
          \end{tabular}
          \caption{Number of stars with $P_{\rm det} \geq 0.99$  and $\deltanu \leq 0.2 ~ \mu$Hz or $0.1 ~ \mu$Hz.}
          \label{table3}
     \end{table}
     
\begin{figure}
          \centering
          \begin{tabular}{cc}
               \includegraphics[width=0.5\linewidth]{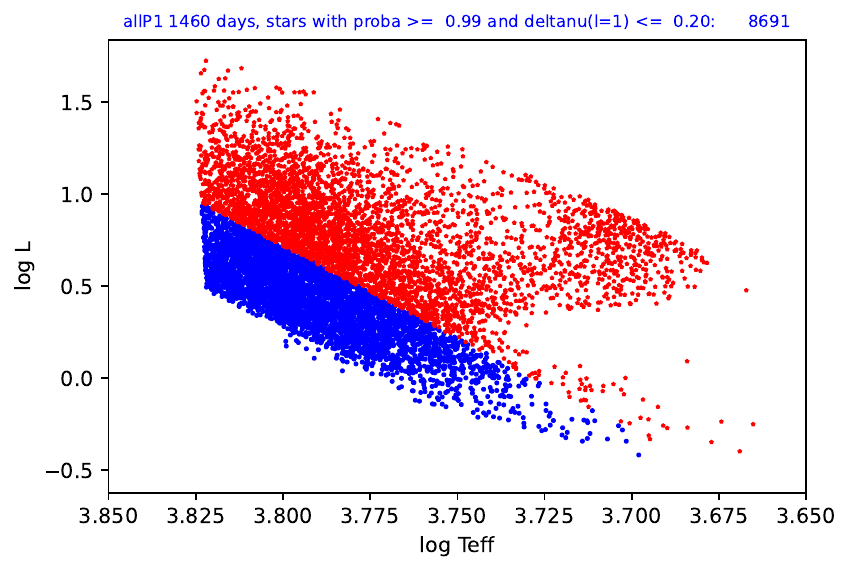} &
               \includegraphics[width=0.5\linewidth]{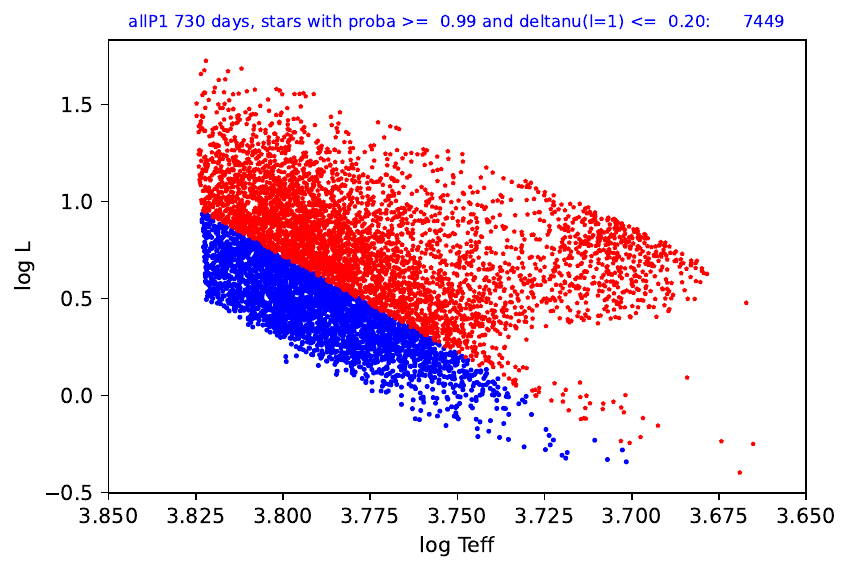} \\
               \includegraphics[width=0.5\linewidth]{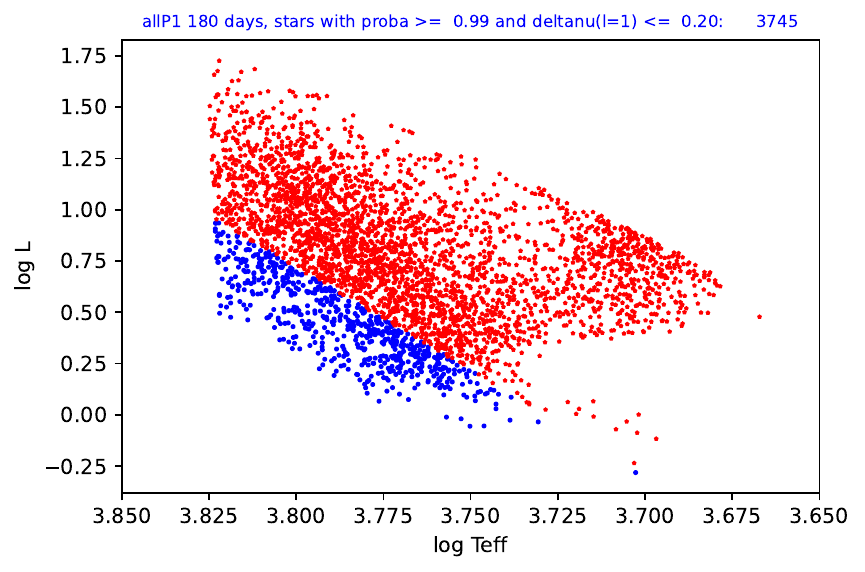} &
               \includegraphics[width=0.5\linewidth]{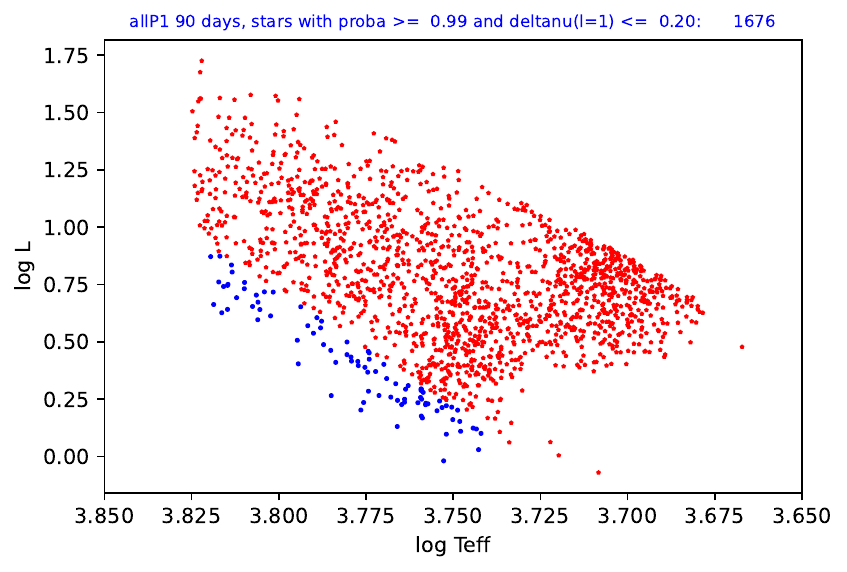}
          \end{tabular}
          \caption{HR diagram of P1 stars with $P_{\rm det} \geq 0.99$ and $\deltanu \leq 0.2~\mu$Hz. From left to right and top to bottom: 4 years, 2 years, 180 days, 90 days. Blue: dwarfs, red: subgiants}
          \label{P1HRD}
     \end{figure}

\begin{figure}
          \centering
          \begin{tabular}{cc}
               \includegraphics[width=0.5\linewidth]{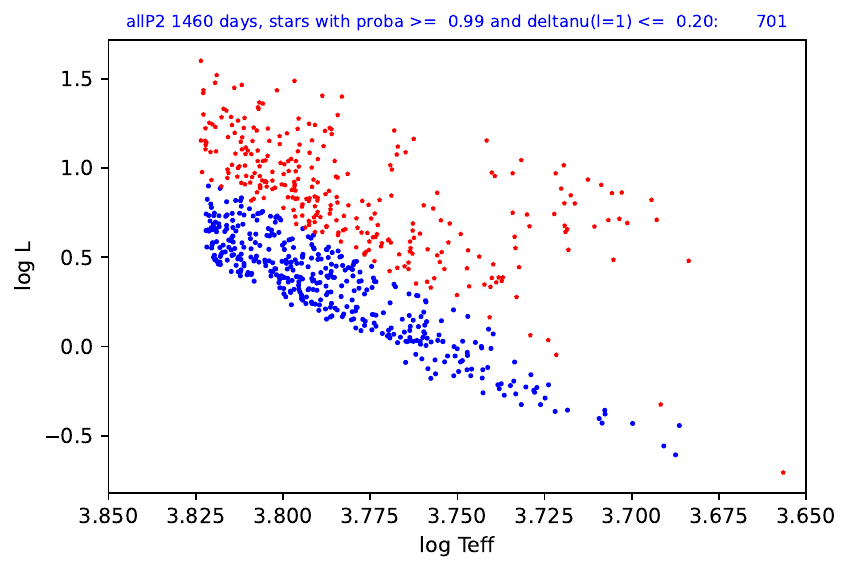} &
               \includegraphics[width=0.5\linewidth]{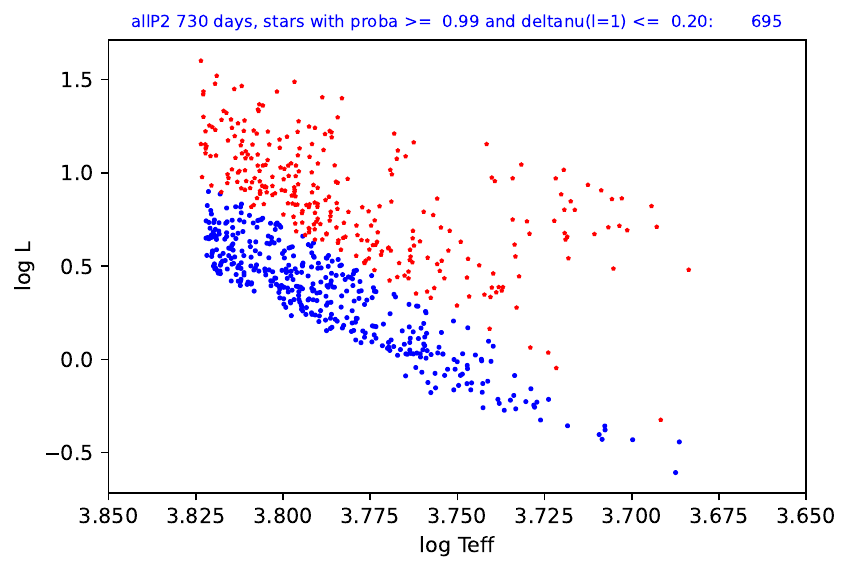} \\
               \includegraphics[width=0.5\linewidth]{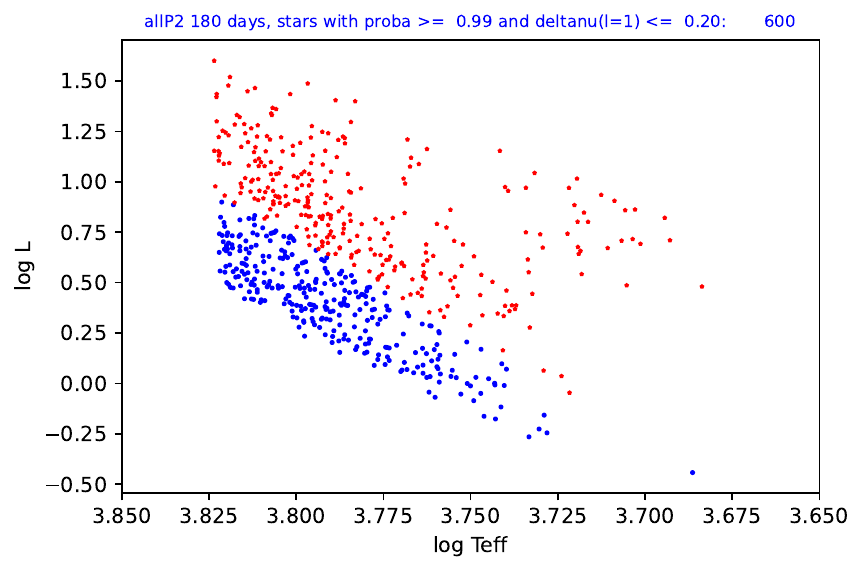} &
               \includegraphics[width=0.5\linewidth]{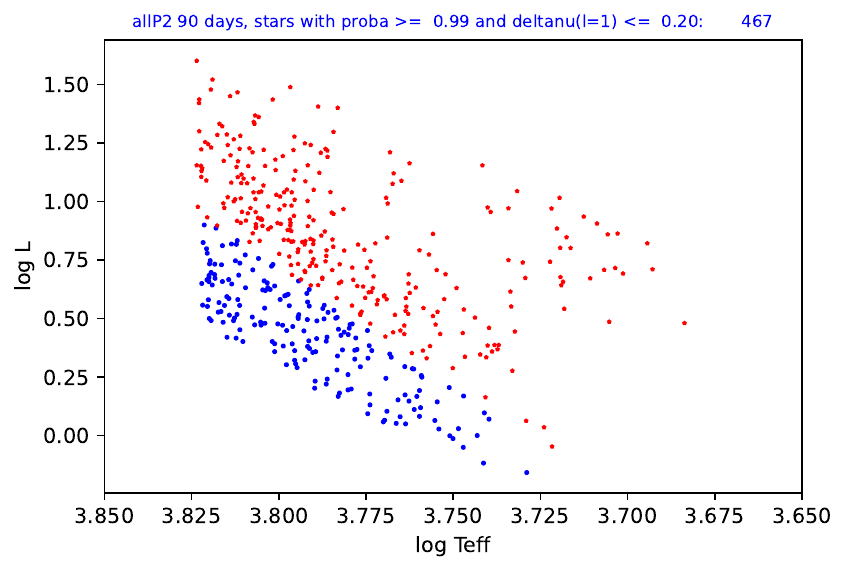}
          \end{tabular}
          \caption{Same as Fig.~\ref{P1HRD}, but for the P2 sample}
          \label{P2HRD}
     \end{figure}

As another illustration of the gain obtained when increasing the 
duration of the monitoring, Fig.~\ref{P1HRD} (resp. \ref{P2HRD}) 
presents, for sample P1 (resp. P2), the location in the HR 
diagram of the stars for which the expected frequency 
uncertainty $\deltanu$ is lower than $0.2~\mu$Hz. 
Each panel correspond to a different monitoring duration, 
respectively 4 years, 2 years, 180 days and 90 days. 
Dwarfs and subgiants are distinguished, using Eq. 16 
of \citet{goupil2024}. 
These figures clearly show the gain in the number of 
dwarfs with positive mode detection and good individual 
frequency accuracy, when increasing the monitoring duration.

Finally, we have calculated the expected performance of the two 
fast cameras, considered either separately or together. 
For these calculations, we have used the expected noise-to-signal 
ratios available in the PIC 2.1 release, which have not changed 
in the meantime. 
For the combination of the data from both blue and red fast cameras, 
we have assumed that the noises in both cameras are uncorrelated, 
leading to a resulting noise-to-signal ratio calculated as:
     \begin{equation}
          N_{\rm tot} = \frac{\sqrt{N_B^2 + r^2 N_R^2}}{1+r}
     \end{equation}
     where $N_B$ (resp. $N_R$ and $N_{tot}$) is the noise in the blue fast camera channel (resp. the red fast camera channel and the combined channel), and $r = F_R/F_B$ is the expected flux ratio between the red and the blue cameras. For the purposes of these calculations, we set $r = 0.81$ for all targets. We checked a posteriori that the resulting yields do not depend significantly on the exact value of $r$. The results for the fast cameras are presented in Tables \ref{table4} and \ref{table5}.

     \begin{table}
          \centering
          \begin{tabular}{|c|c|c|c|c|c|}
               \hline\hline
               \multirow{2}{*}{Type of star} & \multirow{2}{*}{Fast camera channel} & \multicolumn{4}{c|}{Duration of monitoring and star counts} \\
                                             &                                      & ~~ 90d ~~ & ~~ 180d ~~ & ~~ 2y ~~ & ~~ 4y ~~  \\
               \hline\hline
               \multirow{3}{*}{total}        & blue                                 & 20        & 24         & 50       & 70        \\
                                             & red                                  & 15        & 20         & 39       & 60        \\
                                             & both                                 & 29        & 41         & 80       & 94        \\
               \hline
               \multirow{3}{*}{dwarfs}       & blue                                 & 2         & 3          & 11       & 18        \\
                                             & red                                  & 0         & 2          & 7        & 13        \\
                                             & both                                 & 3         & 8          & 20       & 24        \\
               \hline
               \multirow{3}{*}{subgiants}    & blue                                 & 18        & 21         & 39       & 52        \\
                                             & red                                  & 15        & 18         & 32       & 47        \\
                                             & both                                 & 26        & 33         & 60       & 70        \\
               \hline\hline
          \end{tabular}
          \caption{Number of stars (out of 215 total) with $P_{\rm det} \geq 0.99$ for sample P2 and for the fast cameras.}
          \label{table4}
     \end{table}

     \begin{table}
          \centering
          \begin{tabular}{|c|c|c|c|c|c|}
               \hline\hline
               \multirow{2}{*}{Type of star} & \multirow{2}{*}{Fast camera channel} & \multicolumn{4}{c|}{Duration of monitoring and star counts} \\
                                             &                                      & ~~ 90d ~~ & ~~ 180d ~~ & ~~ 2y ~~ & ~~ 4y ~~  \\
               \hline\hline
               \multirow{3}{*}{total}        & blue                                 & 5         & 16         & 42       & 67        \\
                                             & red                                  & 4         & 14         & 35       & 63        \\
                                             & both                                 & 11        & 30         & 71       & 89        \\
               \hline
               \multirow{3}{*}{dwarfs}       & blue                                 & 0         & 2          & 9        & 17        \\
                                             & red                                  & 0         & 2          & 6        & 13        \\
                                             & both                                 & 2         & 6          & 19       & 24        \\
               \hline
               \multirow{3}{*}{subgiants}    & blue                                 & 5         & 14         & 33       & 50        \\
                                             & red                                  & 4         & 12         & 29       & 40        \\
                                             & both                                 & 9         & 24         & 52       & 65        \\
               \hline\hline
          \end{tabular}
          \caption{Number of stars (out of 215 total) with $P_{\rm det} \geq 0.99$ and $\deltanu \leq 0.2~\mu$Hz for sample P2 and for the fast cameras.}
          \label{table5}
     \end{table}

\section{Phase curves of exoplanets in the LOPS2 field}
\label{section:phase_curves}

We checked the transit, occultation, and phase curve properties of all known exoplanet 
(as of 2026, February 3) in the PLATO LOPS2 field. 
For this purpose we downloaded the list of known exoplanets from NASA Exoplanet 
Archive\footnote{https://exoplanetarchive.ipac.caltech.edu/}. 
In the next step we used the code \textsc{Plato utilities} to check which of these 
exoplanets are in the LOPS2 field and how many of them will be in the field of view 
of the fast and normal cameras of \plato. 
We have found 112 transiting exoplanet systems with RV-measured masses in the field, 
50 radial velocity systems without exhibiting transits, 4 transiting exoplanet systems 
with TTV-measured planetary masses and 3 directly imaged planetary systems. 
Note that \citet{nascimbeni2025} found 108 transiting systems in LOPS2. 
The list of these planets in LOPS2 can be found in 
Tables~\ref{tab:list_of_known_planets_part1} - \ref{tab:list_of_known_planets_part3}

\subsection{Transit and occultation probabilities}
\label{subsection:transit_occultation_probabilities(phase_curves)}

We computed the geometric probabilities of the transits and the occultations of already 
known planets in the LOPS2 field of PLATO. 
These probabilities were calculated via the following equations taken from \citet{transit_occ}:
 \begin{equation}
     P_{tra}= \left ( \frac{R_\star\pm R_p}{a} \right )\cdot\left ( \frac{1+e\sin\omega}{1-e^2} \right ) \label{eq:tra}
 \end{equation}
\begin{equation}
    P_{occ}= \left ( \frac{R_\star\pm R_p}{a} \right )\cdot\left ( \frac{1-e\sin\omega}{1-e^2} \right )\label{eq:occ}
\end{equation}
where $a$ is the semi-major axis of the orbit, $R_\star$ is the radius of the star, 
R$_p$ is the radius for the planet, $e$ is the orbital eccentricity and $\omega$ is 
the argument of periastron, respectively. 
The minus sign provides the probability of full transits and the plus sign the 
probability of all transit types (including those grazing). 
The uncertainties were estimated by taking the standard deviation of 5,000 realizations, 
each of which were perturbed by normal distribution of the input parameters. 
The results are reported in Tables~\ref{tab:calc1} - \ref{tab:calc6}.

\subsection{Phase curve amplitudes}
\label{subsection:phase_curve_amplitudes(phase_curves)}

\subsubsection{Input values}

Since some of the data were not reported in the references at the NASA Exoplanet Archive, 
especially for the exoplanets detected by radial velocity measurement, we used the 
available data to make an estimation of these the missing parameters necessary 
for the calculations. When the needed data were provided but without error bars, 
we arbitrarily assumed 15\% error for that. 

If the semi-major axis was not available, we calculated it from Kepler's third law in 
AU from the mass of the star (M$_\star$ in solar units) and the measured period 
(P in years) via Kepler's equation:
\begin{equation}
a = \left(M_{\star} P^2\right)^{1/3}
\end{equation}

The stellar radius (R$_\star$) was also not available in some cases. 
In the omitted cases we estimated it based on the following equation, 
using the mass of the star (M$_\star$) and temperature:
\begin{equation}
R_{\star} = \sqrt{\frac{0.9\, M_{\star}^{0.95}}{T^{4}}}\label{rstar}
\end{equation}
In Equation (\ref{rstar}) every parameters must be substituted in solar units.

For the unknown planetary radius, we estimated the R$_{p}$ based on 
the known masses with the equation from \citet{masses}.
\begin{equation}
R_p =
\begin{cases}
1.02\, M_p^{0.27} & M_p < 4.37 \\
0.56\, M_p^{0.67} & 4.37 < M_p < 127 \\
18.6\, M_p^{-0.06} & M_p > 127
\end{cases}
\end{equation}
In the case of radial velocity-planets, only the minimum mass is given. 
We used this minimum mass estimate in the above equation. If the true mass 
is bigger, then the estimated radius of the planet and its reflection 
amplitude and transit probability are also higher than the values 
reported here.

Using these equations and error estimation method we run an MonteCarlo 
simulation to determine the final probabilities for the transits and 
occultations (Eq. \ref{eq:tra} and Eq. \ref{eq:occ}). 
The table used for the calculations and the table with the results
together with the codes used for calculating the probabilities and 
amplitudes can be found in Zenodo (\url{place_of_the_link}).

\subsubsection{Amplitudes of phase curve components}

The phase angle (denoted by $\alpha$) is the observer - planet - star 
angle and it is related to the inclination ($i$) of the orbit and the 
argument of periastron as
\begin{equation}
    \cos \alpha = \cos(v + \omega) \sin i
\end{equation}
For sake of simplicity we took $\sin i = 1$ for the subsequent 
predictions. In case of transiting exoplanets, inclination is 
necessarily close to $90^\circ$ while in case of non-transiting 
exoplanets we do not know its value. 

The phase curve of the exoplanet system was decomposed as
\begin{eqnarray}
\label{eq:phase_curve}
    F_{\mathrm{phase~curve}} & = & A_{\mathrm{day}} \Phi(\alpha) + A_\mathrm{night} (1 - \Phi(\alpha)) \nonumber \\
    & + & A_{\mathrm{reflection}} \Phi(\alpha) \nonumber \\
    & - & A_{\mathrm{beaming}} (e \cos \omega + \cos(v+\omega)) \\
    & + & A_{\mathrm{ellipsoidal}} \sin^2 (v+\omega) \nonumber
\end{eqnarray}
The first line on the right hand side characterizes the dayside and 
nightside thermal emission of the exoplanet, the second line does the 
reflection effect of the planet, the third and fourth lines yield the 
beaming and the ellipsoidal effect occurring on the star. 
As \citet{csizmadia23}, we use the assumption here that the very same 
phase-function acts for characterising the thermal emission from the 
dayside and the nightside and also for the reflection effect. 
Our goal is to report the expected $A$ amplitudes according to the best 
present knowledge of the systems and using best available estimates 
for the geometric albedo.

The nightside emission is usually very small in close-in exoplanets 
as the heat-redistribution is very inefficient from the dayside to the nightside. 
This is supported by observations and theory, and the reason is lack of 
atmosphere on one side (for instance, in the case of atmosphere-less rocky 
planets. e.g. \citealt{leger11}). On the other side, for strongly irradiated, 
synchronously rotating exoplanets, the weak night-side thermal emission is 
generally interpreted as evidence that advective transport does not efficiently 
redistribute absorbed stellar energy before it is re-radiated, thereby 
producing a large longitudinal temperature contrast; in several cases, 
night-side condensate clouds may additionally mute the escaping thermal 
radiation \citep{2024NatAs...8..879B,2024MNRAS.531.1056R,2025PNAS..12216190K}. 
Planets far-away from their stars have very low nightside emission because 
they receive very small amount of stellar insolation. 
That is why we can safely neglect the nightside emission and take $A_{\mathrm{night}} \approx 0$. 
While \plato\,will definitely be able to detect the nightside emission of 
several exoplanets, it contributes very little to the total phase curve 
amplitude, hence we do not include it into the prediction.

The equilibrium temperature of the planet is given by
\begin{equation}
\label{eq:equilibrium_temperature}
    T_\mathrm{eq} = T_\ast \sqrt{\frac{R_\star}{2 a}} \sqrt{\frac{1+e \cos v}{1-e^2}} (1-A_B)^{1/4}
\end{equation}
where $T_\star$ is the stellar effective temperature, $A_B$ is the Bond-albedo, 
and $v$ is the true anomaly (periastron point - star - planet angle). 
We utilize the Lambertian phase function which yields $A_B = \frac{3}{2} A_g$ where $A_g$ is the geometric albedo. 
Eq. (\ref{eq:equilibrium_temperature}) supposes that the planet reacts immediately to 
the change in stellar instellation which is a good approximation for the present 
purposes. The effect of the eccentric orbits is taken into account in the second 
square root.

The dayside emission is characterized via
\begin{equation}
    A_{\mathrm{day}} (v) = \left( \frac{R_\mathrm{p}}{R_\star} \right)^2 \frac{\int \lambda S(\lambda) B_\lambda (T_{eq}) d\lambda}{\int \lambda S(\lambda) B_\lambda (T_\star) d\lambda}
\end{equation}
in units of the stellar flux. Here $\lambda$ is the wavelength, $S$ is the response 
function of \plato~and we roughly approximate the emission of the star and the planet 
with the Planck-function $B$. Not that this amplitude is phase dependent on eccentric 
orbits as $T_{eq}$ depends on the true anomaly (cf. Eq.~\ref{eq:equilibrium_temperature}). 

The uncertainties of the T$_{eq}$ temperature and the amplitude of thermal emission were estimated via:
\begin{equation}
     \sigma_{T_{eq},rel}\sqrt{\left (  \frac{\sigma T_{eff}}{T_{eff}}\right )^2+\left ( \frac{1}{2}\frac{R_\star}{R_\star}\right )^2+\left ( \frac{1}{2}\frac{\sigma a}{a} \right )^2}
\end{equation}
and, based again on a Monte-Carlo simulation, via
\begin{equation}
    \sigma A_{thermal}= stddev \left[ \left ( \frac{F(T_{eq}+\sigma T_{eq})-F(T_{eq})}{F(T_{eff})} \right )\left ( \frac{R_p}{R_\star} \right )^2 \right]
\end{equation}

\begin{figure}
    \centering
    \includegraphics[width=0.95\linewidth]{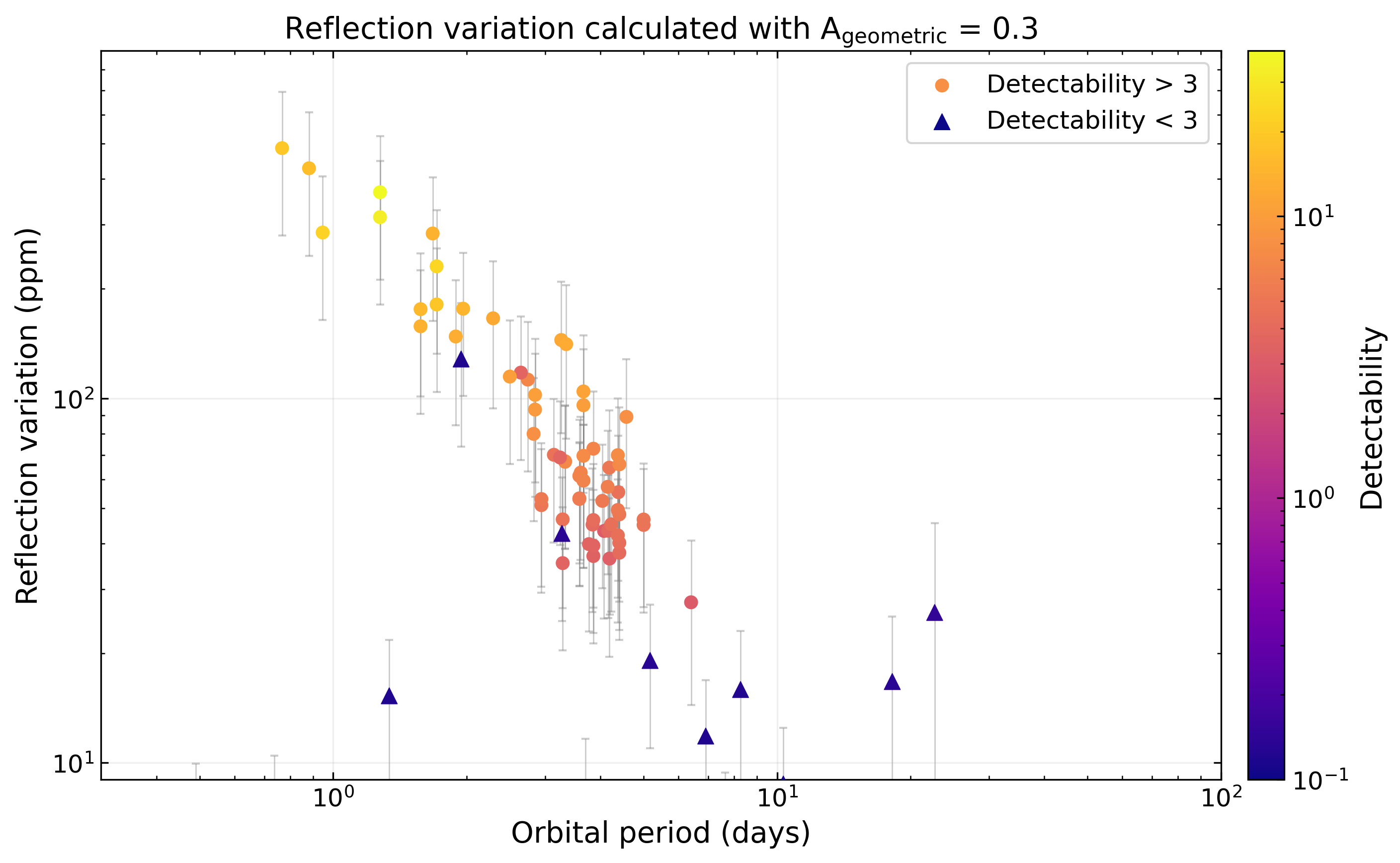}
    \caption{The estimated reflection amplitudes with A$_{\mathrm{geometric}}$=0.3 plotted as a function of the orbital period. 
    The points are colour coded based on the calculated detectability value (but set to blue if it is not detectable). 
    Note that if a system had multiple solutions in the NASA Exoplanet Archive we plotted the expected amplitudes for all solutions.}
    \label{fig:ref03}
\end{figure}

\begin{figure}
    \centering
    \includegraphics[width=0.95\linewidth]{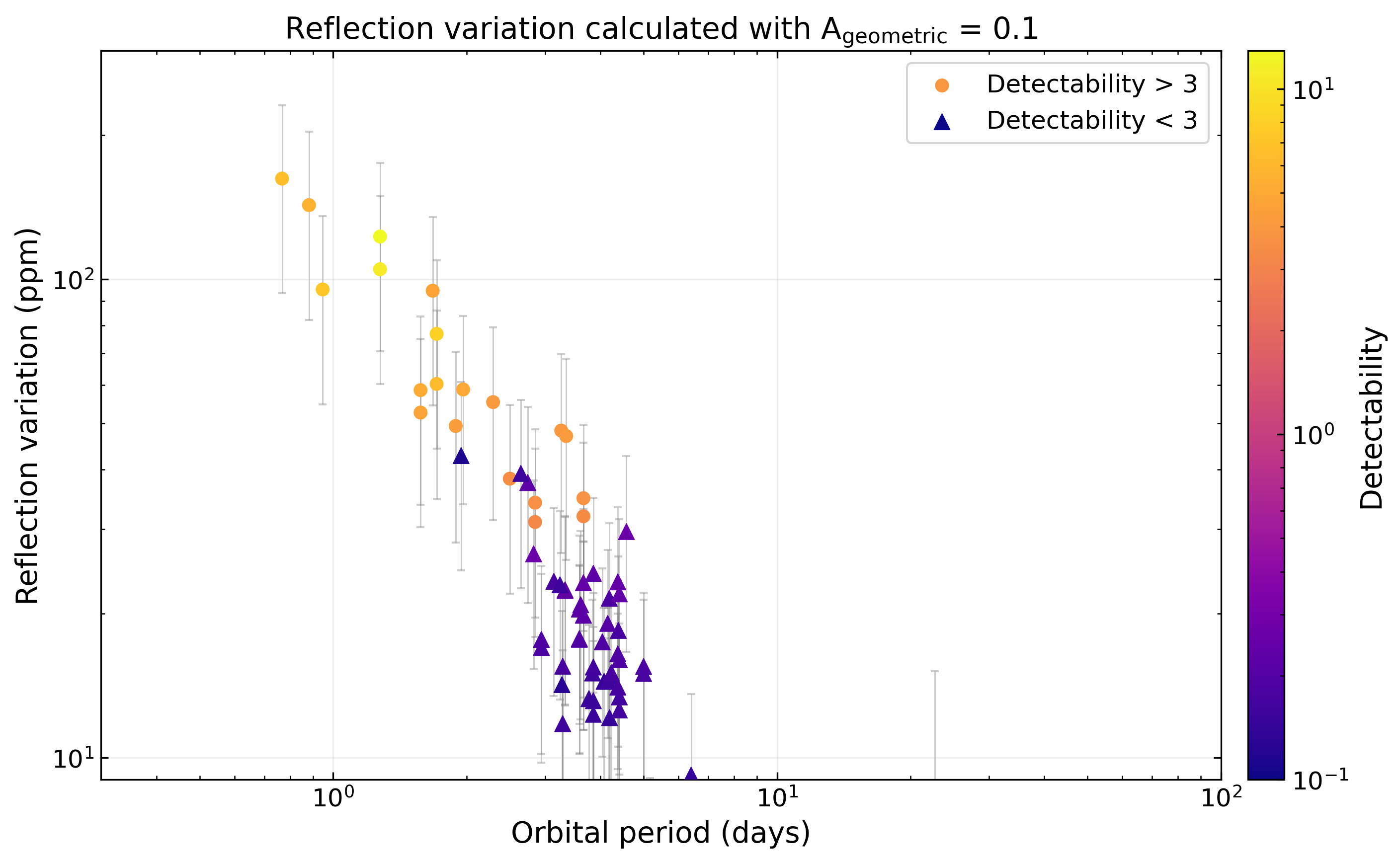}
    \caption{The estimated reflection amplitudes with A$_{\mathrm{geometric}}$=0.1. See Figure~\ref{fig:ref03} for explanation.}
    \label{fig:ref01}
\end{figure}

The amplitudes of the reflection, beaming and ellipsoidal effects were calculated utilizing the equations given in \citet{csizmadia20} and references therein. 
For the reflection amplitude we used the following equation:
\begin{equation}
    A_{\mathrm{refl}} = A_\mathrm{geometric} \left(\frac{R_{p}}{a}\right)^2 \left( \frac{1+e \cos v}{1-e^2}\right)^2
\end{equation}
where we applied two values of the geometric albedos. $A_\mathrm{geometric} = 0.3$ was assumed first, 
independent of their type (rocky planets with atmosphere or atmosphere-less, Neptunes, cold, 
warm or hot gas giants). For hot Jupiters the geometric albedo is lower, usually it is 
around $A_\mathrm{geometric} = 0.1$. That is why we included the reflection amplitudes determined with 
both 0.3 and 0.1 in the Tables~\ref{tab:calc1} - \ref{tab:calc6} and in Figures~\ref{fig:ref03} and \ref{fig:ref01}.

The uncertainty of the estimate was calculated with error propagation:
\begin{equation}
    \sigma_{\mathrm{A_{refl}}} = A_\mathrm{refl}
\sqrt{ 2\left[
\left( \frac{\sigma_{R_p}}{R_{p}}\right)^2 +
\left(\frac{\sigma_a}{a}\right)^2 +
\left(\frac{\sigma_e}{1-e}\right)^2  \right] }
\end{equation}
$\sigma_i$ means the uncertainties of the corresponding parameter. 

\begin{figure}
    \centering
    \includegraphics[width=0.95\linewidth]{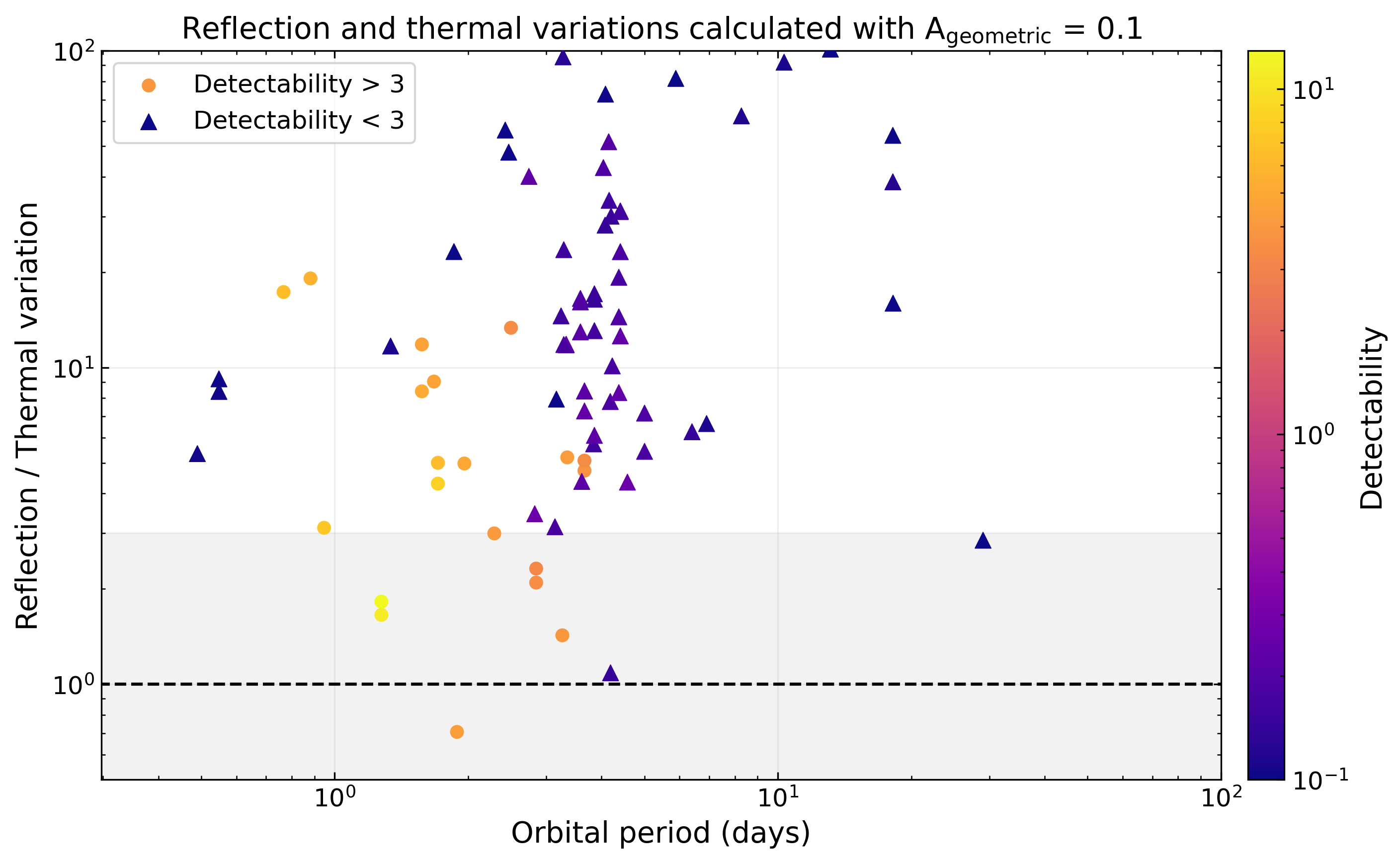}
    \caption{The ratio of the estimated reflection amplitudes with A$_{\mathrm{geometric}}$=0.1 to the thermal emission of the planet 
    in the PLATO passband as a function of the orbital period. See Figure~\ref{fig:ref03} for explanation of symbols. 
    Note that double solutions close to each other at the given period are due to the same system as the system has multiple solutions.}
    \label{fig:th_refl_ratio}
\end{figure}

On eccentric orbits the maximum of the observable thermal emission curve and the reflection curve is not necessarily at phase 0.5. 
This is why we calculate the amplitude of the effect from the expressions of $A \times \Phi(\alpha)$ by searching for the maximum values 
for each term in Eq.~\ref{eq:phase_curve}. See the results in Tables \ref{tab:calc1}-\ref{tab:calc6}.

We also determined the predicted ratio of the reflected light to the thermal emission and the results are plotted in Figure~\ref{fig:th_refl_ratio}. 
As is expected, majority of the exoplanets have much larger reflection effect than thermal emission. This is due to the fact that the thermal 
components is dominating in near-infrared as it is characterized by the equilibrium temperature which is usually between a few hundred Kelvins 
to maximum $\sim 2000$K. The optical wavelength regime where \plato\,will perform its observations is dominated by the reflected light. 
We identified only 7 systems where the thermal emission components is larger than 1/3 of the reflection component - in every other planet 
the thermal components is smaller or much smaller than this ratio. Only 5 of them could be detected (Figure~\ref{fig:th_refl_ratio}). 
There is only one system where the expected thermal emission is stronger than the reflected component. This is HATS-70b. 
This object has a mass of $12.9_{-1.6}^{+1.8}$ Jupiter masses. This puts it into the transition range between giant exoplanets and brown dwarfs. 
Its orbital period is also short ($P<2$ days) and its light collecting area is big (it has $\sim30$ Erth-radii diameter) 
which allows it to receive a lot of stellar instellation. Other systems where the thermal emission contributes significantly to the observable 
phase curve will be HATS-40b, HATS-42b, WASP-100b and WASP-121b.

The beaming amplitude was calculated via the following steps. First, we calculated the amplitude of the radial velocity of the star:
\begin{equation}
    K =
\frac{2\pi a}{P\sqrt{1-e^2}}
\frac{M_p}{M_\star + M_p}
\end{equation}
where $P$ is the period. We assumed that the uncertainty of the radial velocity amplitude ($\sigma$K) is uniformly 15\% of the K value. After that we have got the beaming amplitude as:
\begin{equation}
    A_{\mathrm{beam}} = \alpha \frac{K}{c}
\end{equation}
where $c$ is the speed of light.

The spectral index is considered as a function of the stellar effective temperature in the following form:
\begin{eqnarray}
    \alpha & = & 5 + 72.053(\pm1.579)  \nonumber \\
           & + & 51.998(\pm1.157) \frac{x e^{1.27071(\pm0.001744)x}}{1-e^x} \nonumber \\
           & + & 37.2687(\pm0.9884) x^2\nonumber \\
           & - & 3.7051(\pm0.1309) x^3 \\ 
           & + & 0.748143(\pm0.02727) x^4 \nonumber
\end{eqnarray}
with 
\begin{equation}
    x = \frac{T_\mathrm{eff}}{5775 \mathrm{K}}
\end{equation}
The calculation f of the spectral index have been done in the same way as described in Section 2.6.2 of \citet{csizmadia20}.

The uncertainties of beaming were calculated via:
\begin{equation}
    \sigma_{A_{\mathrm{beam}}} = A_{\mathrm{beam}} \sqrt{\left(\frac{\sigma_\alpha}{\alpha}\right)^2 + \left(\frac{\sigma_K}{K}\right)^2}
\end{equation}

\begin{figure}
    \centering
    \includegraphics[width=0.95\linewidth]{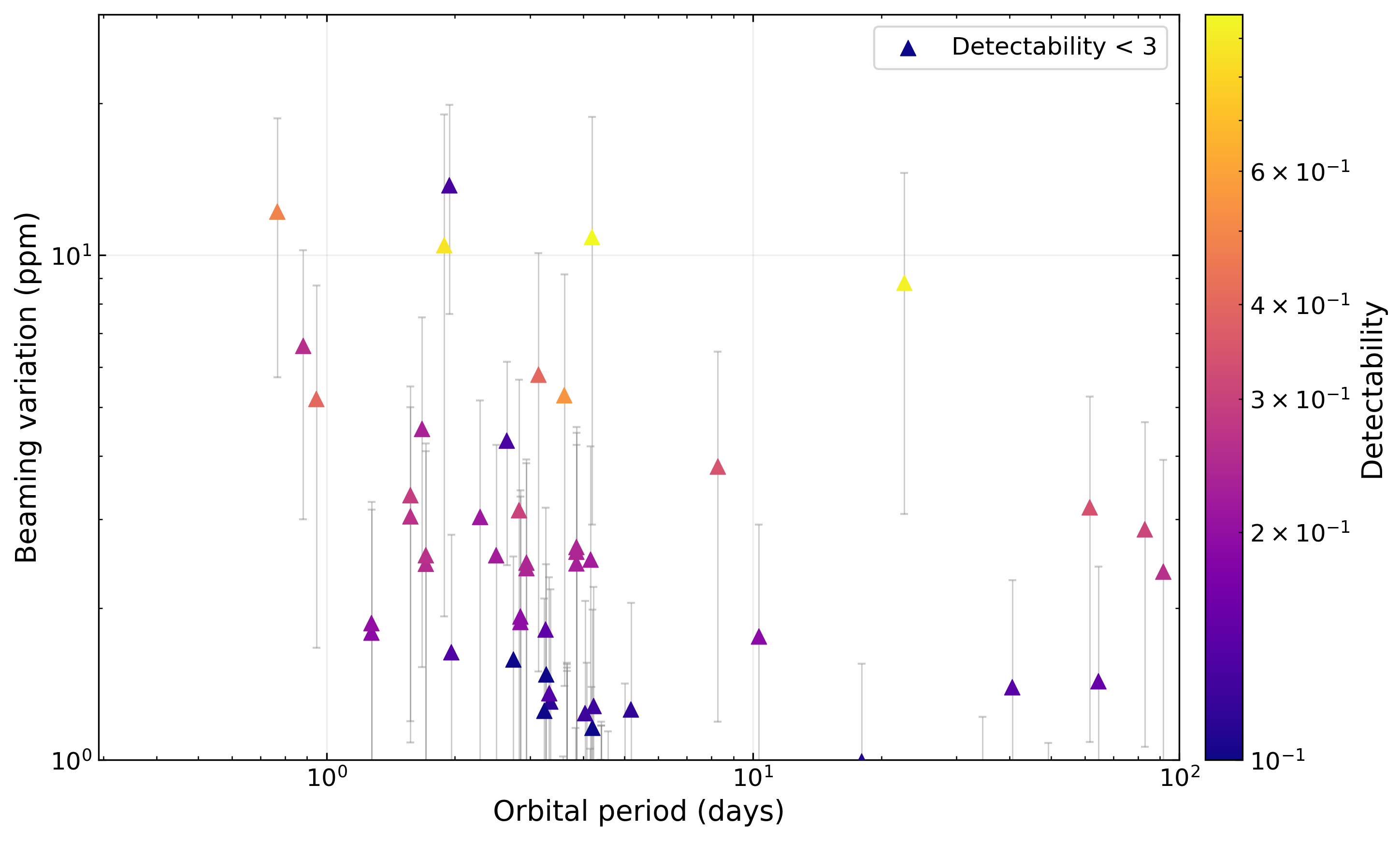}
    \caption{Beaming amplitudes in the function of the orbital period. See Figure~\ref{fig:ref03} for the meaning of symbols.}
    \label{fig:beam}
\end{figure}

The results are plotted in Figure \ref{fig:beam}.

The amplitude of the ellipsoidal effect and its uncertainty was estimated in the following way. We denoted the mass ratio by $q$:
\begin{equation}
    q = \frac{M_p}{M_\ast}
\end{equation}
and its uncertainty as:
\begin{equation}
    \sigma_q =
q\sqrt{
\left(\frac{\sigma_{M_p}}{M_p}\right)^2 +
\left(\frac{\sigma_{M_\ast}}{M_\ast}\right)^2
}
\end{equation}
With this $q$ value, the amplitude of the ellipsoidal effect is the following up to the first order:
\begin{equation}
    A_{\mathrm{ell}} \approx \frac{3}{2}\,q\left(\frac{R_\ast}{a}\right)^3
\end{equation}
Since we need just a first order approximation here, we do not use the more precise expressions presented in \citet{csizmadia23} which takes the effect of limb- and gravity darkening into account. We assumed these are in the order of unity here.

\begin{figure}
    \centering
    \includegraphics[width=0.95\linewidth]{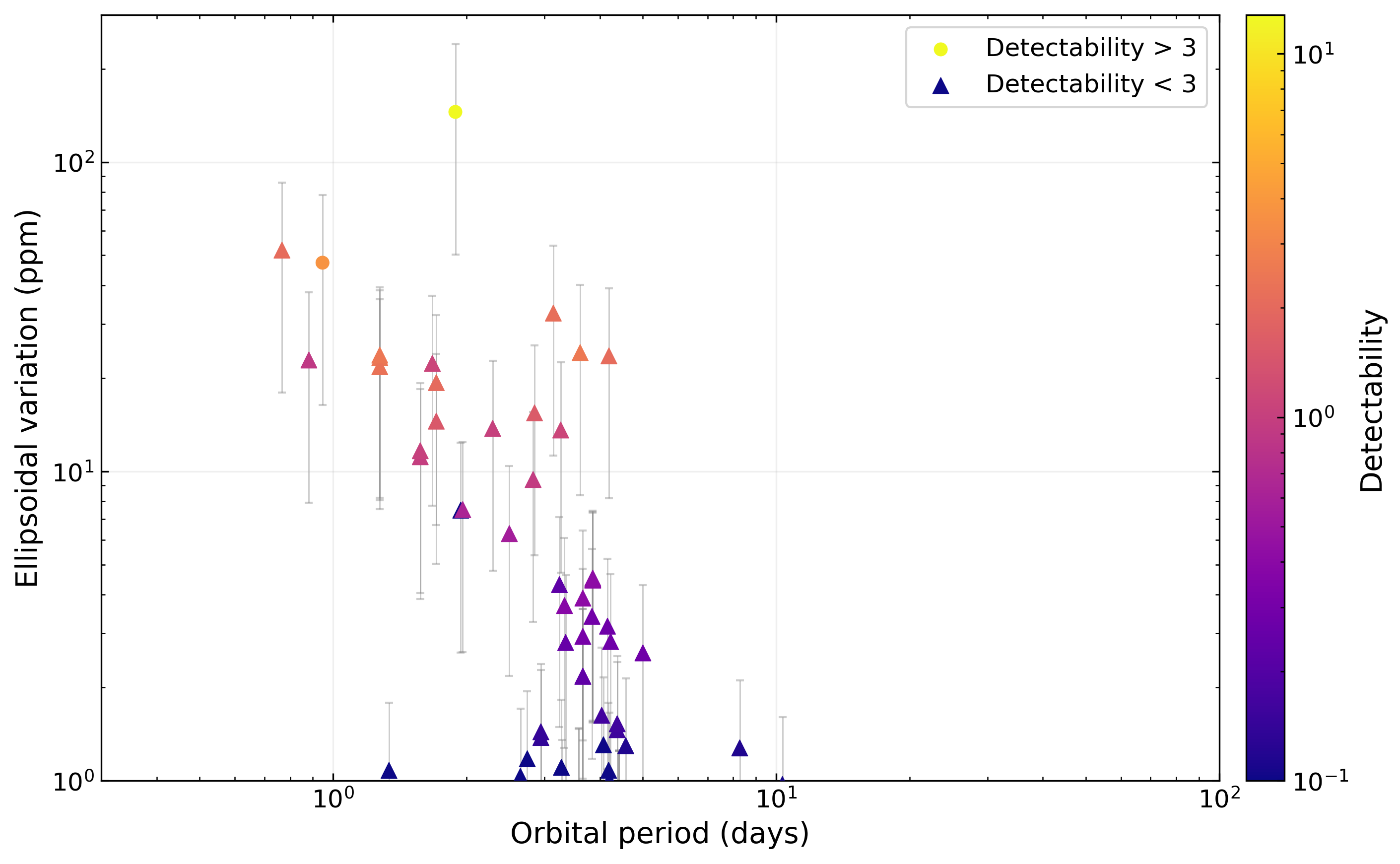}
    \caption{The calculated ellipsoidal amplitudes in the function of the orbital period. We indicated the 3 year period with a vertical gray line - this is the expected length of the observations of PLATO in the LOPS2 field. See Figure~\ref{fig:ref03} for the meaning of symbols.}
    \label{fig:ell}
\end{figure}

The relative error was calculated with the following formula:
\begin{equation}
    \sigma_{A_\mathrm{ell}} = A_\mathrm{ell} \sqrt{\left(
\frac{\sigma_q}{q} \right)^2 + 9 \left(\frac{\sigma_{R_\ast}}{R_\ast} + \frac{\sigma_a}{a}\right)^2}
\end{equation}
The results are plotted for the reflection, beaming and ellipsoidal amplitudes with their 1$\sigma$ uncertainties in Figures~\ref{fig:ref01} - \ref{fig:ell} and presented in Tables  \ref{tab:calc1} - \ref{tab:calc6}.

\subsubsection{Estimation of detectability}

After the aforementioned amplitude-estimations we made an attempt to predict the detectability 
of the the phase curve components. The number of data points during the \plato\,observations are optimistically taken as:
\begin{equation}
    N_D=N_{tr}\cdot\frac{P}{25s}
\end{equation}
where $N_{tr}=3 \times 365.25/P$ and $P$ is the orbital period in days from the original table - here we assumed a 3 years long observational interval.
The following step was the determination of the photometric error based on the know effects like the jitter, the photon and readout noises.
\begin{equation}
    \sigma^2_{phot}=(9\cdot10^{-6})+\left [ \frac{1}{f} +\left ( \frac{200}{f} \right )^2\right ]\cdot\frac{1}{N_{cams}}\cdot\frac{3600s}{25s}
\end{equation}
where $N_{cams}$ is the number of the PLATO cameras which will measure the exoplanet and $f$ is the flux based on the following equation:
\begin{equation}
    f=160 000 \cdot10^{-0.4(V_{mag}-11^m)}
\end{equation}
Here the $V_{mag}$ is the host star's flux measured in the V photometric band from the original table.
We then calculated whether the given effect could be detected by \plato\,based on the given formula:
\begin{equation}
    D=\frac{A_{effect}}{\frac{\sigma_{phot}-9\cdot10^{-6}}{{\sqrt{N_D}}}+9\cdot10^{-6}}
\label{eq:dec}
\end{equation}
If this value is greater than 3, we consider the effect detectable. Here the $A_{effect}$ is the calculated 
amplitude of the effect. The denominator takes into account that we assumed that nothing is detected below 
the jitter noise limit as is probably does not behave as white noise. The detectabilities are indicated in 
Figures \ref{fig:ref03}, \ref{fig:ref01}, \ref{fig:th_refl_ratio}, \ref{fig:beam} and \ref{fig:ell}. 
If the effect is not detectable in the case of the planet, we plotted it as a triangle. 

We performed 5,000 Monte-Carlo simulations for every planet using the input data and their uncertainties. 
We report the mean and the standard deviations of this samples as predicted amplitudes and their $1\sigma$ uncertainties.

We see that the ellipsoidal and the beaming effect is mostly undetectable in the host stars of the known exoplanets 
in the \plato\,LOPS2 field. The reflection effect is observable up to 3 days orbital period amongst the already known planets, the

\section{Summary}
\label{section:summary}

This paper presents an overview of the expected performance of the \plato~mission
before launch, with information available after the calibration of all the flight
model cameras.
We provide a reference to the community of what to expect in terms of planet detection
yield in comparison with \tess, which is a more relevant comparison than \kepler\,in
terms of dimensioning of the follow-up efforts.
The planet yield estimates are highly uncertain not only because of the discrepancies
in the planet occurrence rates and transit detection efficiency, but in terms of
assessing the impact of stellar variability.
Stellar variability will indeed remain a challenge for achieving \plato~results.
Nevertheless, our planet detection estimates show that \plato~has the capabilities 
to boost our knowledge on planet populations and of well-characterised small planets 
up to one astronomical unit orbital distance.

\bmhead{Acknowledgments}

This work presents results from the European Space Agency (ESA) space mission PLATO. The PLATO payload, the PLATO Ground Segment and PLATO data processing are joint developments of ESA and the PLATO Mission Consortium (PMC). Funding for the PMC is provided at national levels, in particular by countries participating in the PLATO Multilateral Agreement (Austria, Belgium, Czech Republic, Denmark, France, Germany, Italy, Netherlands, Portugal, Spain, Sweden, Switzerland, Norway, and United Kingdom) and institutions from Brazil. Members of the PLATO Consortium can be found at https://platomission.com/. The ESA PLATO mission website is https://www.cosmos.esa.int/plato. We thank the teams working for PLATO for all their work. \\
We would like to thank M. Kunimoto for useful discussions that clarified our message and improved our paper.
C. A., A. B., R. H., P. R., J. D. R, N. J., T. M., S. R., D. S., and B. V. acknowledge support from the Belgian Science Policy Office (BELSPO) in the form of PRODEX grants for the development and exploitation of the PLATO and {\it Gaia\/} missions.\\
R.~H. and M.~A.-v.~E. acknowledge support from the German Aerospace Agency (Deutsches Zentrum f{\"u}r Luft- und Raumfahrt) under PLATO Data Center grants 50OO1501 and 50OP1902.\\
G. G. Bal{\'a}zs would like to thank the ERASMUS+ programme and the Municipality of Dabas for their financial support of his work.\\
G\'abor G. Bal\'azs would like to thank the ERASMUS+ programme and the Municipality of Dabas for their financial support of his work. He also thanks for the hospitality of DLR.\\
MSC acknowledge support from Funda\c{c}{\~a}o para a Ci{\^e}ncia e Tecnologia through the grant UID/04434/2025 and work contract doi.org/10.54499/2023.09303.CEECIND/CP2839/CT0003.\\
VN, LM, MM, IP, GP, RR acknowledge support from PLATO ASI-INAF agreements n. 2022-28-HH.0.\\
INTA and CAB Authors would like to thank to Agencia Estatal de Investigaci{\'o}n of Ministerio de Ciencia, Innovaci{\'o}n y Universidades of Spain for the MICIU/AEI/10.13039/501100011033 and ERDF/EU grants PID2019-107061GB-C61/C62 and PID2023-147338NB-C21/C22 as well as to INTA, for the funding support to all the activities described in this paper.\\
A.M. acknowledges funding support from grant PID2023-149439NB-C44 funded by MCIN/AEI/10.13039/501100011033/ and from FEDER "Una manera de hacer Europa, EU", and from Generalitat Valenciana in the frame of the GenT Project ESGENT-CIESGT2025.\\
This work was supported by the Spanish Ministry of Science, Innovation and Universities / State Research Agency (MICIU/AEI/10.13039/501100011033) and by ERDF, a way of making Europe, under project PID2023-149439NB-C42.\\
Authors are thankful to the CNES and OHB teams that supported the \plato~performance studies during Phase B of the project. In particular, to P. Levacher, J. Dall'Amico, M. Klebor, V. Mogulsky, A. Orlandi, M. Schweitzer, D. Serrano-Velarde, M. Wieser, and Th. Wocjan.

%
%

\bibliography{bibl}

\begin{appendices}

\section{PLATO Performance Parameters}
\label{appendix:performance_parameters}

Table~\ref{table:parameters} shows the main parameters driving the performance of \plato.

\begin{table*}
\caption{Table with the values of the main drivers for the \plato~Mission performance. BOL stands for beginning of life. EOL stands for end of life.}             
\label{table:parameters}
\centering
\renewcommand{\arraystretch}{1.5}
\begin{tabular}{p{4cm}cp{4.5cm}}
\hline\hline       
parameter & value & comment\\ 
\hline                    
Number of fast cameras              & 2                             & BOL and EOL \\
Number of normal cameras BOL        & 24                            & nominal design \\
Number of normal cameras EOL        & 22                            & EOL the probability of having lost 2 or more N-CAMs after 4.5 yr operations is less than 21\% (reliability requirement: 79\%) \\
Number of CCDs per camera           & 4                             & 104 in the payload, combined illuminated surface of approximately 0.65 m$^2$ \\
Number of pixels in the CCDs        & 4510x4510 pixels              & The N-CAM CCDs are read in full-frame mode while the F-CAM CCDs are read in frame-transfer mode. \\
Pixel size                          & 18 micron                     & \\
Pixel scale                         & 15.0 arcsec/pixel             & On-axis. \\ 
Pupil size                          & 12 cm                         & The diameter of the telescope optical unit is 20 cm, but the pupil size is 12 cm. \\
Global approx. optical transmission &  ~70\%                        & see Fig.~\ref{figure:instrument_response} \\
Spectral range                      & 500nm-1000nm                  & see Fig.~\ref{figure:instrument_response}, in particular for the F-CAMs \\
PSF Size                            & $\approx$3 pixels diameter    & required more than 77\% enclosed energy in 2x2 pixels, achieved in test more than 80\%, see~\citet{borsa2022}\\
Exposure cadence                    & 25\,s (N-) \& 2.5\,s (F-CAMs) & including CCD readout \\
Exposure time                       & 21\,s (N-) \& 2.3\,s (F-CAMs) & see also~\citet{boerner2024} \\
\hline                  
\end{tabular}
\end{table*}

\clearpage

\section{Instrument Response Functions}
\label{appendix:response_functions}

\begin{figure*}
  \centering
  \includegraphics[%
    width=0.9\linewidth,%
    height=0.5\textheight,%
    keepaspectratio]{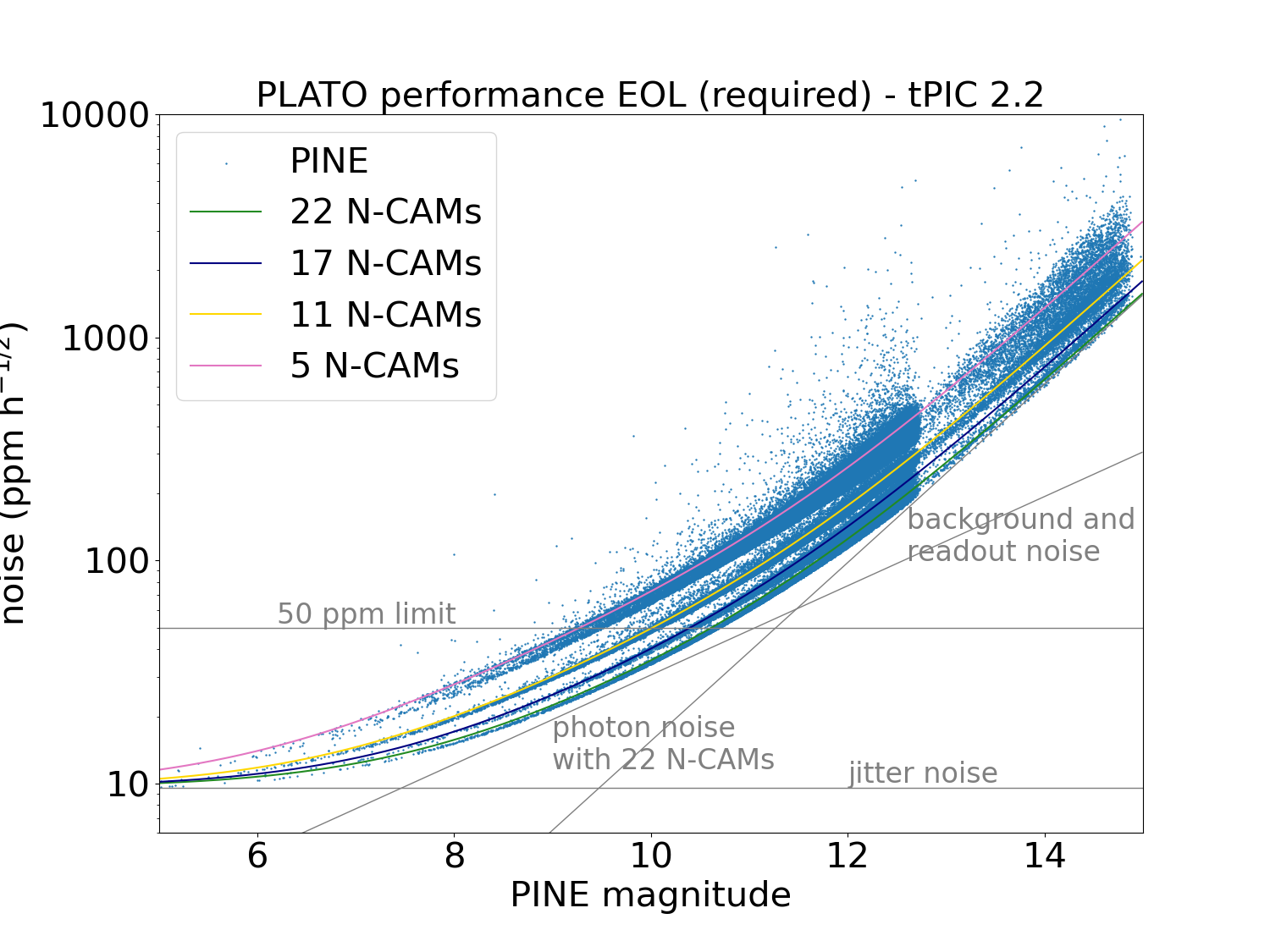}
  \caption{Noise to signal ratio (NSR) for the PIC 2.2 computed with \pine~for N-CAMs EOL.
  Overplotted lines are the quick noise model values for the end of life scenario (see Table~\ref{table:quick_noise_model}).}
  \label{figure:quick_noise_model_eol}
\end{figure*}

\begin{figure*}
  \centering
  \includegraphics[%
    width=0.9\linewidth,%
    height=0.5\textheight,%
    keepaspectratio]{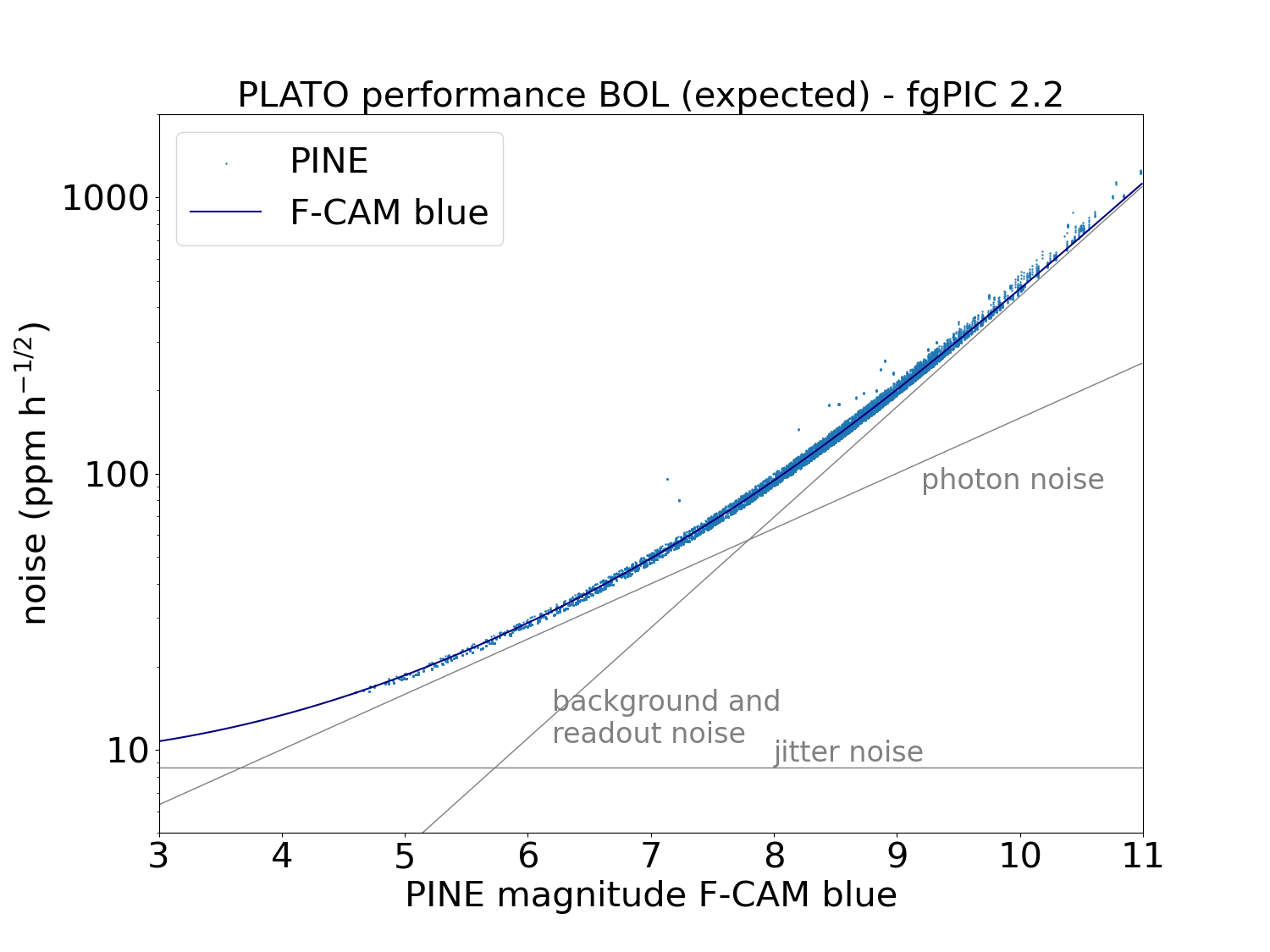}
  \caption{Noise to signal ratio (NSR) for the PIC 2.2 computed with \pine~for the F-CAM blue BOL.
  Overplotted lines are the quick noise model values for the beginning of life   scenario (see Table~\ref{table:quick_noise_model}).}
  \label{figure:quick_noise_model_bol_fcamb}
\end{figure*}

\begin{figure*}
  \centering
  \includegraphics[%
    width=0.9\linewidth,%
    height=0.5\textheight,%
    keepaspectratio]{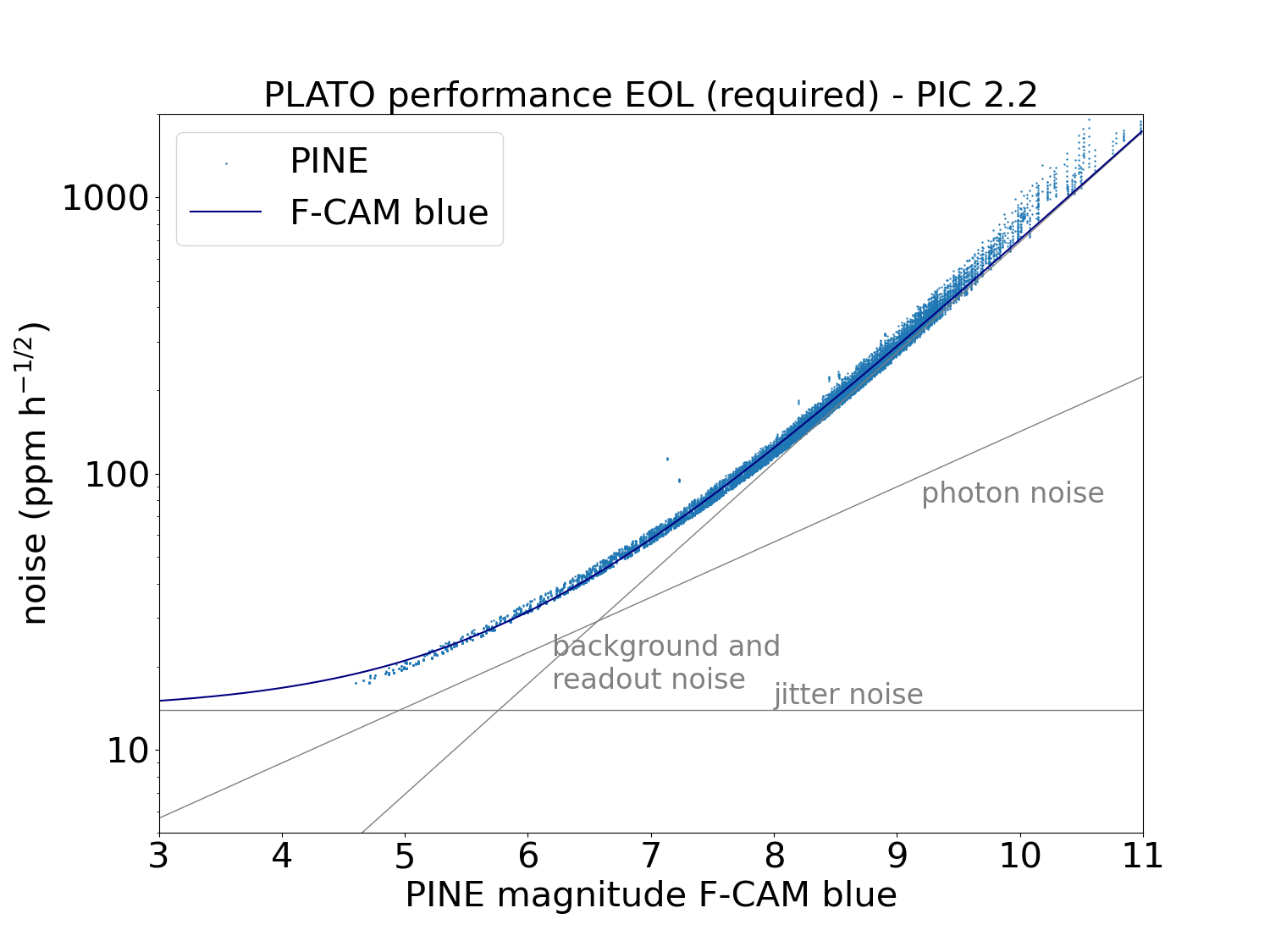}
  \caption{Noise to signal ratio (NSR) for the PIC 2.2 computed with \pine~for the F-CAM blue EOL.
  Overplotted lines are the quick noise model values for the end of life scenario (see Table~\ref{table:quick_noise_model}).}
  \label{figure:quick_noise_model_eol_fcamb}
\end{figure*}

\begin{figure*}
  \centering
  \includegraphics[%
    width=0.9\linewidth,%
    height=0.5\textheight,%
    keepaspectratio]{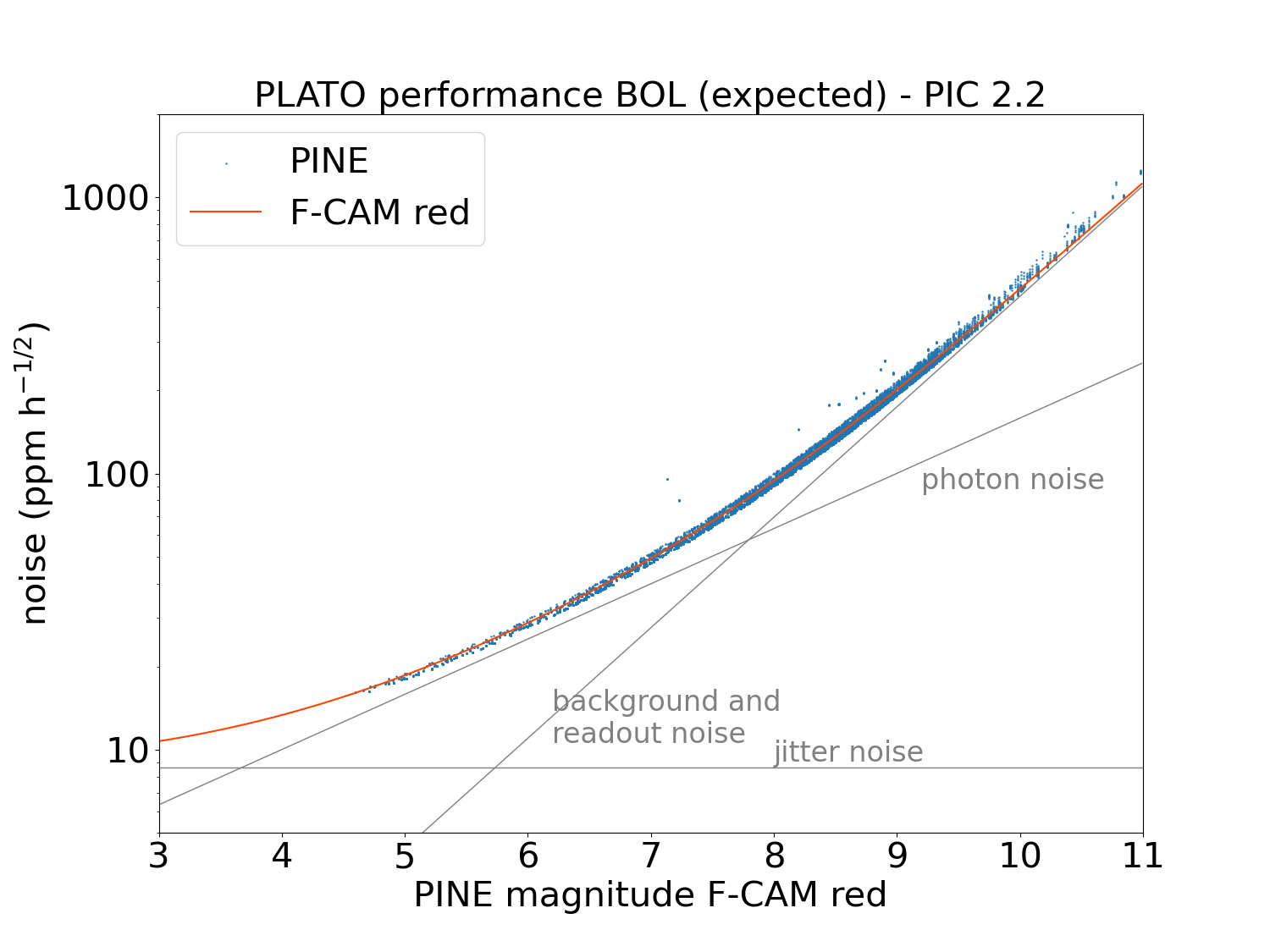}
  \caption{Noise to signal ratio (NSR) for the PIC 2.2 computed with \pine~for the F-CAM red BOL.
  Overplotted lines are the quick noise model values for the beginning of life scenario (see Table~\ref{table:quick_noise_model}).}
  \label{figure:quick_noise_model_bol_fcamr}
\end{figure*}

\begin{figure*}
  \centering
  \includegraphics[%
    width=0.9\linewidth,%
    height=0.5\textheight,%
    keepaspectratio]{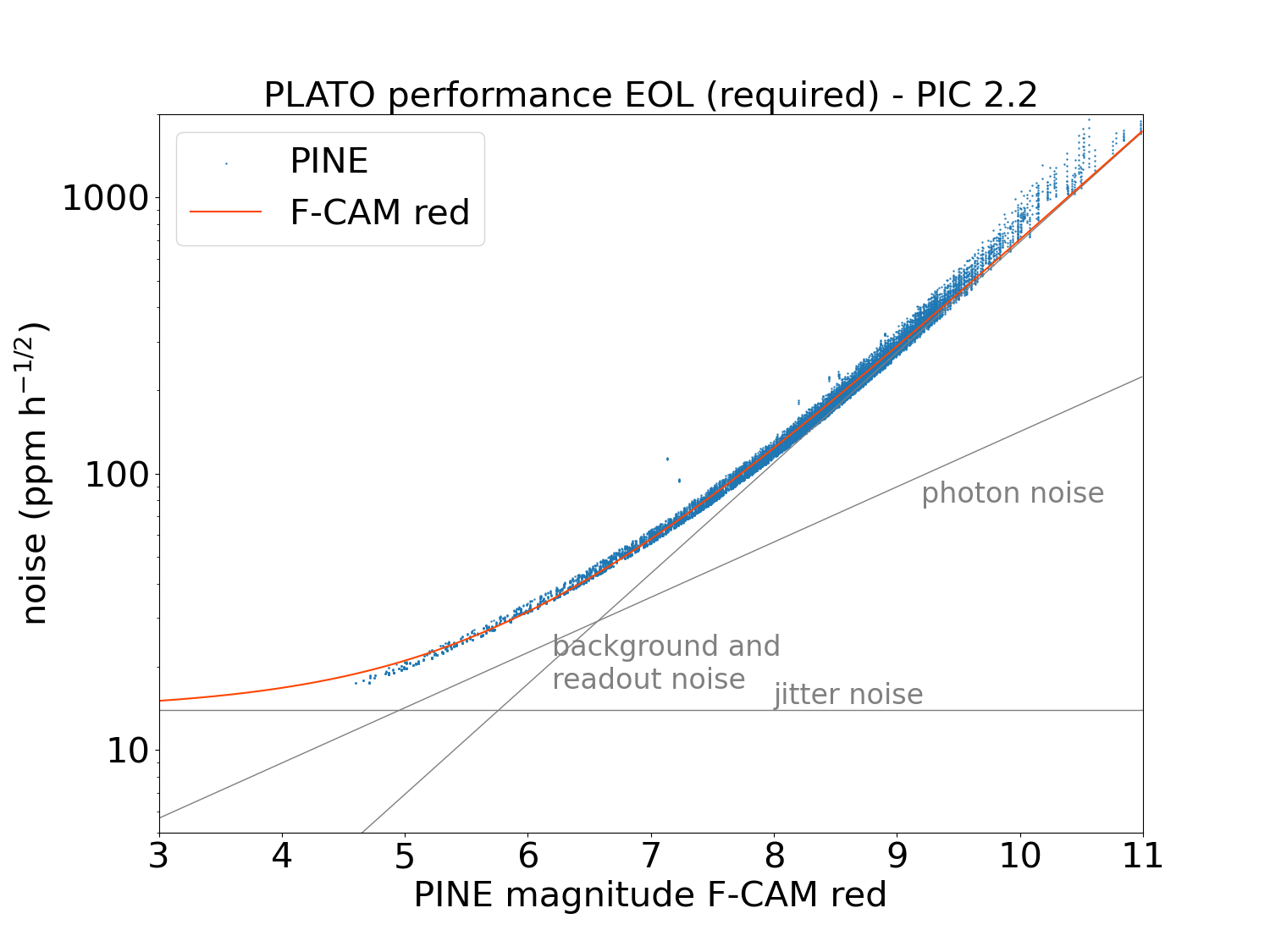}
  \caption{Noise to signal ratio (NSR) for the PIC 2.2 computed with \pine~for the F-CAM red EOL.
  Overplotted lines are the quick noise model values for the end of life scenario (see Table~\ref{table:quick_noise_model}).}
  \label{figure:quick_noise_model_eol_fcamr}
\end{figure*}

In this section we present the figures describing the instrument response function for the N-CAMs
EOL (Fig.~\ref{figure:quick_noise_model_eol}) and for the 
F-CAM blue (Fig.~\ref{figure:quick_noise_model_bol_fcamb} for BOL and 
Fig.~\ref{figure:quick_noise_model_eol_fcamb} for EOL) and
F-CAM red (Fig.~\ref{figure:quick_noise_model_bol_fcamr} for BOL and 
Fig.~\ref{figure:quick_noise_model_eol_fcamr} for EOL), complementing the 
information given in Section~\ref{subsection:instrument_response}.

\clearpage

\section{Additional planet yield estimates}
\label{appendix:yield_estimates}

We present in Table~\ref{table:planet-yield-hsu19hz040} planet yield results
using occurrence rates and detectability criteria from~\citep{hsu2019} and in 
Table~\ref{table:planet-yield-kunimoto20hz040} estimates using~\citep{kunimoto2020}.
The estimates for the Prime Sample yield are given in 
Tables~\ref{table:planet-yield-hsu19hz040_PS} and~\ref{table:planet-yield-kunimoto20hz040_PS} respectively.
The comparison with \tess\,is presented in 
Tables~\ref{table:planet-yield-hsu19hz040_table4kunimoto2022} and \ref{table:planet-yield-kunimoto20hz040_table4kunimoto2022}.

\begin{table}[ht]
\tabcolsep=1pt
    \begin{tabular}{c|c|c|c|c|c} \hline
                               & known transiting & \multicolumn{4}{c}{2+2 scenario}      \\
 Samples                       & planets   & Red Book        & Heller  & This work      & Matuszewski    \\ \hline
all planets orbiting stars     &           &                 &         &                & \\
$<$13 mag in P1+P5 samples     & 1\,550    & $\approx$4\,600 & n/a     & 8\,400-8\,700  & 4\,500-46\,000 \\ \hline
all planets orbiting stars     &           &                 &         &                & \\
V$<$11 mag in P1+P5 samples    & 520       & $\approx$1\,200 & n/a     & 1\,250-1\,400  & 1\,700-11\,000 \\ \hline
planets $<$2 \rearth\, in HZ   &           &                 &         &                & \\
orbiting P1+P5 stars $<$11 mag &      0    & 6 - 280         & 11 - 34 & 0 - 70         & $\approx$45    \\ \hline
         \hline
                               & known transiting & \multicolumn{4}{c}{3+1 scenario}       \\
 Samples                       & planets   & Red Book         & Heller  & This work        & Matuszewski     \\ \hline
all planets orbiting stars     &           &                  &         &                  & \\
$<$13 mag in P1+P5 samples     & 1\,550    & $\approx$11\,000 & n/a     & 10\,400-11\,000  & 12\,000-68\,000 \\ \hline
all planets orbiting stars     &           &                  &         &                  & \\ 
V$<$11 mag in P1+P5 samples    & 520       & $\approx$2\,700  & n/a     & 1\,750-2\,000    & 4\,000-42\,000  \\ \hline
planets $<$2 \rearth\, in HZ   &           &                  &         &                  & \\
orbiting P1+P5 stars $<$11 mag &        0  & 3-140            & 8-25    & 0 - 60           & $\approx$30 \\ \hline
    \end{tabular}
    \caption{Estimated PLATO planet yields. Red Book: ESA-SCI(2017)1; Rauer:~\citet{rauer2025}; Heller:~\citet{heller2022}; This work: using occurrence rates and detectability criterion as per \citet{hsu2019} on PIC 2.2; Matuszewski: \citet{matuszewski2023}. 2+2 means 2 long pointings of 2 years duration; 3+1 means one 3-year observation followed by one year with six target fields for 60 days each, as in the Red Book. Known (confirmed) transiting planets are taken from the NASA exoplanet archive in Feb.\,2026 (https://exoplanetarchive.ipac.caltech.edu/) for all planet radii and orbits.}
    \label{table:planet-yield-hsu19hz040}
\end{table}

\begin{table}[ht]
\renewcommand{\arraystretch}{1.1}
\begin{tabular}{r|*{2}{c|}} \hline
\multicolumn{3}{l}{2 years}\\\hline\hline
after 2 years in LOPS2       & Total       & smaller than 2\rearth \\\hline
Planets in the Prime Sample  & 648 - 719   & 432 - 483             \\
\end{tabular}
    \caption{Estimated planet yields on the Prime Sample using occurrence rates by~\citep{hsu2019}.}
    \label{table:planet-yield-hsu19hz040_PS}
\end{table}

\begin{table}[ht]
\renewcommand{\arraystretch}{1.1}
\begin{tabular}{r|*{5}{c|}} \hline
\multicolumn{6}{l}{2 years}\\\hline\hline
by spectral type            & Total       & F                & G                    & K                     & M                \\ \hline
\tess~Prime Mission         & 4\,719      &           1\,209 &               2\,134 &                   859 & 261             \\
\plato                      & 4\,646      &           1\,800 &               1\,990 &                   486 & 370             \\ \hline\hline
by size (values in \rearth) & Total       & $R_p < 2$ & $2 < R_p < 4$ & $ 4 < R_p < 8$ & $R_p > 8$ \\ \hline
\tess~Prime Mission         & 4\,719      &              152 &                  770 &                   673 &           3\,124 \\
\plato                      & 4\,647      &           1\,874 &               2\,150 &                   292 &              331 \\ \hline\hline
by period                   & $>$ 20 days &     $>$ 100 days & \multicolumn{3}{c}{} \\ \cmidrule{1-3}
\tess~Prime Mission         &         398 &               48 & \multicolumn{3}{c}{} \\
\plato                      &      1\,360 &              272 & \multicolumn{3}{c}{} \\ \cmidrule{1-3}
\end{tabular}
    \caption{Estimated PLATO planet yields compared with the values in Table 4 of~\citet{hsu2019} using occurrence 
    rates by~\citep{hsu2019}. 
    In the table we do not give explicit uncertainties in the values and refer to the text for details. 
    The \tess~total numbers include A stars that we do not include for \plato~because they do not belong to PIC.}
    \label{table:planet-yield-hsu19hz040_table4kunimoto2022}
\end{table}

\begin{table}[ht]
\tabcolsep=1pt
    \begin{tabular}{c|c|c|c|c|c} \hline
                               & known transiting & \multicolumn{4}{c}{2+2 scenario}      \\
 Samples                       & planets   & Red Book        & Heller  & This work     & Matuszewski    \\ \hline
all planets orbiting stars     &           &                 &         &               & \\
$<$13 mag in P1+P5 samples     & 1\,550    & $\approx$4\,600 & n/a     & 5\,450-5\,650 & 4\,500-46\,000 \\ \hline
all planets orbiting stars     &           &                 &         &               & \\
V$<$11 mag in P1+P5 samples    & 520       & $\approx$1\,200 & n/a     &       740-830 & 1\,700-11\,000 \\ \hline
planets $<$2 \rearth\, in HZ   &           &                 &         &               & \\
orbiting P1+P5 stars $<$11 mag &      0    & 6 - 280         & 11 - 34 & 0 - 25        & $\approx$45    \\ \hline
         \hline
                               & known transiting & \multicolumn{4}{c}{3+1 scenario}       \\
 Samples                       & planets   & Red Book         & Heller  & This work     & Matuszewski     \\ \hline
all planets orbiting stars     &           &                  &         &               & \\
$<$13 mag in P1+P5 samples     & 1\,550    & $\approx$11\,000 & n/a     & 5\,900-6\,300 & 12\,000-68\,000 \\ \hline
all planets orbiting stars     &           &                  &         &               & \\ 
V$<$11 mag in P1+P5 samples    & 520       & $\approx$2\,700  & n/a     &    900-1\,100 & 4\,000-42\,000  \\ \hline
planets $<$2 \rearth\, in HZ   &           &                  &         &               & \\
orbiting P1+P5 stars $<$11 mag &        0  & 3-140            & 8-25    & 0 - 20        & $\approx$30 \\ \hline
    \end{tabular}
    \caption{Estimated PLATO planet yields. Red Book: ESA-SCI(2017)1; Rauer:~\citet{rauer2025}; Heller:~\citet{heller2022}; This work: using occurrence rates and detectability criterion as per \citet{kunimoto2020} on PIC 2.2; Matuszewski: \citet{matuszewski2023}. 2+2 means 2 long pointings of 2 years duration; 3+1 means one 3-year observation followed by one year with six target fields for 60 days each, as in the Red Book. Known (confirmed) transiting planets are taken from the NASA exoplanet archive in Feb.\,2026 (https://exoplanetarchive.ipac.caltech.edu/) for all planet radii and orbits.}
    \label{table:planet-yield-kunimoto20hz040}
\end{table}

\begin{table}[ht]
\renewcommand{\arraystretch}{1.1}
\begin{tabular}{r|*{2}{c|}} \hline
\multicolumn{3}{l}{2 years}\\\hline\hline
after 2 years in LOPS2       & Total       & smaller than 2\rearth \\\hline
Planets in the Prime Sample  & 336 - 379   & 163 - 191             \\
\end{tabular}
    \caption{Estimated planet yields on the Prime Sample using occurrence rates by~\citep{kunimoto2020}.}
    \label{table:planet-yield-kunimoto20hz040_PS}
\end{table}

\begin{table}[ht]
\renewcommand{\arraystretch}{1.1}
\begin{tabular}{r|*{5}{c|}} \hline
\multicolumn{6}{l}{2 years}\\\hline\hline
by spectral type            & Total       & F                & G                    & K                     & M                \\ \hline
\tess~Prime Mission         & 4\,719      &           1\,209 &               2\,134 &                   859 &              261 \\
\plato                      & 3\,121      &           1\,220 &               1\,278 &                   276 &              347 \\ \hline\hline
by size (values in \rearth) & Total       & $R_p < 2$ & $2 < R_p < 4$ & $ 4 < R_p < 8$ & $R_p > 8$ \\ \hline
\tess~Prime Mission         & 4\,719      &              152 &                  770 &                   673 &           3\,124 \\
\plato                      & 3\,121      &           1\,006 &               1\,624 &                   220 &              271 \\ \hline\hline
by period                   & $>$ 20 days &     $>$ 100 days & \multicolumn{3}{c}{} \\ \cmidrule{1-3}
\tess~Prime Mission         &         398 &               48 & \multicolumn{3}{c}{} \\
\plato                      &         813 &              109 & \multicolumn{3}{c}{} \\ \cmidrule{1-3}
\end{tabular}
    \caption{Estimated PLATO planet yields compared with the values in Table 4 of~\citet{kunimoto2022} using occurrence 
    rates by~\citep{kunimoto2020}. 
    In the table we do not give explicit uncertainties in the values and refer to the text for details. 
    PM stands for Prime Mission of \tess. The \tess~total numbers include A stars, that we do not include for \plato~because
    they do not belong to PIC.}
    \label{table:planet-yield-kunimoto20hz040_table4kunimoto2022}
\end{table}

\begin{table}[ht]
    \centering
    \begin{tabular}{c||c|c|c}
        Sample & Kunimoto 2020 & Hsu 2019 & Fressin 2013\\
        \hline \hline
        PLATO P5    & 4\,107$\pm$152 & 6\,598$\pm$314 & 3\,546$\pm$127\\
        PLATO P1    &    444$\pm$16  &    996$\pm$47  &    427$\pm$17 \\
        PLATO Prime &    373$\pm$13  & 1\,016$\pm$52  &    376$\pm$17 \\ 
    \end{tabular}
    \caption{Estimated PLATO yields using sensitivities from \citet{eschen2024} re-binned for different occurrence rates \citep{kunimoto2020, hsu2019, fressin2013}. }
    \label{tab:sensitivity_yields_total}
\end{table}

\begin{table}[ht]
    \centering
    \begin{tabular}{c||c|c|c}
        Sample & 1\% & 40\% & 100\%\\
        \hline \hline
        PLATO P5 $<$2\,R$_\oplus$ HZ    & 0 & 10 & 25 \\
        PLATO P1 $<$2\,R$_\oplus$ HZ    & 0 &  3 &  7 \\
        PLATO Prime $<$2\,R$_\oplus$ HZ & 0 &  9 & 23 \\
        PLATO P4 $<$2\,R$_\oplus$ HZ    & 0 & 10 & 26 \\
    \end{tabular}
    \caption{Estimated Yields of planets $<$2\,R$_\oplus$ orbiting stars of the different samples in their habitable zone using the sensitivities from \citet{eschen2024}. The habitable zone was computed per star following \citet{kopparapu2014} and occurrence rates of 1\%, 40\% and 100\% were assumed.}
    \label{tab:sensitivity_yields_hz}
\end{table}

\begin{figure*}
  \centering
  \includegraphics[%
    width=0.9\linewidth,%
    height=0.5\textheight,%
    keepaspectratio]{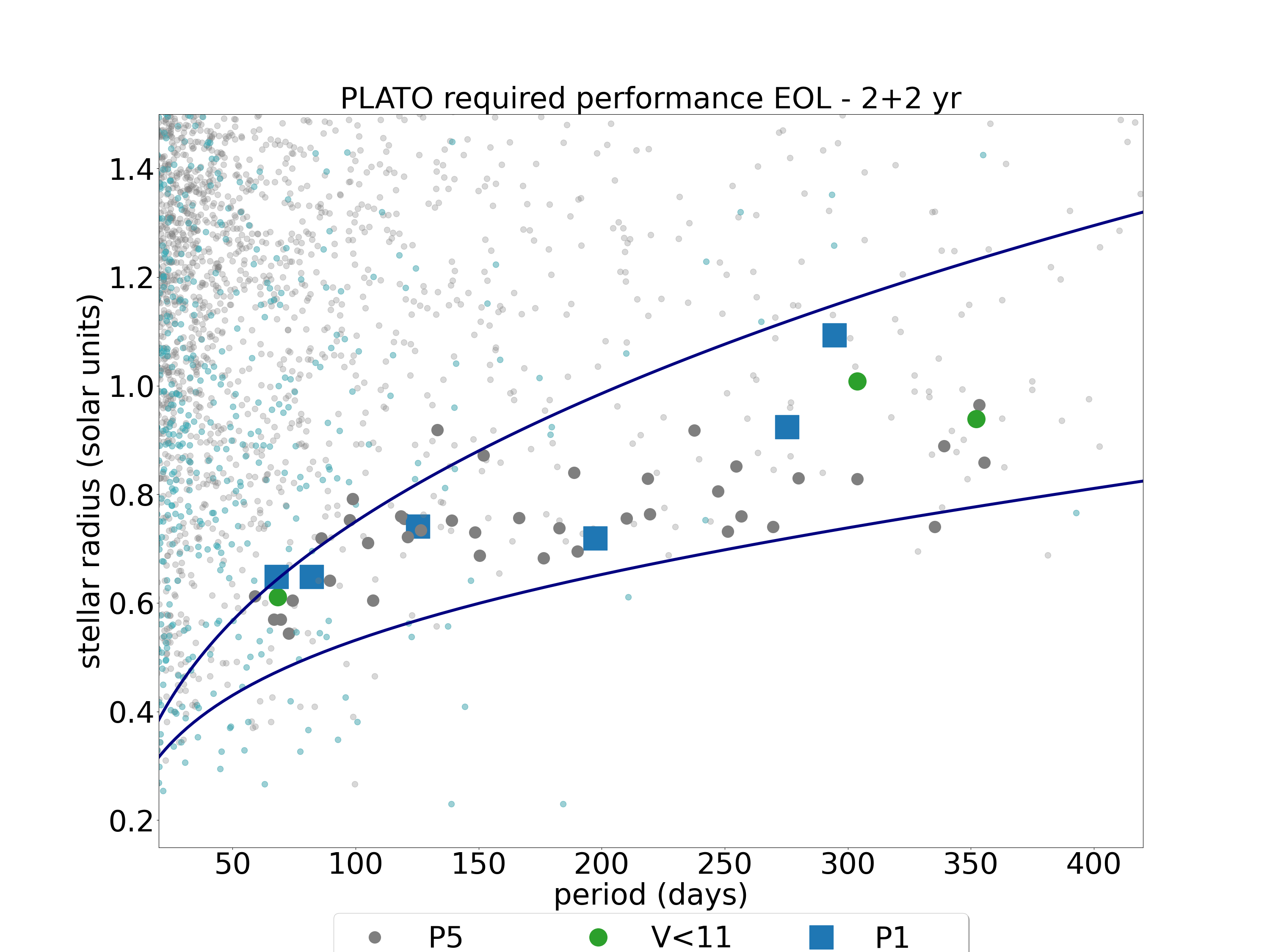}
  \caption{Distribution of detected planets in the habitable zone assuming 40\% occurrence rate 
  and detectability criteria as per~\citet{hsu2019}. The habitable zone is computed for each star
  according to its stellar properties in the PIC (mass, radius, effective temperature). In the 
  habitable zone, grey dots design are stars from the P5 sample fainter than magnitude 11, where
  follow-up efforts will be challenging. Green dots design stars from the P5 sample brighter than 
  magnitude 11, where follow-up efforts might be feasible. Blue squares design stars of the P1 
  sample, where full characterization shall be possible. Here we present two realizations of a
  2 year simulation (representative of a 2+2 scenario). To account in a more realistic way 
  for the dispersion of values in the number of planets expected in the habitable zone, refer to 
  Table~\ref{table:planet-yield-hsu19hz040_table4kunimoto2022}.}
  \label{figure:yield_hz_hsu19hz040}
\end{figure*}

\begin{figure*}
  \centering
  \includegraphics[%
    width=0.9\linewidth,%
    height=0.5\textheight,%
    keepaspectratio]{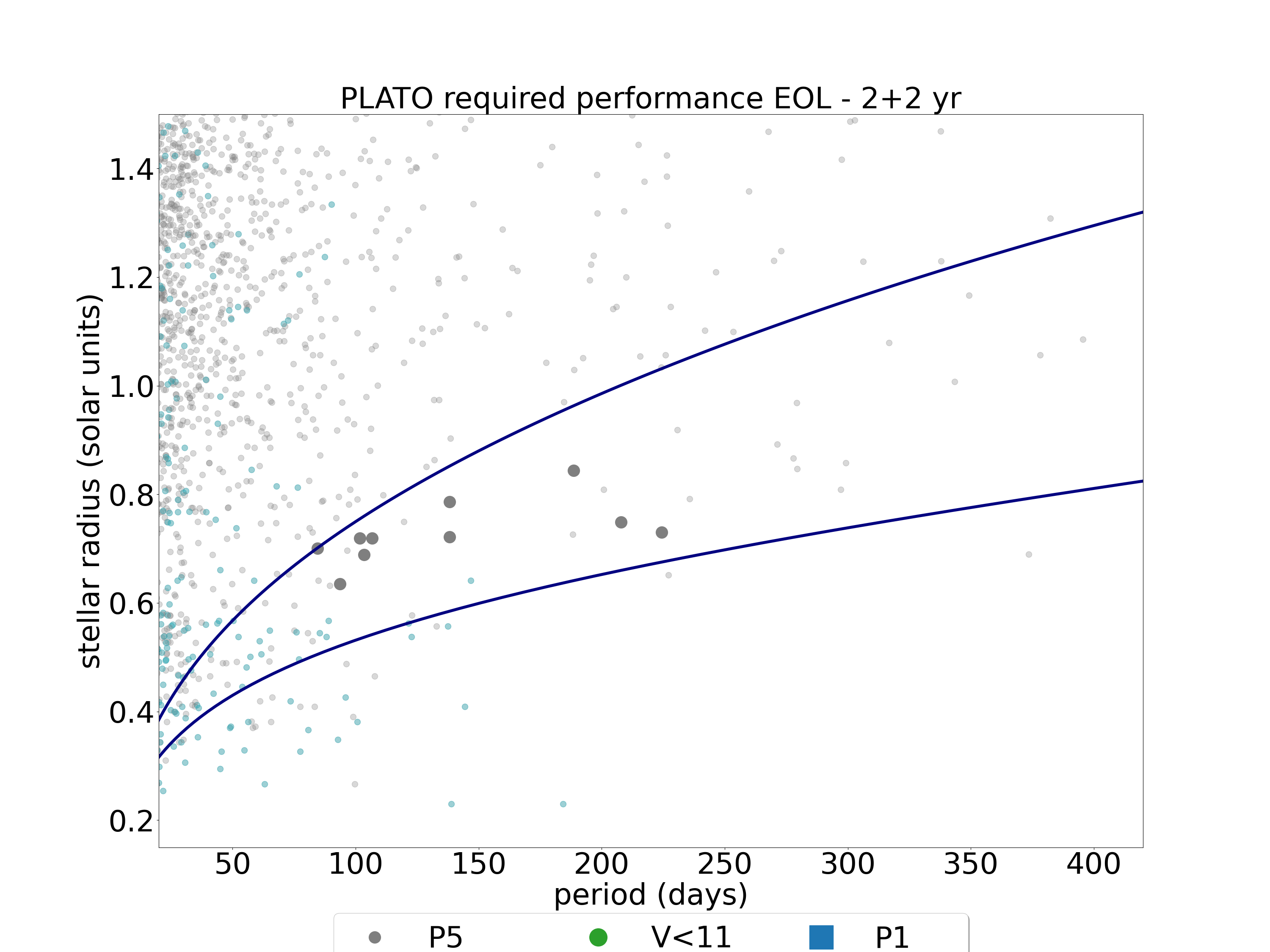}
  \caption{Distribution of detected planets in the habitable zone assuming 40\% occurrence rate 
  and detectability criteria as per~\citet{kunimoto2020}. The habitable zone is computed for each star
  according to its stellar properties in the PIC (mass, radius, effective temperature). In the 
  habitable zone, grey dots design are stars from the P5 sample fainter than magnitude 11, where
  follow-up efforts will be challenging. Green dots design stars from the P5 sample brighter than 
  magnitude 11, where follow-up efforts might be feasible. Blue squares design stars of the P1 
  sample, where full characterization shall be possible. Here we present two realizations of a
  2 year simulation (representative of a 2+2 scenario). To account in a more realistic way 
  for the dispersion of values in the number of planets expected in the habitable zone, refer to 
  Table~\ref{table:planet-yield-kunimoto20hz040_table4kunimoto2022}.}
  \label{figure:yield_hz_kunimoto20hz040}
\end{figure*}

\begin{figure*}[ht]
\includegraphics[%
    width=0.9\linewidth,%
    height=0.5\textheight,%
    keepaspectratio]{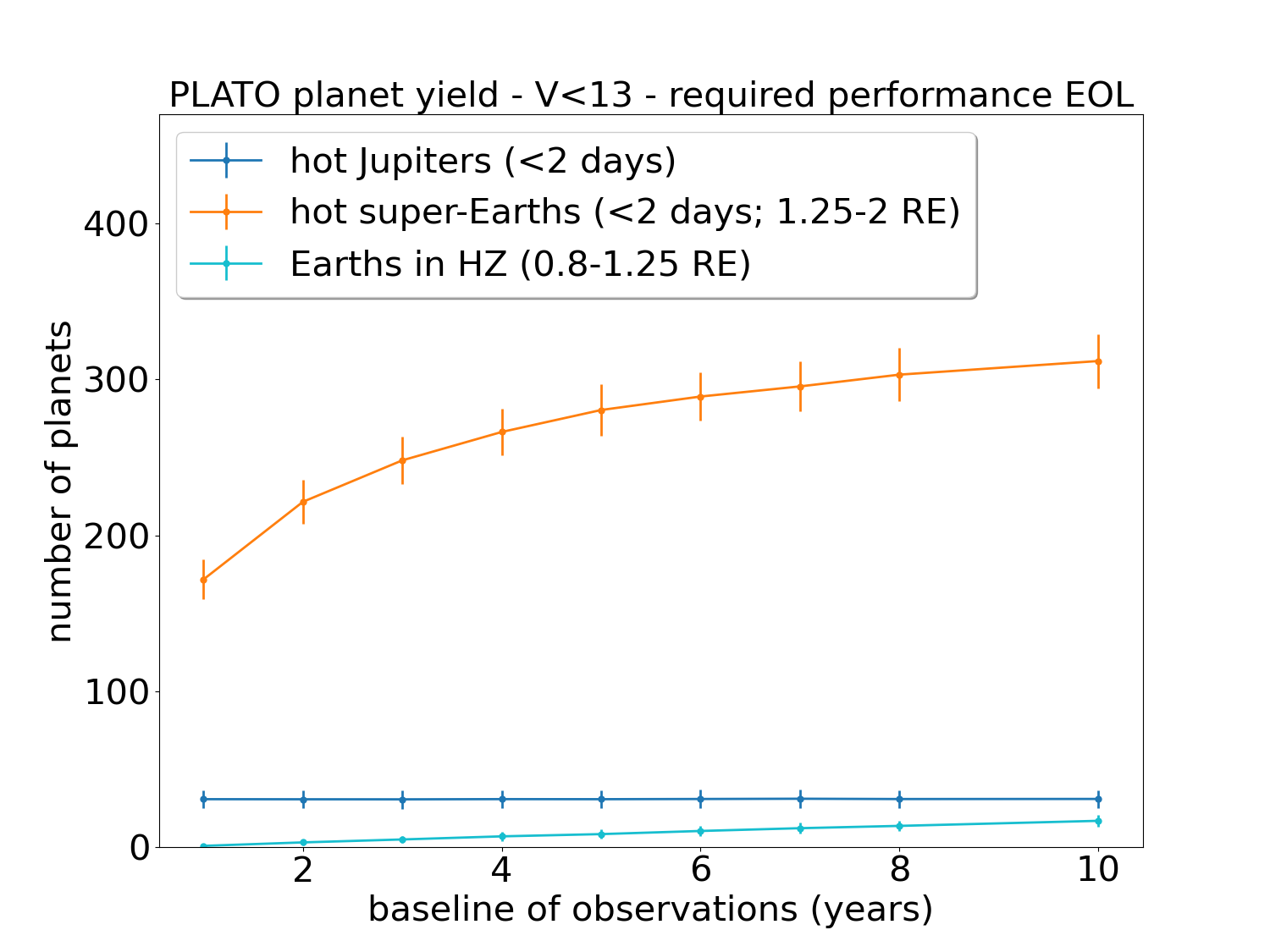}
    \caption{Number of planets anticipated to be found as a function of the observing baseline for hot-Jupiter planets 
    (defined as planets with 6 to 22 \rearth\, and orbital period $<2$ days), hot super-earths 
    (defined as planets with 1.25 to 2 \rearth\, and orbital period $<2$ days), and temperate Earths 
    (defined as planets with 0.8 to 1.25 \rearth\, and orbital period between 245 and 418 days). 
    The vertical lines represent the expected uncertainty in the number of planets. 
    We have considered the end-of-life (EOL) with PIC 2.2 and occurrence rates and detectability
    criteria as per~\citet{hsu2019}.}
\label{figure:yield_time_hsu19hz040}
\end{figure*}

\begin{figure*}[ht]
\includegraphics[%
    width=0.9\linewidth,%
    height=0.5\textheight,%
    keepaspectratio]{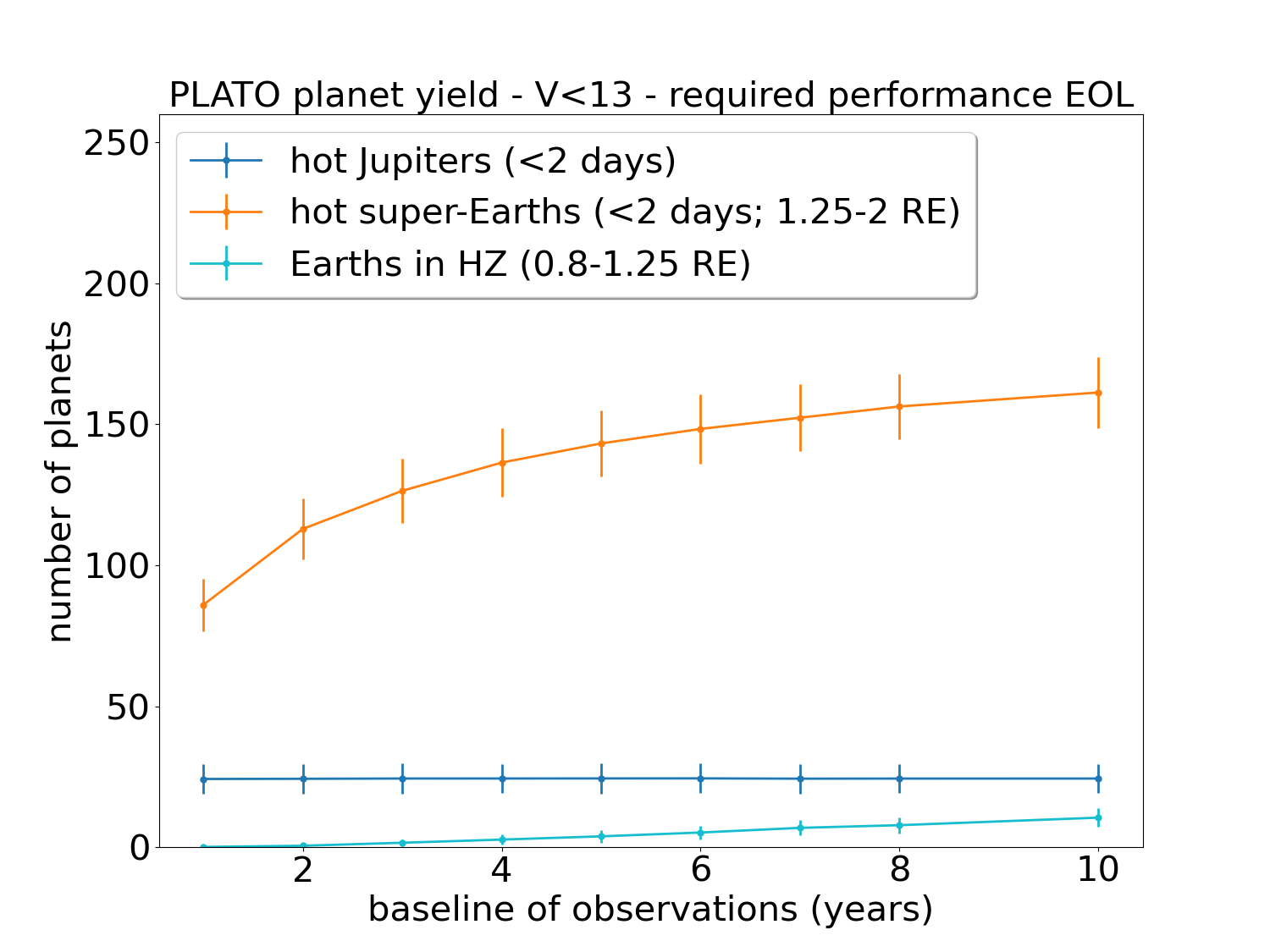}
    \caption{Number of planets anticipated to be found as a function of the observing baseline for hot-Jupiter planets 
    (defined as planets with 6 to 22 \rearth\, and orbital period $<2$ days), hot super-earths 
    (defined as planets with 1.25 to 2 \rearth\, and orbital period $<2$ days), and temperate Earths 
    (defined as planets with 0.8 to 1.25 \rearth\, and orbital period between 245 and 418 days). 
    The vertical lines represent the expected uncertainty in the number of planets. 
    We have considered the end-of-life (EOL) with PIC 2.2 and occurrence rates and detectability
    criteria as per~\citet{kunimoto2020}.}
\label{figure:yield_time_kunimoto20hz040}
\end{figure*}

\clearpage

\section{Additional estimates of synergies with Ariel}
\label{appendix:ariel_estimates}

\begin{figure*}
  \centering
  \includegraphics[%
    width=0.9\linewidth,%
    height=0.5\textheight,%
    keepaspectratio]{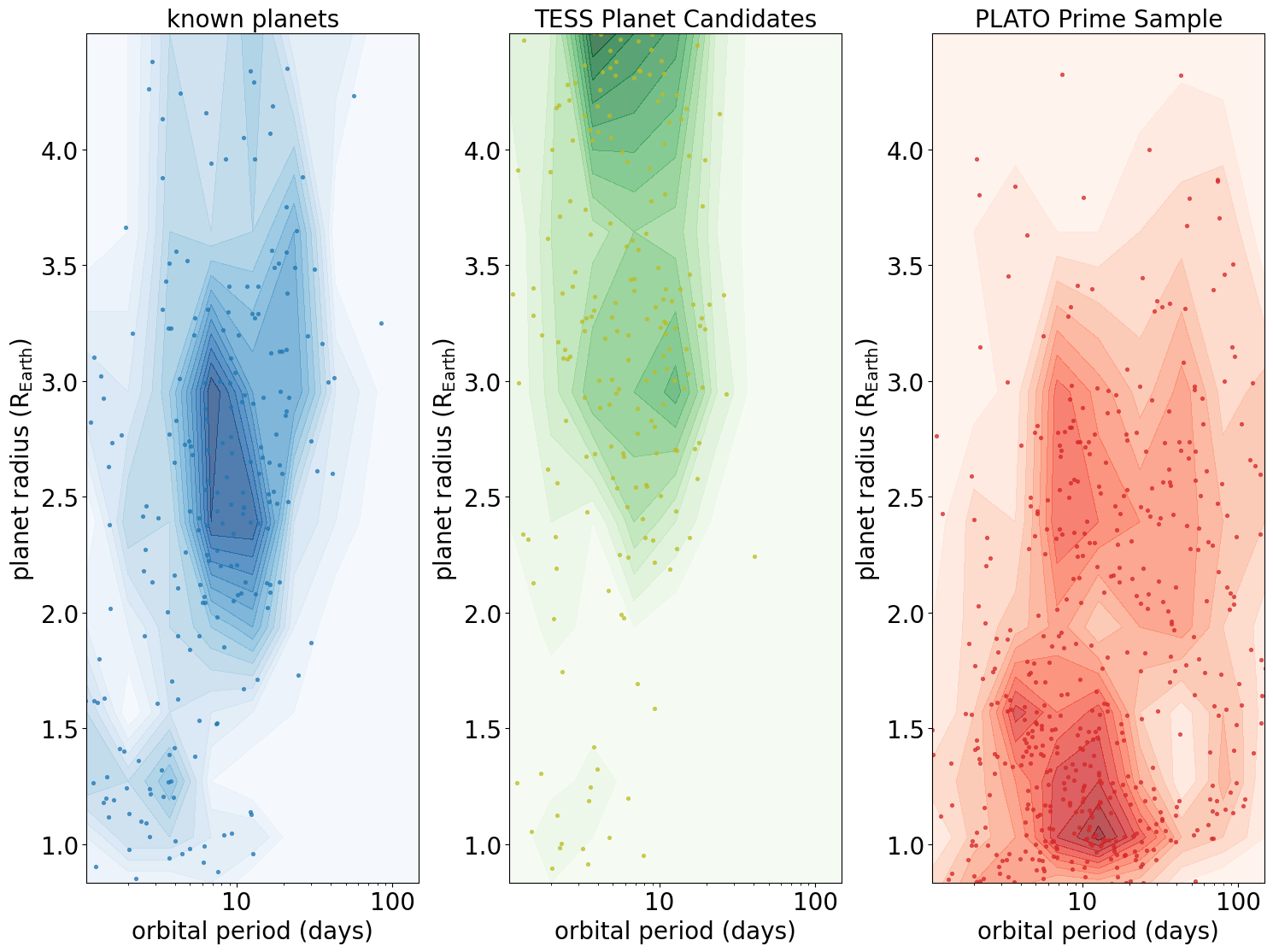}
  \caption{Density plots showing the distribution of known planets considered for follow-up with \ariel,
  \tess\,planet candidates considered for \ariel, and the distribution of Prime Sample targets expected
  to be detected with \plato~for occurrence rates by \citet{hsu2019}.}
  \label{figure:ariel_target_list_hsu19}
\end{figure*}

\begin{figure*}
  \centering
  \includegraphics[%
    width=0.9\linewidth,%
    height=0.5\textheight,%
    keepaspectratio]{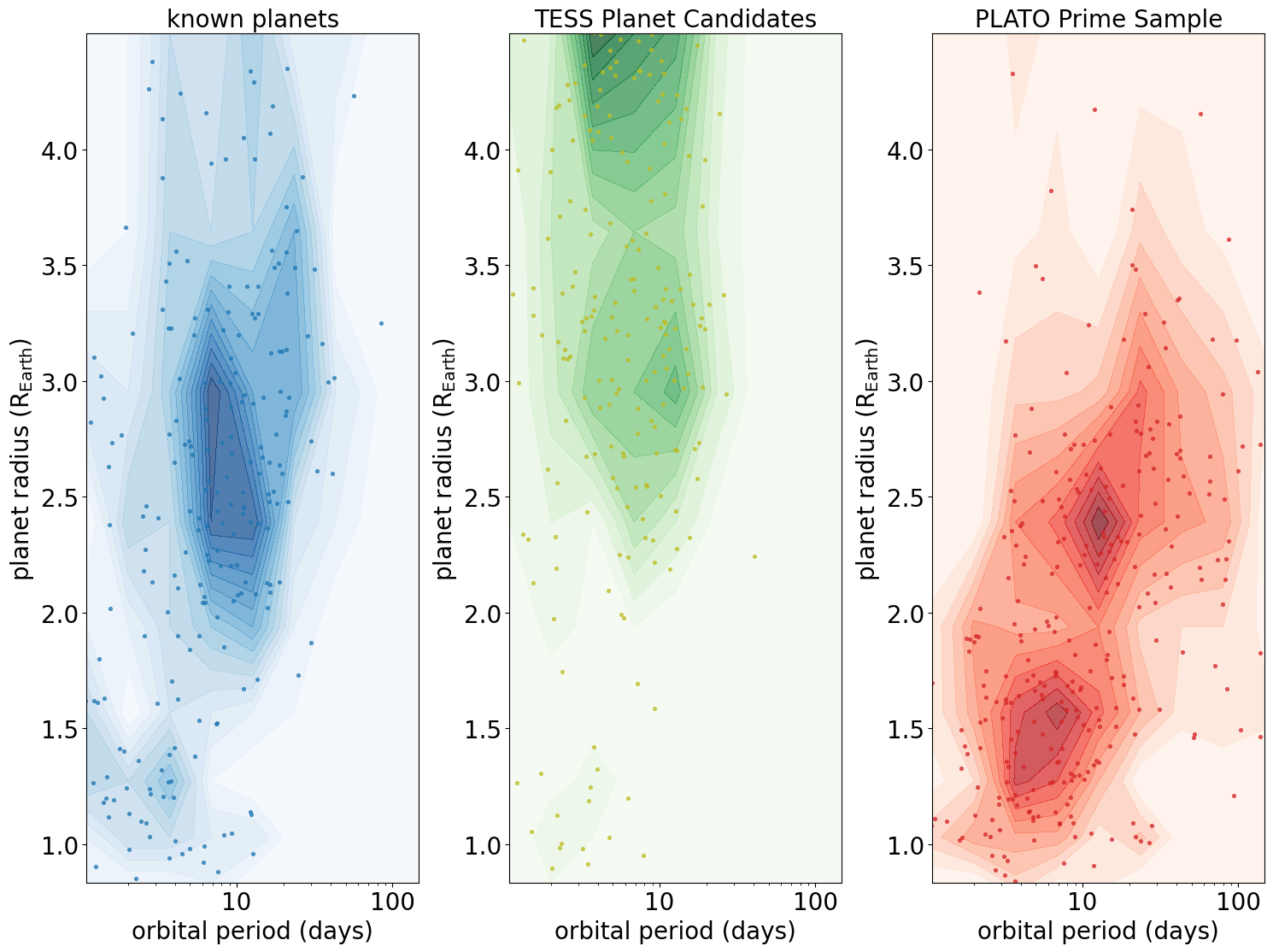}
  \caption{Density plots showing the distribution of known planets considered for follow-up with \ariel,
  \tess\,planet candidates considered for \ariel, and the distribution of Prime Sample targets expected
  to be detected with \plato~for occurrence rates by \citet{kunimoto2020}.}
  \label{figure:ariel_target_list_kunimoto20}
\end{figure*}

Comparison of the distribution of planets selected for \ariel\,observations and planets
in the Prime Sample (see Section~\ref{subsection:ariel} using occurrence rates from 
\citet{hsu2019} (Fig.~\ref{figure:ariel_target_list_hsu19}) and
\citet{kunimoto2020} (Fig.~\ref{figure:ariel_target_list_kunimoto20}).

\section{Initial parameters of the known exoplanets in the LOPS2}

\begin{table*}
\centering
\begin{tabular}{llrrrr}
\hline
Planet name & Detection method & Eccentricity & Period [d] & T$_\text{eff}$ & $V_{mag}$ \\
\hline
CD-35 2722 b&Imaging&-&-&$3680^{+100}_{-100}$&11.041\\
DMPP-3 A b&Radial Velocity&$0.14^{+0.091}_{-0.053}$&$6.67^{+0.0011}_{-0.0003}$&$5138^{+99}_{-99}$&nan\\
GJ 238 b&Transit&-&$1.74^{+1.9e-06}_{-1.8e-06}$&$3485^{+140}_{-140}$&11.62\\
GJ 3341 b&Radial Velocity&$0.31^{+0.11}_{-0.11}$&$14.2^{+0.007}_{-0.007}$&$3526^{+49}_{-49}$&12.062\\
HATS-39 b&Transit&$0.275$&$4.58^{+7.3e-06}_{-7.3e-06}$&$6572^{+83}_{-83}$&12.745\\
HATS-40 b&Transit&$0.312$&$3.26^{+5.8e-06}_{-5.8e-06}$&$6460^{+130}_{-130}$&13.478\\
HATS-41 b&Transit&$0.38^{+0.11}_{-0.11}$&$4.19^{+1.3e-05}_{-1.3e-05}$&$6424^{+91}_{-91}$&12.678\\
\hline
\end{tabular}
\caption{Fundamental characteristics of the known exoplanets in the LOPS2 (1. part)}
\label{tab:list_of_known_planets_part1}
\end{table*}

\newpage

\begin{table*}
\centering
\begin{tabular}{llrrrr}
\hline
Planet name & Detection method & Eccentricity & Period [d] & T$_\text{eff}$ & $V_{mag}$ \\
\hline
HD 50499 c&Radial Velocity&$0.241^{+0.089}_{-0.075}$&$10400^{+3200}_{-1300}$&$5978^{+103.7}_{-103.7}$&7.21\\
HD 51608 b&Radial Velocity&$0.09^{+0.04}_{-0.04}$&$14.1^{+0.0016}_{-0.0016}$&$5358^{+22}_{-22}$&8.17\\
HD 51608 c&Radial Velocity&$0.14^{+0.07}_{-0.07}$&$95.9^{+0.1555}_{-0.1366}$&$5358^{+22}_{-22}$&8.17\\
HD 55696 b&Radial Velocity&$0.705^{+0.022}_{-0.022}$&$1827^{+10}_{-10}$&$6012$&7.95\\
HD 56414 b&Transit&$0.68$&$29^{+0.00021}_{-0.00021}$&$8500^{+150}_{-150}$&9.217\\
HD 63765 b&Radial Velocity&$0.24^{+0.043}_{-0.043}$&$358^{+1}_{-1}$&$5432^{+19}_{-19}$&8.1\\
HD 64121 b&Radial Velocity&$0.11^{+0.07}_{-0.07}$&$623^{+3.4}_{-3.4}$&$5078^{+22}_{-22}$&7.43\\
\hline
\end{tabular}
\caption{Fundamental characteristics of the known exoplanets in the LOPS2 (2. part)}
\label{tab:list_of_known_planets_part2}
\end{table*}

\newpage

\begin{table*}
\centering
\begin{tabular}{llrrrr}
\hline
Planet name & Detection method & Eccentricity & Period [d] & T$_\text{eff}$ & $V_{mag}$ \\
\hline
TOI-2803 A b&Transit&$0$&$1.96^{+8.2e-07}_{-8.2e-07}$&$6280^{+99}_{-96}$&12.537\\
TOI-2818 b&Transit&$0$&$4.04^{+2.4e-06}_{-2.3e-06}$&$5721^{+88}_{-83}$&11.937\\
TOI-283 b&Transit&$0$&$17.6^{+2e-05}_{-1e-05}$&$5213^{+70}_{-70}$&10.41\\
TOI-286 b&Transit&$0$&$4.51^{+3.1e-06}_{-2.7e-06}$&$5152^{+12}_{-12}$&9.866\\
TOI-286 c&Transit&$0$&$39.4^{+7e-05}_{-8.1e-05}$&$5152^{+12}_{-12}$&9.866\\
TOI-431 b&Transit&$0$&$0.49^{+1e-05}_{-7e-06}$&$4850^{+75}_{-75}$&9.12\\
TOI-431 c&Radial Velocity&$0$&$4.85^{+0.0003}_{-0.0002}$&$4850^{+75}_{-75}$&9.12\\
\hline
\end{tabular}
\caption{Fundamental characteristics of the known exoplanets in the LOPS2 (3. part)}
\label{tab:list_of_known_planets_part3}
\end{table*}

\section{Calculated amplitudes and probabilities}\label{sec:calculated}

\newpage
\begin{table*}
\centering
\rotatebox{90}{
\begin{tabular}{lllcccccccc}
\hline
Planet & M$_{\star}$ [M$_{\odot}$] & M$_{planet}$ [M$_{Earth}$] & A$_{ell}$ [ppm] & A$_{beaming}$ [ppm] & A$_{ref}$ A=0.3 [ppm] & A$_{ref}$ A=0.1 [ppm] & P$_{tr}$ & P$_{occ}$ \\
\hline
DMPP-3 A b* & 0.87 & $2.58^{+0.35}_{-0.58}$ & $0.0035 \pm 0.0023$ & $0.0159 \pm 0.0095$ & $0.157 \pm 0.067$ & $0.0522 \pm 0.022$ & $7.4 \pm 0.66$ & $5.59 \pm 0.38$ \\
GJ 238 b* & 0.419 &  &  &  & $0.388 \pm 0.16$ & $0.129 \pm 0.055$ & $9.34 \pm 0.17$ & $9.34 \pm 0.17$ \\
GJ 3341 b* & 0.47 & $6.6^{+0.1}_{-0.1}$ & $0.000767 \pm 0.0005$ & $0.0889 \pm 0.034$ & $0.522 \pm 0.23$ & $0.174 \pm 0.077$ & $3.1 \pm 0.72$ & $2 \pm 0.35$ \\
HATS-39 b* & 1.38 & $200^{+41}_{-41}$ & $1.3 \pm 0.85$ & $0.652 \pm 0.49$ & $89 \pm 39$ & $29.7 \pm 13$ & $12.2 \pm 2.5$ & $12.3 \pm 2.5$ \\
HATS-40 b* & 1.56 & $505^{+76}_{-76}$ & $13.6 \pm 8.9$ & $1.81 \pm 1.3$ & $145 \pm 64$ & $48.2 \pm 21$ & $21.6 \pm 5.1$ & $21.8 \pm 5.1$ \\
HATS-41 b* & 1.5 & $3.08e+03^{+5.1e+02}_{-5.1e+02}$ & $23.6 \pm 15$ & $10.9 \pm 7.9$ & $36.4 \pm 17$ & $12.1 \pm 5.6$ & $19.1 \pm 5$ & $11 \pm 2.1$ \\
HATS-42 b* & 1.27 & $598^{+48}_{-48}$ & $13.8 \pm 9$ & $3.03 \pm 2.1$ & $166 \pm 72$ & $55.3 \pm 24$ & $17.9 \pm 3.7$ & $17.8 \pm 3.7$ \\
\hline
\end{tabular}
}
\caption{Calculated amplitudes for each effects and transit and occultation probabilities for each record in our table.}
\label{tab:calc1}
\end{table*}

\newpage

\begin{table*}
\centering
\rotatebox{90}{
\begin{tabular}{lllcccccccc}
\hline
Planet & M$_{\star}$ [M$_{\odot}$] & M$_{planet}$ [M$_{Earth}$] & A$_{ell}$ [ppm] & A$_{beaming}$ [ppm] & A$_{ref}$ A=0.3 [ppm] & A$_{ref}$ A=0.1 [ppm] & P$_{tr}$ & P$_{occ}$ \\
\hline
 & 0.988 & $2.57e+03^{+32}_{-32}$ &  & $1.72 \pm 1.1$ &  &  & $0.133 \pm 0.0014$ & $0.11 \pm 0.0004$ \\
 & 0.93 & $3.01e+03^{+2.7e+02}_{-2.7e+02}$ & $3.83e-05 \pm 2.5e-05$ & $2.03 \pm 1.3$ & $0.00758 \pm 0.0033$ & $0.00253 \pm 0.0011$ & $0.13 \pm 0.019$ & $0.13 \pm 0.019$ \\
HD 30177 c* & 0.988 & $2.42e+03^{+9.9e+02}_{-9.9e+02}$ &  & $0.973 \pm 0.65$ &  &  & $0.0496 \pm 0.0083$ & $0.0427 \pm 0.005$ \\
HD 30669 b* & 0.92 & $149^{+19}_{-19}$ & $2.86e-06 \pm 1.9e-06$ & $0.126 \pm 0.08$ & $0.017 \pm 0.0073$ & $0.00565 \pm 0.0024$ & $0.172 \pm 0.035$ & $0.116 \pm 0.014$ \\
HD 33283 b* & 1.24 & 105 & $0.0218 \pm 0.014$ & $0.312 \pm 0.22$ & $3.87 \pm 2$ & $1.29 \pm 0.65$ & $5.62 \pm 1.1$ & $3.76 \pm 0.72$ \\
 & 1.47 & $118^{+13}_{-13}$ & $0.0568 \pm 0.037$ & $0.28 \pm 0.2$ & $16.7 \pm 8.5$ & $5.57 \pm 2.8$ & $6.58 \pm 2.3$ & $6.56 \pm 2.3$ \\
 & 1.38 & $105^{+23}_{-23}$ & $0.0768 \pm 0.05$ & $0.278 \pm 0.2$ & $3.06 \pm 1.4$ & $1.02 \pm 0.48$ & $8.01 \pm 0.96$ & $5.72 \pm 0.55$ \\
\hline
\end{tabular}
}
\caption{Calculated amplitudes for each effects and transit and occultation probabilities for each record in our table.}
\label{tab:calc2}
\end{table*}

\newpage

\begin{table*}
\centering
\rotatebox{90}{
\begin{tabular}{lllcccccccc}
\hline
Planet & M$_{\star}$ [M$_{\odot}$] & M$_{planet}$ [M$_{Earth}$] & A$_{ell}$ [ppm] & A$_{beaming}$ [ppm] & A$_{ref}$ A=0.3 [ppm] & A$_{ref}$ A=0.1 [ppm] & P$_{tr}$ & P$_{occ}$ \\
\hline
HD 51608 c* & 0.8 & $14.3^{+1.6}_{-1.5}$ &  & $0.0357 \pm 0.022$ &  &  & $1.16 \pm 0.049$ & $1.25 \pm 0.07$ \\
HD 55696 b* & 1.29 & $1.23e+03^{+2.3e+02}_{-2.3e+02}$ & $4.73e-05 \pm 3.1e-05$ & $0.99 \pm 0.7$ & $0.026 \pm 0.022$ & $0.00868 \pm 0.0072$ & $0.612 \pm 0.064$ & $0.214 \pm 0.018$ \\
 & 1.29 & $1.36e+03^{+3.2e+02}_{-2.2e+02}$ &  & $1.07 \pm 0.74$ &  &  & $0.37 \pm 0.061$ & $0.143 \pm 0.013$ \\
HD 56414 b* & 1.89 &  &  &  & $1.4 \pm 1.1$ & $0.466 \pm 0.36$ & $7.43 \pm 6.4$ & $7.36 \pm 6.2$ \\
HD 63765 b* & 0.865 & $203^{+16}_{-16}$ &  & $0.312 \pm 0.2$ &  &  & $0.556 \pm 0.03$ & $0.367 \pm 0.011$ \\
 & 0.65 & $168^{+9.5}_{-9.5}$ & $8.11e-05 \pm 5.3e-05$ & $0.336 \pm 0.22$ & $0.2 \pm 0.087$ & $0.0666 \pm 0.029$ & $0.372 \pm 0.065$ & $0.371 \pm 0.065$ \\
HD 64121 b* & 1.64 & $814^{+60}_{-60}$ & $0.0105 \pm 0.0069$ & $0.662 \pm 0.39$ & $0.0467 \pm 0.02$ & $0.0156 \pm 0.0066$ & $1.68 \pm 0.042$ & $1.66 \pm 0.033$ \\
\hline
\end{tabular}
}
\caption{Calculated amplitudes for each effects and transit and occultation probabilities for each record in our table.}
\label{tab:calc3}
\end{table*}

\newpage

\begin{table*}
\centering
\rotatebox{90}{
\begin{tabular}{lllcccccccc}
\hline
Planet & M$_{\star}$ [M$_{\odot}$] & M$_{planet}$ [M$_{Earth}$] & A$_{ell}$ [ppm] & A$_{beaming}$ [ppm] & A$_{ref}$ A=0.3 [ppm] & A$_{ref}$ A=0.1 [ppm] & P$_{tr}$ & P$_{occ}$ \\
\hline
TOI-1338 b* & 1.13 & $33^{+20}_{-20}$ & $0.000321 \pm 0.00021$ & $0.0545 \pm 0.039$ & $0.12 \pm 0.051$ & $0.04 \pm 0.017$ & $1.18 \pm 0.032$ & $1.4 \pm 0.038$ \\
 & 1.13 & $21.8^{+0.9}_{-0.9}$ & $0.000219 \pm 0.00014$ & $0.0367 \pm 0.026$ & $0.0602 \pm 0.026$ & $0.0201 \pm 0.0085$ & $1.33 \pm 0.098$ & $1.33 \pm 0.098$ \\
TOI-1338 c* & 1.13 & $65.2^{+12}_{-12}$ & $0.000128 \pm 8.4e-05$ & $0.0839 \pm 0.058$ & $0.0616 \pm 0.026$ & $0.0205 \pm 0.0088$ & $0.681 \pm 0.03$ & $0.838 \pm 0.04$ \\
TOI-163 b* & 1.44 & $388^{+38}_{-38}$ & $2.82 \pm 1.8$ & $1.28 \pm 0.92$ & $45.2 \pm 19$ & $15.1 \pm 6.4$ & $12 \pm 0.35$ & $12 \pm 0.35$ \\
TOI-1937 A b* & 1.07 & $639^{+54}_{-51}$ & $47.3 \pm 31$ & $5.19 \pm 3.5$ & $285 \pm 1.2e+02$ & $95.1 \pm 40$ & $23 \pm 0.75$ & $23 \pm 0.75$ \\
TOI-199 b* & 0.936 & $54^{+6.4}_{-6.4}$ & $0.000188 \pm 0.00012$ & $0.121 \pm 0.075$ & $0.299 \pm 0.13$ & $0.0998 \pm 0.042$ & $0.8 \pm 0.0049$ & $0.826 \pm 0.0068$ \\
TOI-199 c* & 0.936 & $89^{+6.4}_{-3.2}$ & $4.53e-05 \pm 3e-05$ & $0.145 \pm 0.09$ & $0.131 \pm 0.056$ & $0.0438 \pm 0.019$ & $0.423 \pm 0.0035$ & $0.411 \pm 0.0032$ \\
\hline
\end{tabular}
}
\caption{Calculated amplitudes for each effects and transit and occultation probabilities for each record in our table.}
\label{tab:calc4}
\end{table*}

\newpage

\begin{table*}
\centering
\rotatebox{90}{
\begin{tabular}{lllcccccccc}
\hline
Planet & M$_{\star}$ [M$_{\odot}$] & M$_{planet}$ [M$_{Earth}$] & A$_{ell}$ [ppm] & A$_{beaming}$ [ppm] & A$_{ref}$ A=0.3 [ppm] & A$_{ref}$ A=0.1 [ppm] & P$_{tr}$ & P$_{occ}$ \\
\hline
TOI-4507 b* & 1.11 & 20 & $9.72e-05 \pm 6.4e-05$ & $0.0293 \pm 0.021$ & $0.164 \pm 0.07$ & $0.0548 \pm 0.023$ & $0.912 \pm 0.069$ & $1.12 \pm 0.17$ \\
TOI-451 b* & 0.95 &  &  &  & $2.48 \pm 1.1$ & $0.828 \pm 0.35$ & $14.2 \pm 0.79$ & $14.2 \pm 0.79$ \\
TOI-451 c* & 0.95 &  &  &  & $0.774 \pm 0.33$ & $0.258 \pm 0.11$ & $4.82 \pm 0.28$ & $4.82 \pm 0.28$ \\
TOI-451 d* & 0.95 &  &  &  & $0.619 \pm 0.26$ & $0.206 \pm 0.088$ & $3.25 \pm 0.18$ & $3.25 \pm 0.18$ \\
TOI-4562 b* & 1.19 & $732^{+1.5e+02}_{-1.5e+02}$ & $0.000941 \pm 0.00062$ & $1.37 \pm 0.99$ & $1.97 \pm 2.1$ & $0.657 \pm 0.68$ & $2.48 \pm 0.24$ & $0.51 \pm 0.026$ \\
TOI-4562 c* & 1.19 & $1.83e+03^{+1.2e+02}_{-1.8e+02}$ & $7.52e-06 \pm 4.9e-06$ & $0.765 \pm 0.55$ & $0.00365 \pm 0.0016$ & $0.00122 \pm 0.00052$ & $0.0951 \pm 0.0043$ & $0.0939 \pm 0.0042$ \\
TOI-470 b* & 0.87 &  &  &  & $0.725 \pm 0.31$ & $0.242 \pm 0.1$ & $3.11 \pm 0.23$ & $3.11 \pm 0.23$ \\
\hline
\end{tabular}
}
\caption{Calculated amplitudes for each effects and transit and occultation probabilities for each record in our table.}
\label{tab:calc5}
\end{table*}

\newpage

\begin{table*}
\centering
\rotatebox{90}{
\begin{tabular}{lllcccccccc}
\hline
Planet & M$_{\star}$ [M$_{\odot}$] & M$_{planet}$ [M$_{Earth}$] & A$_{ell}$ [ppm] & A$_{beaming}$ [ppm] & A$_{ref}$ A=0.3 [ppm] & A$_{ref}$ A=0.1 [ppm] & P$_{tr}$ & P$_{occ}$ \\
\hline
 & 1.36 & $368^{+22}_{-22}$ & $21.8 \pm 14$ &  &  &  & $22.9 \pm 0.72$ & $22.9 \pm 0.72$ \\
 & 1.35 & $376^{+20}_{-20}$ & $23.8 \pm 16$ & $1.87 \pm 1.4$ & $368 \pm 1.6e+02$ & $123 \pm 52$ & $23.2 \pm 0.72$ & $23.2 \pm 0.72$ \\
WASP-126 b* & 1.12 & $89^{+13}_{-13}$ & $0.817 \pm 0.53$ & $0.467 \pm 0.32$ & $46.6 \pm 20$ & $15.5 \pm 6.7$ & $12.6 \pm 1.8$ & $12.6 \pm 1.8$ \\
 & 1.12 & 90.3 &  &  & $35.3 \pm 15$ & $11.8 \pm 5$ & $12.1 \pm 2$ & $12.1 \pm 2$ \\
WASP-126 c* & 1.12 & $64.2^{+25}_{-25}$ &  &  & $6.61 \pm 2.8$ & $2.2 \pm 0.94$ & $7.01 \pm 1.1$ & $7.05 \pm 1.1$ \\
WASP-159 b* & 1.41 & $175^{+25}_{-25}$ & $3.4 \pm 2.2$ & $0.68 \pm 0.48$ & $45.1 \pm 19$ & $15 \pm 6.4$ & $17 \pm 0.98$ & $17 \pm 0.98$ \\
WASP-160 B b* & 0.87 & $88.4^{+14}_{-14}$ & $0.327 \pm 0.21$ & $0.605 \pm 0.38$ & $39.8 \pm 17$ & $13.3 \pm 5.6$ & $7.78 \pm 0.37$ & $7.78 \pm 0.37$ \\
\hline
\end{tabular}
}
\caption{Calculated amplitudes for each effects and transit and occultation probabilities for each record in our table.}
\label{tab:calc6}
\end{table*}

\end{appendices}

\end{document}